\documentclass[a4paper,11pt]{article}
\usepackage{jheppub} 
\usepackage{xcolor}
\usepackage{amsmath,soul}
\usepackage{comment}
\usepackage{subfigure}
\usepackage{graphicx}
 
\usepackage{tikz}
\usepackage{appendix}

\title{\boldmath Holographic Extended Thermodynamics of deformed AdS-Schwarzschild black hole}

\author[]{Kamal L. Panigrahi,}
\author[]{Balbeer Singh}
\affiliation{Department of Physics,\\
Indian Institute of Technology Kharagpur,\\
Kharagpur 721 302, India.}

\emailAdd{panigrahi@phy.iitkgp.ac.in}
\emailAdd{curiosity1729@kgpian.iitkgp.ac.in}
\abstract{
We investigate the thermodynamics and phase structure of the deformed AdS-Schwarzschild black hole, generated via the gravitational decoupling (GD) method.
In the bulk canonical ensemble, our results exhibit a van der Waals-type first-order phase transition in addition to the Hawking-Page transition, in the suitable parameter regime. Further, we compute the critical exponents characterising the bulk transition, confirming their consistency with mean-field theory predictions. Exploiting the exact holographic dictionary between extended black hole thermodynamics and the dual conformal field theory (CFT), we extend this analysis to the boundary and uncover a rich array of phase transitions and critical phenomena across three distinct thermodynamic ensembles. In particular, in the fixed 
$(\mathcal{V},C)$ ensemble, the dual CFT exhibits a Hawking–Page–type transition. However, in the fixed 
$(p,C)$ ensemble, the deformation parameter leads to a distinct thermodynamic behaviour in which multiple branches become unstable, leaving a single thermodynamically stable phase, thus marking a clear departure from the standard van der Waals scenario.
Throughout, we emphasise the pivotal influence of the GD deformation parameter on the thermodynamic behaviour, and we elucidate its role in the confinement-deconfinement transitions characteristic of the deformed AdS-Schwarzschild geometry.
}

\makeatletter
\gdef\@fpheader{}
\makeatother

\begin{document}
\maketitle
\flushbottom
\section{Introduction}
Black hole thermodynamics \cite{Bekenstein:1973ur,Bardeen:1973gs,Wald:1999vt,Witten:2024upt} finds broad applicability beyond the framework of Einstein gravity with particular interest in phase transitions emerging from modified gravity theories~\cite{Kastor:2010gq,Zou:2013owa}. 
Numerous studies in the literature have revealed novel phenomena and a rich phase structure~\cite{Altamirano:2013ane,Altamirano:2013uqa,Cai:2013qga,Xu:2014tja,Dolan:2014vba,Zou:2016sab}. The non-linear nature of gravitational field equations motivates approximate and analytical techniques, among which the gravitational decoupling (GD) method 
has proven effective for constructing black hole solutions in extended settings~\cite{Ovalle:2017fgl,Ovalle:2018gic,Ovalle:2023ref}. The GD method facilitates the extension of the standard gravitational action to include additional matter sources and explore broader classes of modified gravity theories. In a recent study \cite{Khosravipoor:2023jsl}, the authors applied the GD method realised through the minimal geometric deformation by taking the AdS-Schwarzschild vacuum as the seed solution and constructed a new geometry which is referred to as the deformed AdS-Schwarzschild black hole, in the presence of an additional gravitational source that satisfies the weak energy condition. They also analysed the horizon structure, thermodynamics of the deformed black hole and the Hawking-Page phase transition for the suitable choice of the parameters. 
In this work, we begin by extending the analysis to explore the van der Waals (vdW)–type behaviour in the deformed AdS-Schwarzschild black hole. 
We give concrete evidence that the deformed AdS-Schwarzschild black hole indeed exhibits the vdW liquid-gas phase transition. 

The holographic principle \cite{Susskind:1994vu,Gubser:1998bc,Witten:1998qj} has significantly influenced modern theoretical physics by establishing a correspondence between gravitational dynamics in a higher-dimensional bulk and quantum field theories on the boundary. 
A prominent and well-studied realization of this duality is the anti-de Sitter/conformal field theory (AdS/CFT) correspondence \cite{Maldacena:1997re}, wherein the thermal properties of black holes in asymptotically AdS spacetime are mapped onto those of a CFT at finite temperature. This line of investigation was pioneered by Witten, who demonstrated that the Hawking-Page first-order phase transition \cite{Hawking:1982dh} between thermal AdS space at low temperatures and the AdS–Schwarzschild black hole at high temperatures corresponds to a confinement–deconfinement transition in the boundary $\mathcal{N}=4$ supersymmetric Yang–Mills (SYM) theory with gauge group $SU(N_c)$  \cite{Witten:1998zw}.

In the extended version of black hole thermodynamics, the (negative) cosmological constant $\Lambda$ is promoted to a dynamical quantity and interpreted as a (positive) thermodynamic pressure via the relation \cite{Kastor:2009wy,Dolan:2011xt,Dolan:2010ha,Cvetic:2010jb,Kubiznak:2014zwa} (in $D= d+1= \# \text{spacetime dimensions}=4$) 
\begin{align}
    P = -\frac{\Lambda}{8 \pi G_{N}} 
     ,\qquad \Lambda = -\frac{d(d-1)}{2 l^2}=- \frac{3}{l^2},
\end{align}
where $l$ is the AdS curvature radius, and $G_{N}$ is the Newton's constant.
This reinterpretation introduces a new pressure–volume term, $+V\,dP$, into the first law of bulk thermodynamics \cite{Kastor:2009wy,Kastor:2014dra,Cong:2021fnf}. This is known as the \textit{black hole chemistry} \cite{Kubiznak:2016qmn,Karch:2015rpa}. 
In particular, the thermodynamic variables associated with black holes—namely energy $E$, entropy $S$, and temperature $T_{H}$—are given (in units where $\hbar = c = k_B = 1$) by
\begin{align}
    E = M, \qquad S = \frac{A}{4 G_{N}}, \qquad T_{H} = \frac{\kappa}{2 \pi},
\end{align}
where $A$ is the horizon area, and $\kappa$ is the surface gravity at the event horizon. These quantities correspond to the energy, entropy, and temperature of thermal states in the dual boundary field theory, as identified via the generalised dictionary \cite{Cong:2021jgb,Visser:2021eqk}:
\begin{align}
    \mathcal{E} = \frac{M}{\omega}, \qquad \mathcal{S} = S, \qquad T = \frac{T_H}{\omega}, \qquad C \sim \frac{l^{d-1}}{G_{N}}, \qquad \mathcal{V} \sim \omega^{\,d-1} l^{d-1}
    \label{bulk-boundary-duality}
\end{align}
where $\omega \equiv \frac{R}{l}$ denotes the appropriate conformal factor relating bulk and boundary time coordinates. This makes the
boundary volume $\mathcal{V}$ and the central charge $C$ completely independent, since the volume depends on R and the central charge on $l$ \cite{Cong:2021jgb}.
Here, the central charge $C$ characterises the number of degrees of freedom in the dual theory and serves as a thermodynamic variable, with $\mu$ acting as its conjugate chemical potential. 

 The purpose of this paper is to conduct an extensive investigation of the thermodynamics of the deformed AdS-Schwarzschild black hole in the extended phase space and the dual CFT thermodynamics, applying the formalism of the holographic extended thermodynamics outlined in \cite{Cong:2021jgb} for the charged AdS black holes. We construct and analyse the thermodynamics of the deformed AdS–Schwarzschild solution obtained via gravitational decoupling \cite{Khosravipoor:2023jsl}. We compute the bulk phase structure, identify van der Waals–type behaviour and the Hawking–Page transition, and determine the critical exponents. On the holographic side, we map the bulk thermodynamics to boundary variables, compare three physically relevant boundary ensembles, and clarify the role of the deformation parameters \(\xi,\eta\). For concreteness (and to remain consistent with the approach followed in \cite{Khosravipoor:2023jsl}) we treat \(\xi,\eta\) as fixed deformation parameters in the main analysis.

On the bulk side, the thermodynamic behaviour of the deformed AdS-Schwarzschild black hole in the extended phase space is critically regulated by the deformation parameter $\xi$ in a canonical ensemble, with distinct phase transitions emerging under specific parameter choices. When the control parameter $\eta \neq 0$, numerical analysis reveals two critical values, $\xi_{c_{1}}$ and $\xi_{c_{2}}$, bounding an interval $\xi \in (\xi_{c_{1}}, \xi_{c_{2}})$ where the system exhibits a first-order van der Waals (vdW) phase transition. This is characterized by a non-monotonic temperature profile (Fig.~\ref{T-rh-new}, left panel) and the canonical swallowtail structure in the $P$–$v$ diagram (Fig.~\ref{P-v-Tc1-Tc2}). Outside this interval ($\xi < \xi_{c_{1}}$ or $\xi > \xi_{c_{2}}$), no vdW transition is observed.
Moreover, when the deformation is turned off, we observe no vdW phase structure (Fig.~\ref{P-v-not-vdW}, right panel). In the limit $\eta = 0$, the solution resembles to a charged AdS black hole, where $\xi$ acts as the square of the electric charge. Here, a single critical value $\xi_c = 6.25$ triggers a standard vdW transition (Fig.~\ref{P-v-not-vdW}, left panel), similar to the charged-AdS-like. Conversely, when $\xi < \xi_{c_{1}}$ (with $\eta \neq 0$), the system undergoes a Hawking-Page-type transition between thermal radiation and black hole phases (Fig.~\ref{beta-1-lit}). Near criticality, the critical exponents universally satisfy $\alpha = 0$, $\beta \approx 1/2$, $\gamma \approx 1$, $\delta = 3$, matching mean-field theory predictions for vdW system.

In the dual boundary side, we first obtain the various thermodynamical quantities on the dual CFT states by assuming the variation of the cosmological constant. At the same time we allow the central charge $C$ to vary.
The boundary thermodynamics of the dual conformal field theory exhibits distinctive phase behaviour governed by the dimensionless deformation parameter $\tilde{\xi} \equiv \xi/l^{2}$. In the fixed-volume and central-charge ensemble $(\mathcal{V}, C)$, a Hawking-Page transition occurs when $\tilde{\xi} < \tilde{\xi}_c$ for fixed $b \equiv \eta/l$, manifesting as characteristic feature in free energy $F$ (Fig.~\ref{CFT-vary-para-alpha-b}). Increasing $\tilde{\xi}$ elevates both $F$ and the Hawking temperature $T_{\rm HP}$ (Fig.~\ref{CFT-vary-para-alpha-b}, right panel), while increasing $b$ suppresses these quantities (Fig.~\ref{CFT-vary-para-alpha-b}, left panel). In the undeformed limit ($\tilde{\xi} \to 0$), we recover the standard AdS-Schwarzschild result $T_{\rm HP} = (\pi R)^{-1}$. 
The ensemble $(p, C)$ exhibits a novel phase transition feature with 
Gibbs free energy $G$ (Figs.~\ref{new-G-T-plot}--\ref{G-T-bdry-all-diff}) having a single stable phase confined exclusively to the intermediate-entropy branch, revealed from the specific-heat plots (Fig. \ref{C-plots-G-T-last}). Further, this departure from the swallowtail-like structure is strengthened from the $(p,\mathcal{V})$ plots (Fig. \ref{p-V-for-G-T}), which do not qualitatively match with the vdW phase transition. \\

The rest of the paper is organised as follows. In Section \ref{review}, we review the deformed AdS Schwarzschild black hole solution and its associated first law. In Section~\ref{bulk thermo}, we establish the existence of a van der Waals-like liquid–gas phase transition in the bulk and compute the universal ratio, and discuss the Hawking–Page transition; 
Section~\ref{exponent} presents the computation of the critical exponents, shown to match those of the van der Waals system. In Section~\ref{boundary thermo}, we study the thermodynamics of the holographic boundary ensembles dual to these black holes: we define the boundary thermodynamic variables, analyse the phase structure across three physically relevant ensembles. 
Section~\ref{conclusion} summarises our conclusions. Unless stated otherwise we set \(k_B=\ell_P=1\).

\section{Deformed AdS-Schwarzschild black hole and first law }\label{review}
In this section, we will briefly review the deformed AdS-Schwarzschild black hole solution based on paper \cite{Khosravipoor:2023jsl}.
The bulk action is given as
\begin{equation}\label{action}
S \;=\; \int d^4x\sqrt{-g}\Big(\frac{1}{2\kappa}(R-2\Lambda)+\mathcal L_m+\mathcal L_X\Big)\,,\qquad \kappa=8\pi G_N,
\end{equation}
where $\mathcal{L}_{m}$ is the usual matter Lagrangian and \(\mathcal L_X\) is an additional matter sector (e.g. Lovelock gravity and/or new other scalar/vector/tensor
fields). On employing the gravitational decoupling framework to construct an analytic family of deformed AdS--Schwarzschild solutions \cite{Khosravipoor:2023jsl} 
\begin{equation}
ds^2
=
-\,F(r)\,dt^2
+ \frac{1}{F(r)}\,dr^2
+ r^2\bigl(d\theta^2 + \sin^2\theta\,d\phi^2\bigr),
\label{metric}
\end{equation}
where, 
\begin{equation}
F(r)\equiv\;1 - \frac{2M}{r}
+ \frac{r^2}{l^2}
+ \xi\,
\frac{\eta^2 + 3r^2 + 3\eta r}{3\,r\,(\eta + r)^3}
\;.
\end{equation}
Here, $M$ represents the ADM mass, $l$ is the AdS radius, and $\xi$ and $\eta$ are parameters arising from the GD method. The deformation parameter $\xi$, a positive constant, modifies the standard AdS-Schwarzschild geometry,
with $ \xi = 0 $ restoring the undeformed case.
The energy density associated to the $\mathcal{L}_{X}$ takes the profile \(E_X(r)=\xi/[\kappa(\eta+r)^4]\). For generic \(\eta\), this extra sector is \emph{not} the Maxwell field; hence \(\xi\) and \(\eta\) should be regarded as deformation parameters of the theory rather than conserved charges.
In the special limit \(\eta\to0\) the stress tensor of \(\mathcal L_X\) becomes traceless and the metric function reduces to the Reissner--Nordström--AdS like form
\begin{equation}\label{rn-limit}
F(r)\xrightarrow{\;\eta\to0\;}1-\frac{2M}{r}+\frac{r^2}{l^2}+\frac{\xi}{r^2},
\end{equation}
so that \(\sqrt{\xi}\) takes the usual electric charge \(Q\) (i.e. \(\xi=Q^2\)).
We emphasise that the identification \(\xi\leftrightarrow Q^2\) in \eqref{rn-limit} is a matching of asymptotic falloffs; it does not imply the original action \eqref{action} contains a Maxwell term. 
Beyond producing new exact solutions, deformations of AdS black holes obtained via gravitational decoupling provide a controlled laboratory to probe how non-Maxwell matter sources alter both bulk thermodynamic universality and the holographically dual finite-temperature dynamics. From the bulk perspective, such deformations test the robustness of the van der Waals/Hawking–Page phenomenology when the stress tensor departs from standard Maxwell falloffs and when additional length scales (here set by $\eta$) are present. From the boundary viewpoint, these geometries model deformations of the dual CFT state 
allowing one to study how confinement–deconfinement phenomena and central-charge-dependent thermodynamics respond to non-trivial operator profiles. Moreover, because the $\eta\!\to\!0$ limit resembles RN–AdS asymptotics, the model interpolates between charged-like and genuinely non-Maxwell deformations, making it an ideal setting to study them at the same time under a unified treatment.
We note that one may formally include deformation conjugates in the extended first law,
however, if $\xi,\eta$ are regarded as fixed parameters (the choice adopted in this work and in \cite{Khosravipoor:2023jsl}), the first law reduces to the usual form
\[
dM = T_H\,dS + V\,dP  \qquad(\xi,\eta\ \text{fixed}),
\]
where the thermodynamic quantities $T_H,S,V$ are given by the corresponding partial derivatives of $M$ with $\xi,\eta$ held fixed,
\[
T_H=\Big(\frac{\partial M}{\partial S}\Big)_{P},\qquad
V=\Big(\frac{\partial M}{\partial P}\Big)_{S}    \qquad(\xi,\eta\ \text{fixed})
\]
The deformation contributions instead appear in the Smarr scaling relation as additive (state-dependent) terms (see Appendix \ref{app:smarr}),
\[
M=2T_H S - 2PV + 2\xi\Psi_\xi + \eta\Psi_\eta,
\]
and are thereby present as numbers when $\xi,\eta$ are fixed. If one instead promotes $\xi,\eta$ to thermodynamic variables, $\Psi_\xi,\Psi_\eta$ must be treated as conjugate variables and the same partial-derivative definitions apply with the appropriate held-fixed variables \cite{Khosravipoor:2023jsl}. 

Using the holographic dictionary \eqref{bulk-boundary-duality},
the extended CFT first law \cite{Ahmed:2023snm,Cong:2021jgb} becomes 
\begin{align} \label{boundary-law}
    d \mathcal{E} &= T \, d S + \mu \, d C - p \, d \mathcal{V},
\end{align}
with the chemical potential (Euler relation for holographic CFTs)
\begin{align} 
     \mu &= \frac{1}{C} (\mathcal{E} - T S ),
     \label{mu-bdry}
\end{align}
where \cite{Ahmed:2023dnh}
\begin{align} 
    C \sim
    \frac{l^2}{ G_N},
    \label{C-eq}
\end{align}
and the equation of state for conformal theories is
\begin{align}
p &= \frac{\mathcal{E}}{2\mathcal{V}}.
\end{align}
In the following section, we initiate the holographic analysis of the deformed AdS–Schwarzschild black hole by first examining its bulk thermodynamics.

\section{ Extended thermodynamics for the deformed AdS-Schwarzschild black hole}\label{bulk thermo}
In this section, we first derive the relevant thermodynamic quantities in the canonical ensemble and demonstrate the presence of a first-order phase transition of van der Waals type. Additionally, we show that the system exhibits a Hawking–Page phase transition under appropriate conditions, as also discussed in \cite{Khosravipoor:2023jsl}. 
\subsection{Thermodynamic quantities}
We begin by reviewing the thermodynamic properties of the deformed AdS–Schwarzschild black hole. Imposing the horizon condition $F(r_h) = 0$, the black hole mass can be expressed in terms of the horizon radius $r_h$ as
\begin{align}
    M = \frac{1}{6} \left[
    3 r_h + \frac{3 r_h^3}{l^2} +
    \xi \frac{\eta^2 + 3 r_h^2 + 3 \eta r_h}{(\eta + r_h)^3}
    \right].
    \label{mass-bulk}
\end{align}
The entropy $S$ is computed via the Bekenstein–Hawking area law,
\begin{align}
    S = \frac{A}{4 l_{\text{P}}^2} = \pi r_h^2,
    \label{entropy-bulk}
\end{align}
where the Planck length $l_P$ is taken in natural units. 
The corresponding expression for the Hawking temperature is given by
\begin{align}
    T_{H} &=  \left. \frac{F'(r)}{4\pi} \right|_{r = r_h} \nonumber \\
    &=\frac{1}{12 \pi  r_{h}^2} \Big[ \frac{6 r_h^3}{l^2}-\frac{\xi  \left(\eta ^3+4 \eta ^2 r_h+6 \eta  r_h^2+6 r_h^3\right)}{\left(\eta +r_h\right){}^4}+6 M \Big] \nonumber \\
    &= \frac{1+r_{h}^2 \left(\frac{3}{l^2}-\frac{\xi }{(\eta +r_{h})^4}\right)}{4 \pi  r_{h}}. \label{temp}
\end{align}
The pressure associated with the AdS background, defined in terms of the (negative) cosmological constant, takes the form
\begin{align}
    P = \frac{3}{8 \pi l^2},
\end{align}
therefore, 
\begin{align}
     P = \frac{4 \pi r_h T_{H} - 1 + \frac{\xi r_h^2}{(\eta + r_h)^4}}{8 \pi r_h^2}.
\end{align}
Following \cite{Kubiznak:2012wp}, we identify the geometric quantities $P$ and $T_{H}$ with the physical pressure and temperature of the system by using dimensional analysis
\begin{align}
    [\text{Press}] = \frac{\hbar c}{l_{p}^{2}}[P], \qquad [\text{Temp}] = \frac{\hbar c}{k}[T_{H}],
\end{align}
with Planck length $l_{P}^2= \frac{\hbar G_{N}}{c^3}$.
Therefore, the physical pressure and physical temperature are given by
\begin{align}
    [\text{Press}]&= \frac{\hbar c}{l_{P}^{2}} P\\
    &= \frac{\hbar c}{l_{P}^{2}} \frac{1}{2 r_{h}} T_{H}+..\\
    &= k \frac{1}{l_{P}^{2} 2 r_{h}} [\text{Temp}] + ...
\end{align}
Now, one could compare them with the van der Waals equation, and identify the specific volume $v$ of the fluid with the horizon radius as $v= 2  r_{h}$ (in units $l_{p}=1$). Then the $P$ can be re-expressed as
\begin{align}
    P= \frac{T_{H}}{v} - \frac{1}{2 \pi v^2} + \frac{\xi}{8 \pi \left( \eta + \frac{v}{2} \right)^4},
     \label{P-v}
\end{align}
which we analyse numerically due to its analytical complexity in the next section. 
 Thus, after some algebra, the first law of black hole thermodynamics in the extended phase-space becomes as \cite{Khosravipoor:2023jsl}, 
\begin{align}
     dM=T_{H}dS+ V dP ,
 \end{align}
The thermodynamic volume $V$ is obtained as
\begin{align}
    V= \frac{4}{3}\pi r_{h}^3 \,.
\end{align}
\subsection{Phase transition}
\begin{figure}[!h]
\begin{subfigure}
\centering
    \includegraphics[width=0.5\linewidth]{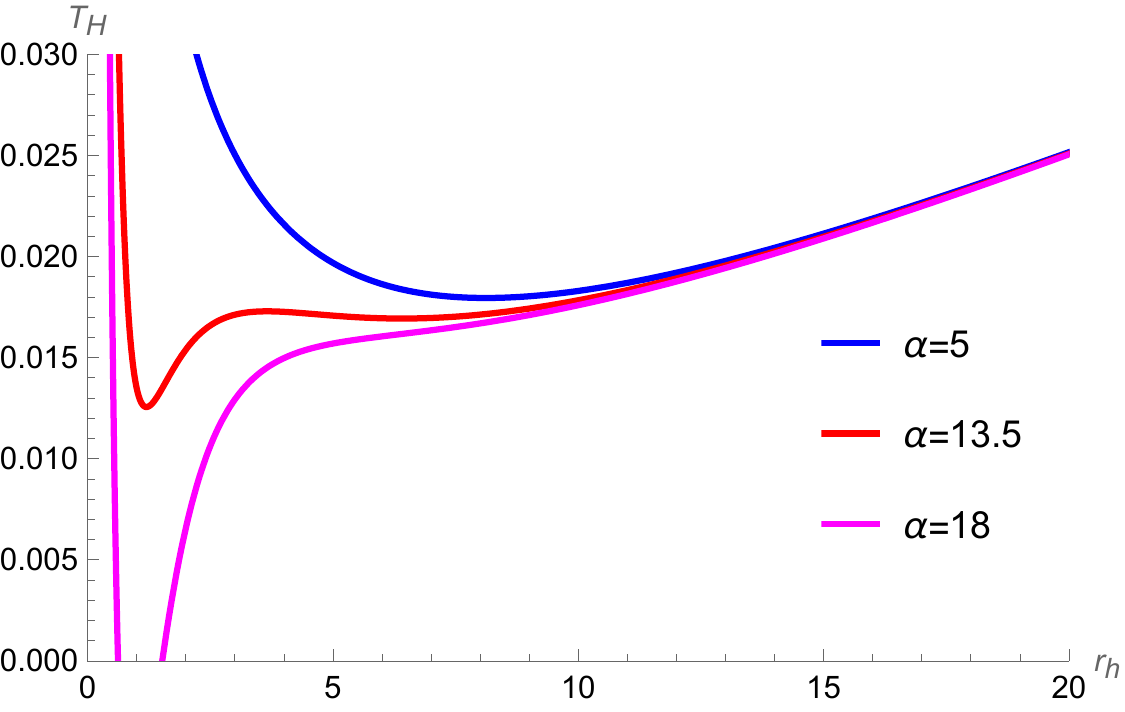}
\end{subfigure}
    \begin{subfigure}
    \centering
   \includegraphics[width=0.5\linewidth]{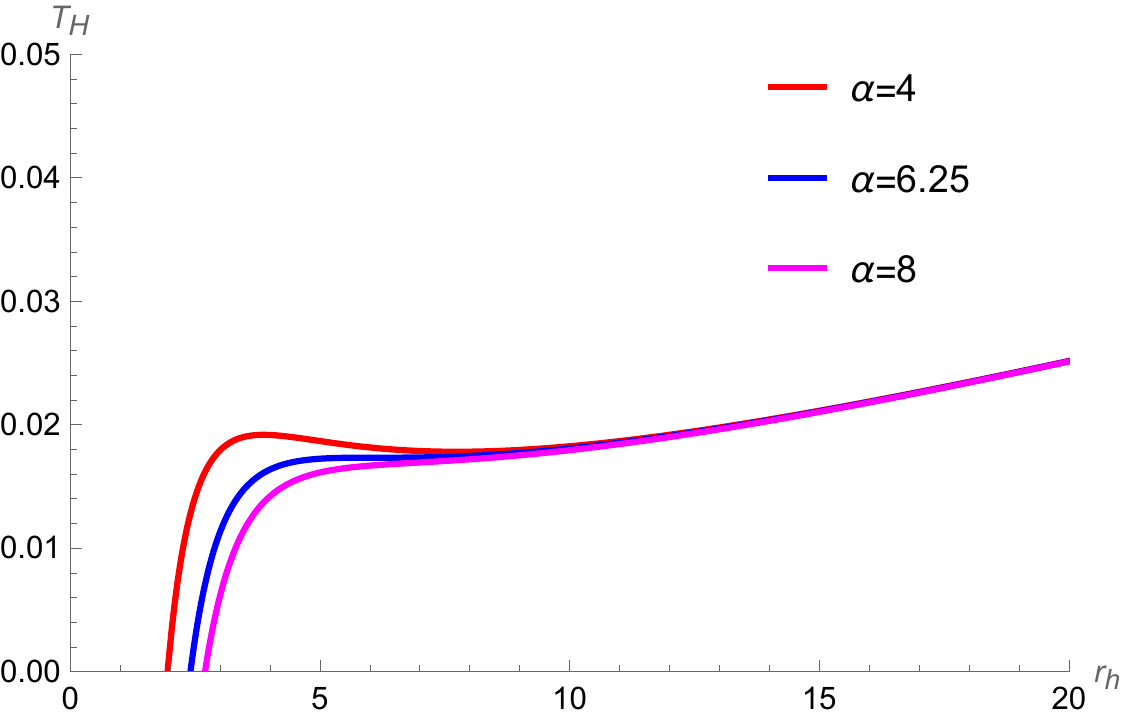}
\end{subfigure}
     \caption{ 
    Plot of the $T$–$r_{h}$ relation for two distinct scenarios. \textbf{(Left)}: For $\eta = 1$, two critical values of the deformation parameter are observed:  $\xi_{c_{1}} = 10.3$ and $\xi_{c_{2}} = 16.2$.
    The red curve exhibits two inflection points, the blue curve exhibits one, and the magenta curve corresponds to the case with no inflection points. Here,  $\xi_{c}^{min} = 11.4998$ and $\xi_{c}^{max} = 14.8394$. \textbf{(Right)}: For $\eta = 0$, the critical value is $\xi_{c} = 6.25$. In both cases, the AdS curvature scale is fixed at $l = 15$. 
}
    \label{T-rh-new}
\end{figure}

We investigate the relation between the Hawking temperature $T_H$ and the horizon radius $r_h$ to identify the critical points. This involves solving the conditions
\begin{align}
    \frac{dT_{H}}{dr_h} = 0, \qquad \frac{d^2 T_{H}}{dr_h^2} = 0.
\end{align}
Due to the complexity of the resulting expressions, we resort to numerical methods. The critical horizon radius $r_{hc}$ and the corresponding critical deformation parameter $\xi_{c}$ are found to depend on the specific values of the AdS radius $l$ and the parameter $\eta$. Fig~\ref{T-rh-new} illustrates representative cases exhibiting the two characteristic scenarios: $\eta \neq 0$ and $ \eta=0$.
It has been observed that for $\eta \neq 0$, the temperature profile may exhibit one, two, or no inflection points depending on the value of the deformation parameter $\xi$, as illustrated in Fig.~\ref{T-rh-new}(Left). For fixed $\eta$ and $l$, there exist two threshold (critical) values of $\xi$, denoted by $\xi_{c_{1}}, \xi_{c_{2}}$, between which the profile develops two inflection points $\xi_{c}^{\text{min}}$ and $\xi_{c}^{\text{max}}$. For $\xi < \xi_{c_{1}}$, the temperature decreases to a minimum and then increases monotonically. Conversely, for $\xi > \xi_{c_{2}}$, the profile displays a single local maximum; however, it also goes into unphysical regions where the temperature becomes negative for certain values of the horizon radius $r_{h}$. These unphysical branches have been disregarded in our analysis.
In contrast, for the case $\eta = 0$, a single critical value $\xi_{c}$ is observed, as illustrated in Fig.~\ref{T-rh-new} (Right).
\newline
\begin{figure}
    \centering
    \includegraphics[width=0.6\linewidth]{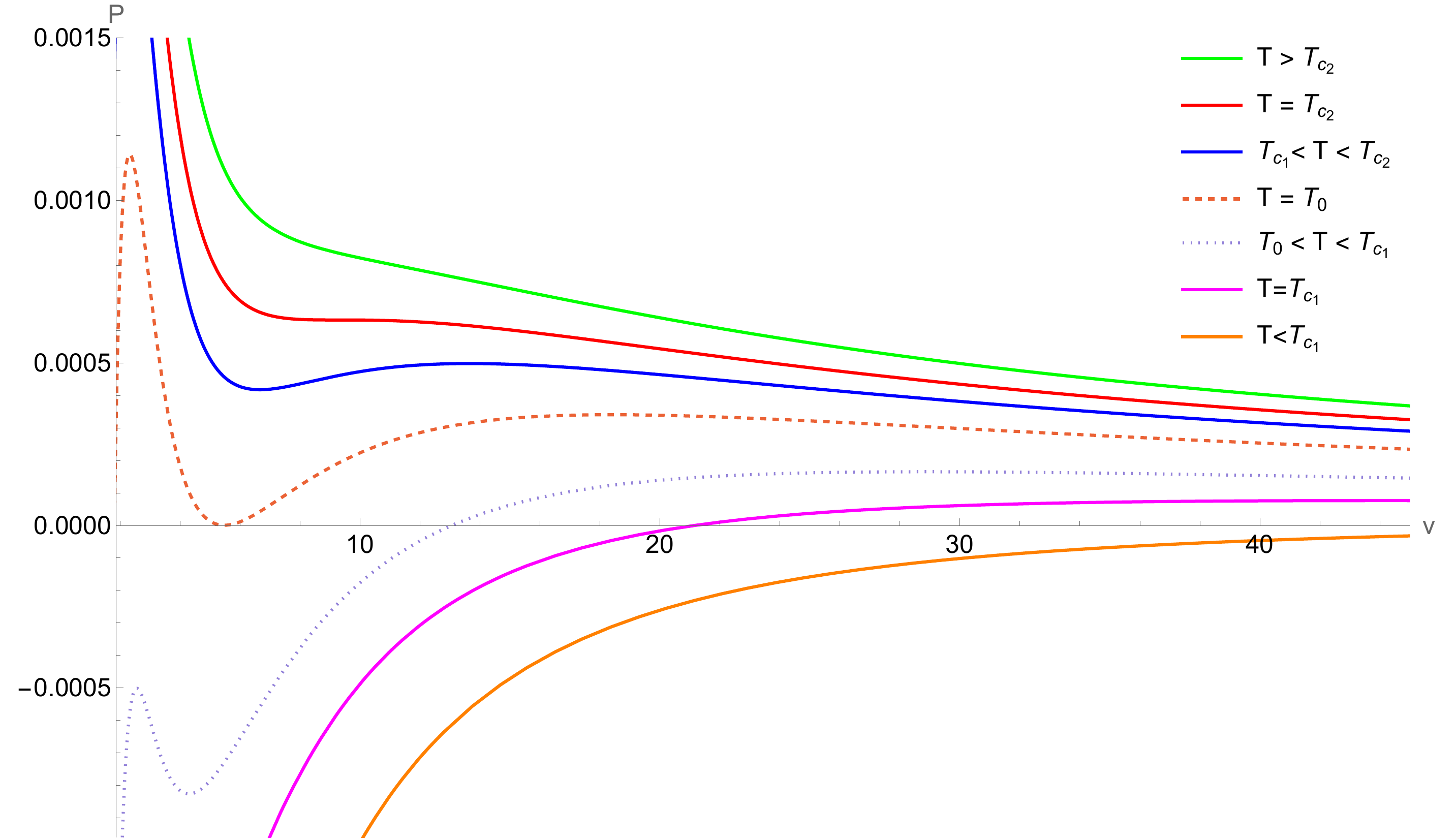}
    \caption{Without loss of generality, for fixed values of $\xi = 13.5$ and $\eta = 1$, there exist two critical temperatures, $T_{c_{1}} = 0.0068$ and $T_{c_{2}} = 0.0180$, such that the system exhibits a van der Waals-like phase transition only in the intermediate range $T \in (T_{c_{1}}, T_{c_{2}})$. Outside this interval, i.e., for $T < T_{c_{1}}$ or $T > T_{c_{2}}$, no first-order phase transition occurs. The characteristic swallowtail structure in the $P-v$ diagram appears within this temperature window. Furthermore, there exists a lower threshold temperature, $T_{0} = 0.014$, below which the pressure $P$ starts to become negative, i.e., for $T \leq T_{0}$ within some range of $v$ as shown. 
    }
    \label{P-v-Tc1-Tc2}
\end{figure}
We numerically analyse the equation of state \eqref{P-v}. For a fixed pair of parameters $(\xi, \eta)$, the critical points are determined by the conditions
\begin{align}
    \frac{dP}{dv} = 0, \qquad \frac{d^2P}{dv^2} = 0,
\end{align}
which yield two distinct critical temperatures, denoted as $T_{c_{1}}$ and $T_{c_{2}}$. The associated $P$–$v$ diagram for the deformed AdS–Schwarzschild black hole is shown in Fig.~\ref{P-v-Tc1-Tc2}. As evident from the plot, for a fixed value of the deformation parameter $\xi$, there exist two critical temperatures, $T_{c_{1}}$ and $T_{c_{2}}$, such that a van der Waals-like first-order phase transition occurs within the temperature range $T \in (T_{c_{1}}, T_{c_{2}})$. Outside this interval, i.e., for $T < T_{c_{1}}$ or $T > T_{c_{2}}$, no first-order phase transition is observed. A similar qualitative behaviour arises when the temperature is held fixed and $\xi$ is varied instead. Additionally, there exists a temperature $T_0$, below which the pressure $P$ becomes negative, as illustrated by the three lowermost curves in Fig.~\ref{P-v-Tc1-Tc2}.

\begin{figure}[!h]
   \begin{subfigure}
       \centering
    \includegraphics[width=0.5\linewidth]{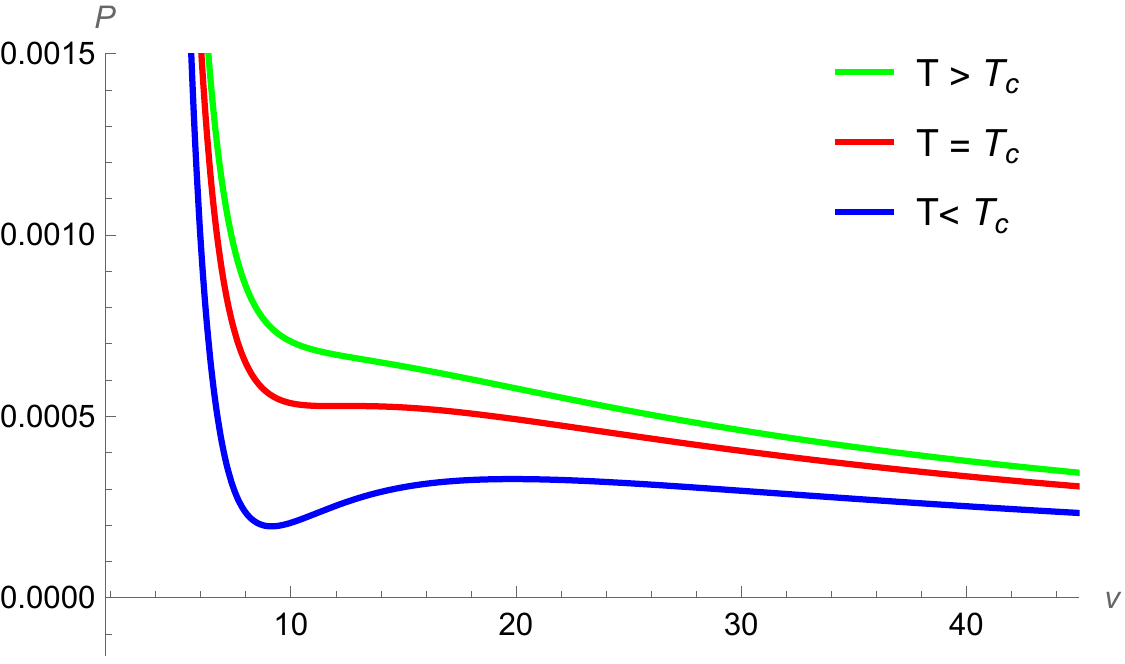}
   \end{subfigure}
   \begin{subfigure}
        \centering
    \includegraphics[width=0.5\linewidth]{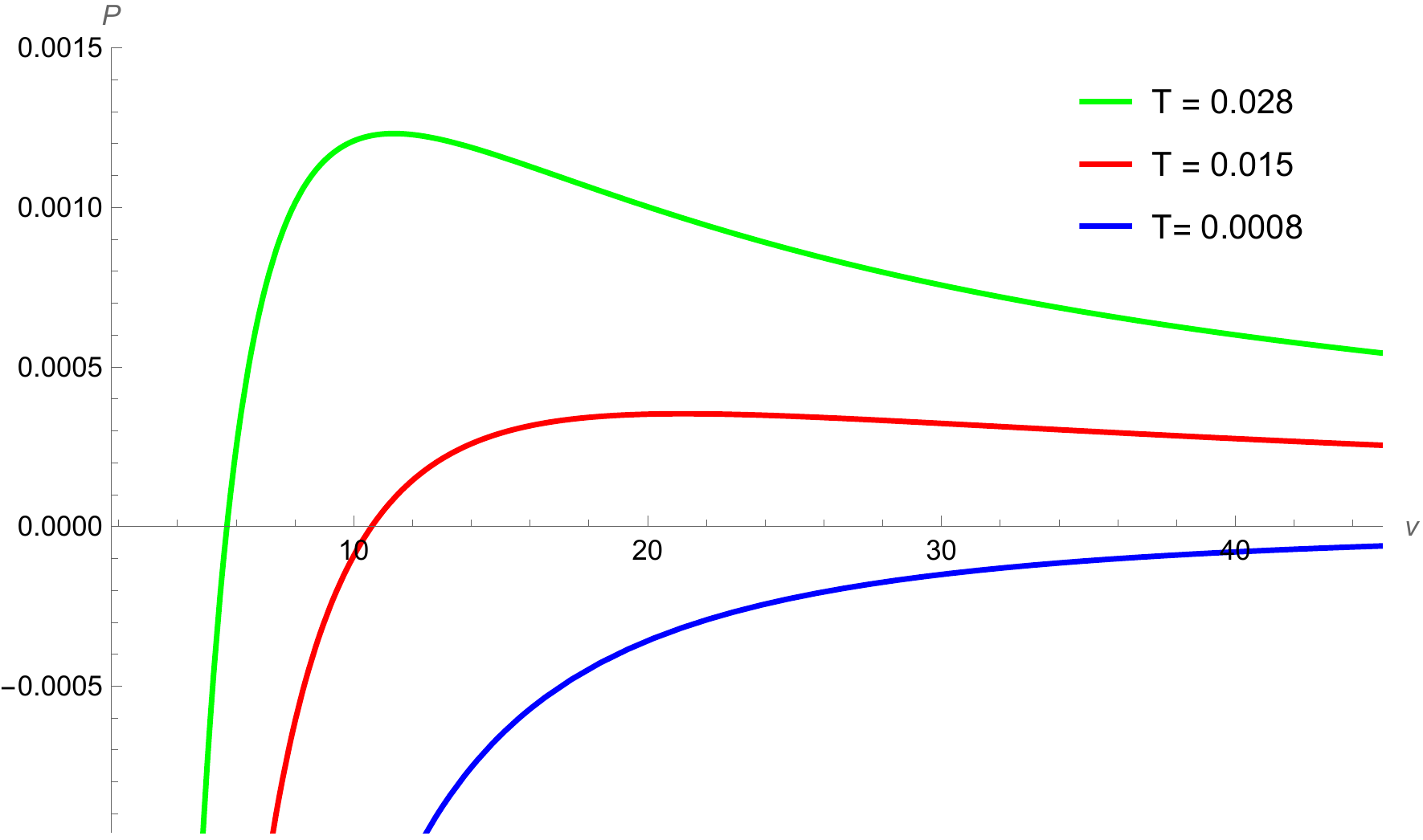}
   \end{subfigure}
    \caption{\textbf{(Left)}: $P$--$v$ diagram for the charged AdS black hole with $\xi = 6.25$, $\eta = 0$, showing a van der Waals-like first-order phase transition at the critical temperature $T_{c} = 0.0173$, as expected.  
\textbf{(Right)}: $P$--$v$ diagram for the standard AdS-Schwarzschild black hole ($\xi = 0$) at three different temperatures with $\eta = 1$, illustrating the absence of van der Waals phase transition. 
}
\label{P-v-not-vdW}
\end{figure}

\begin{figure}[!h]
    \begin{subfigure}
        \centering
        \includegraphics[width=0.5\linewidth]{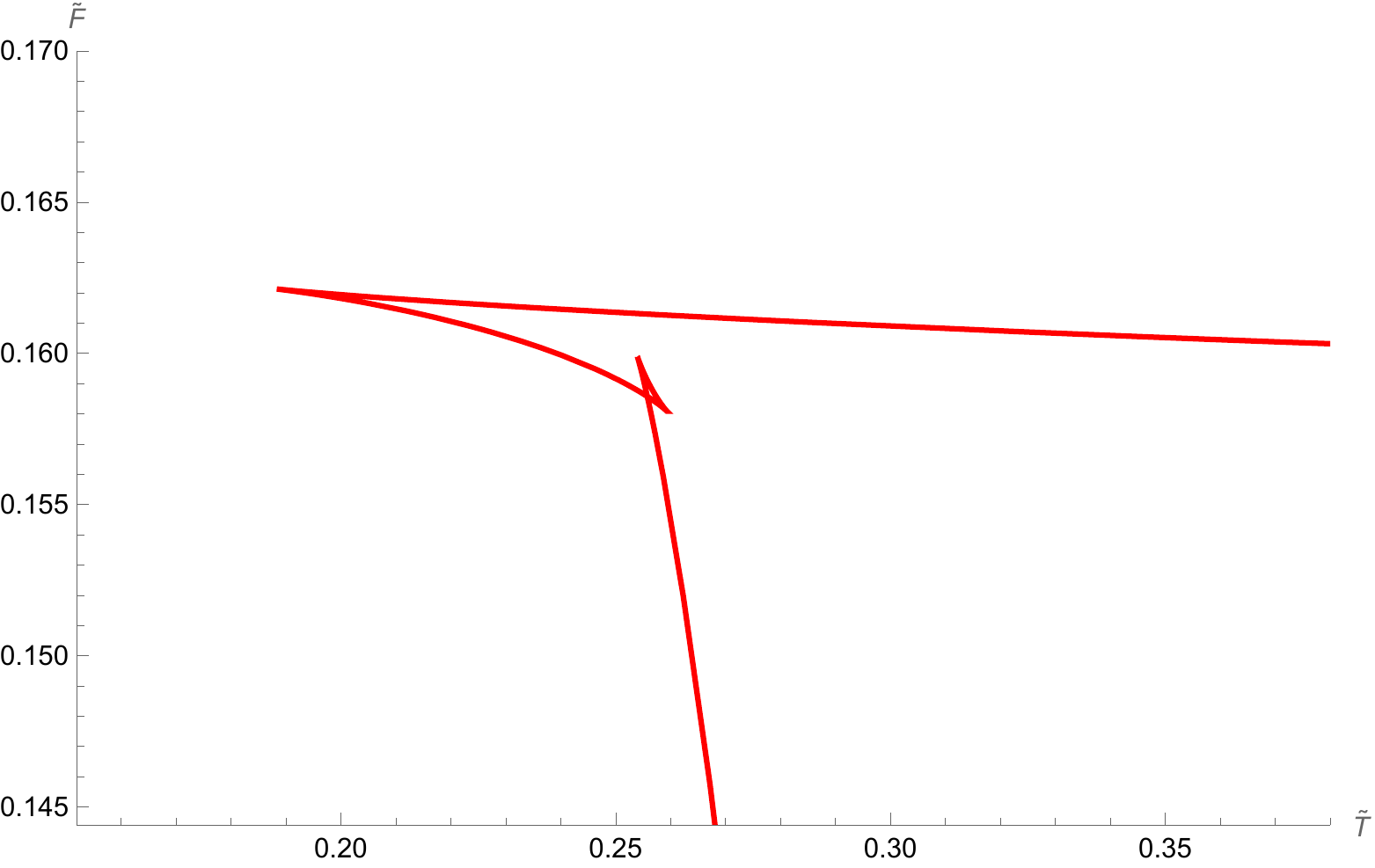}
    \end{subfigure}
     \begin{subfigure}
        \centering
        \includegraphics[width=0.5\linewidth]{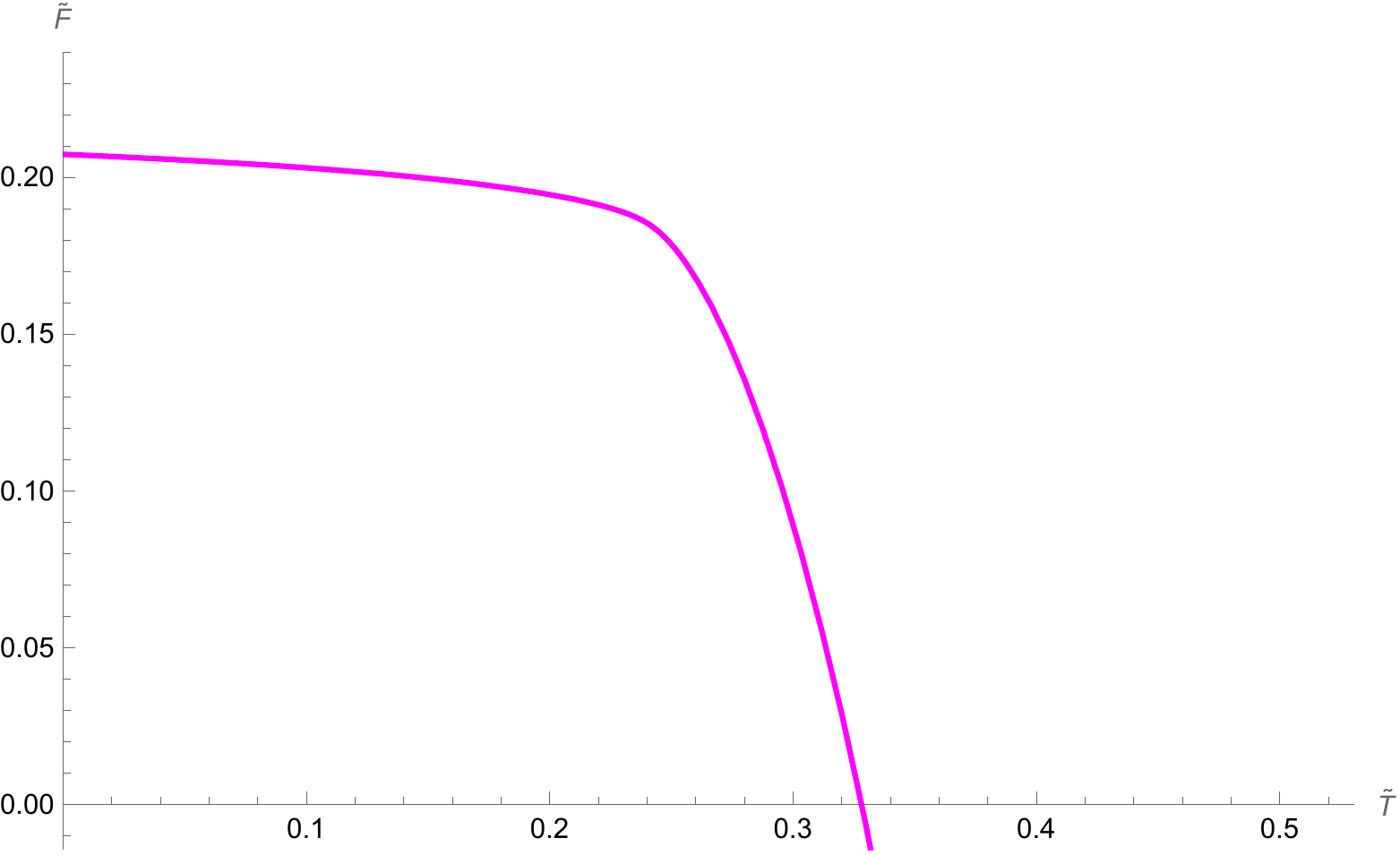}
    \end{subfigure}
    \caption{Free energy versus temperature plots for two cases: \textbf{(Left)}: $\xi_{c_{1}}=11.4998, \,\,\xi_{c_{2}}=14.8394,\,\,\xi = 13.5$, $l = 15$, $\eta = 1$, showing a van der Waals-like first-order phase transition. \textbf{(Right)}: $\xi = 18$, $l = 15$, $\eta = 1$, where no such phase transition is observed. These plots correspond to the {\color{red}red} and {\color{magenta}magenta} curves shown in Fig.~\ref{T-rh-new} (Left).}
    \label{tilde-F-t}
\end{figure}

\begin{figure}[h!]
    \begin{subfigure}
    \centering
        \includegraphics[width=0.5\linewidth]{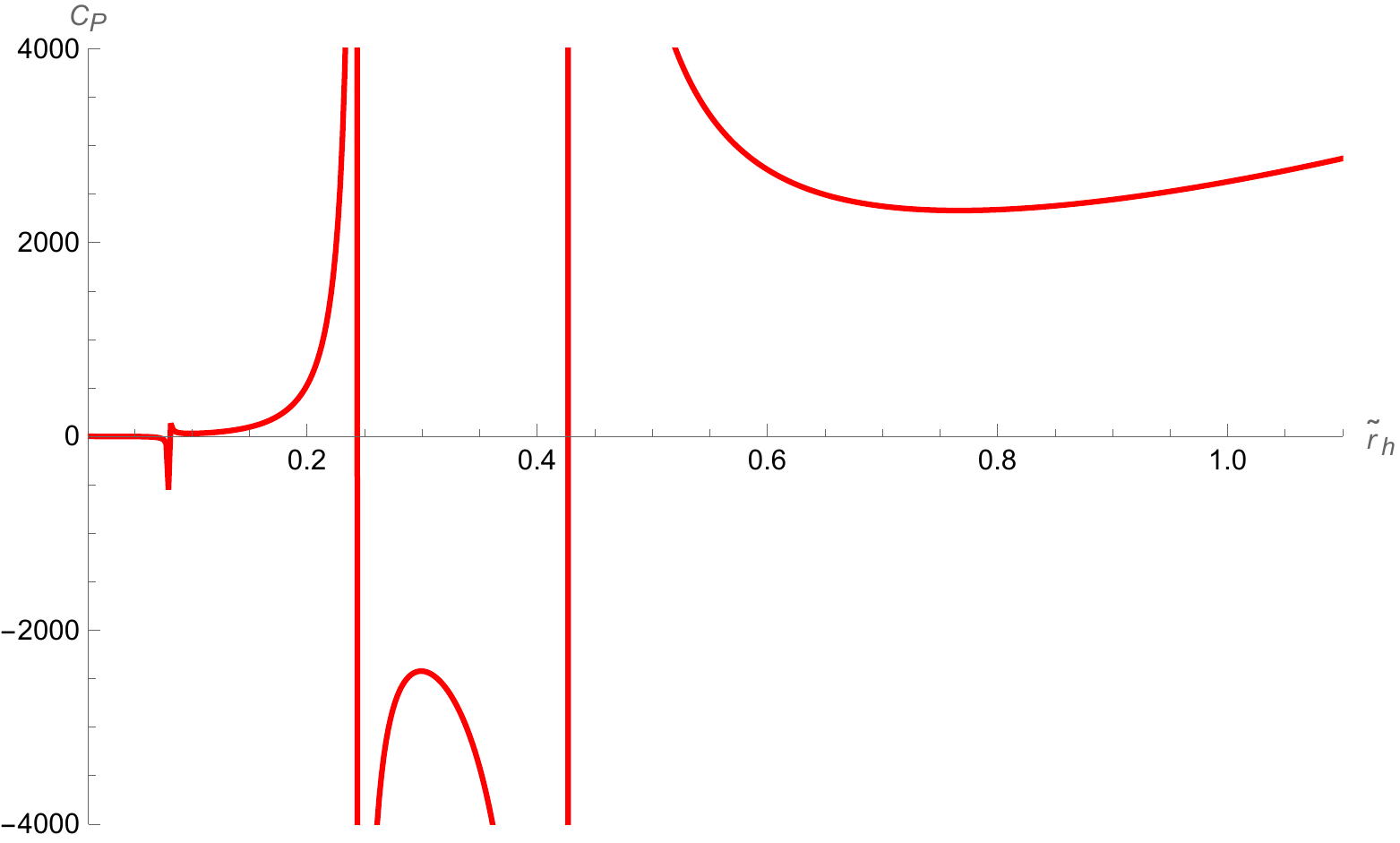}
    \end{subfigure}
    \begin{subfigure}
    \centering
        \includegraphics[width=0.5\linewidth]{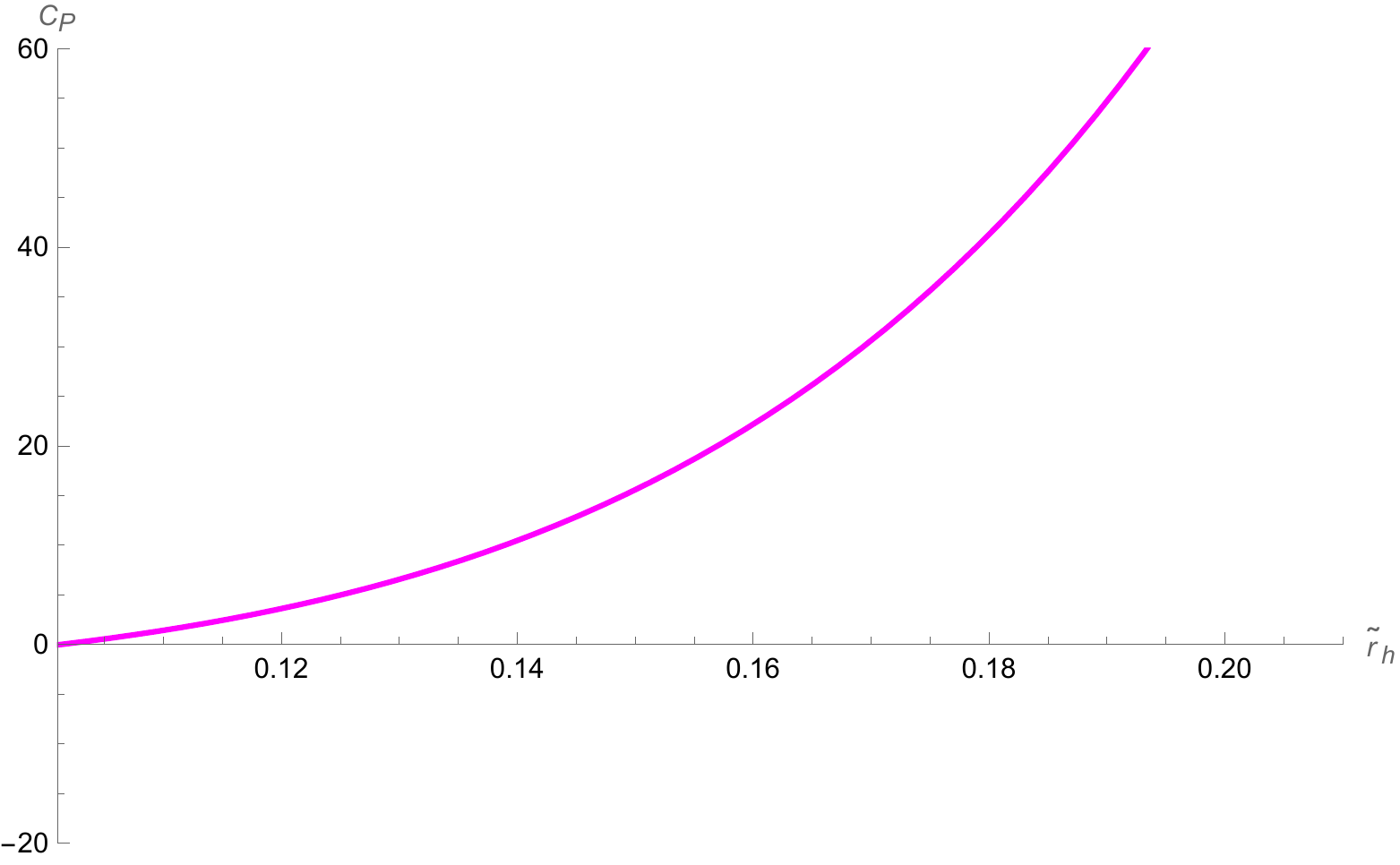}
    \end{subfigure}
    \caption{Plot of the heat capacity as a function of $\tilde{r}_{h}$, evaluated at the same parameter values used in Fig.~\ref{tilde-F-t}. The left panel illustrates the van der Waals-like behaviour, while the right panel corresponds to a regime where the first-order phase transition is absent.}
    \label{C-rh-bulk}
\end{figure}

\begin{figure}
    \begin{subfigure}
        \centering
        \includegraphics[width=0.5\linewidth]{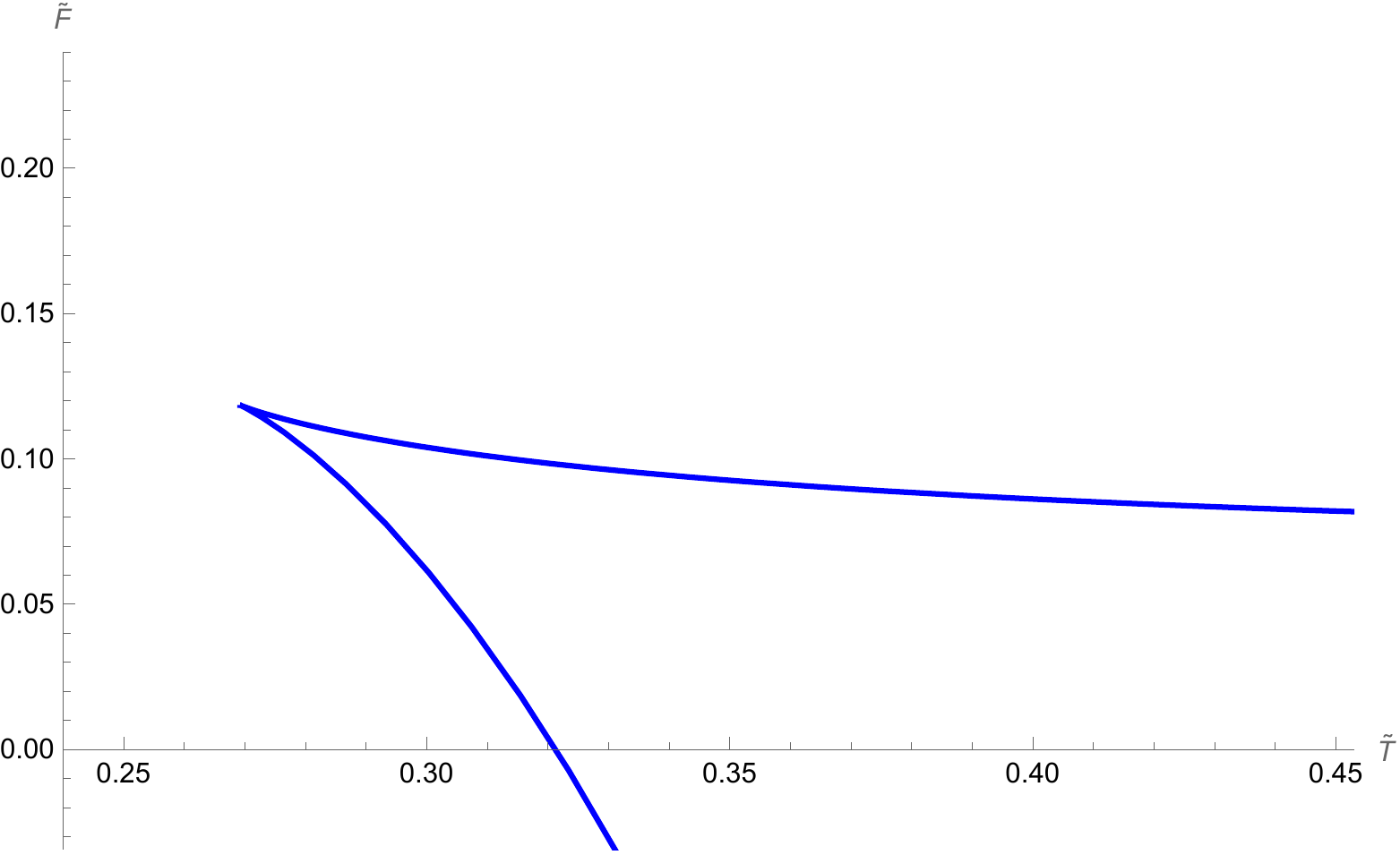}
    \end{subfigure}
    \begin{subfigure}
        \centering
        \includegraphics[width=0.5\linewidth]{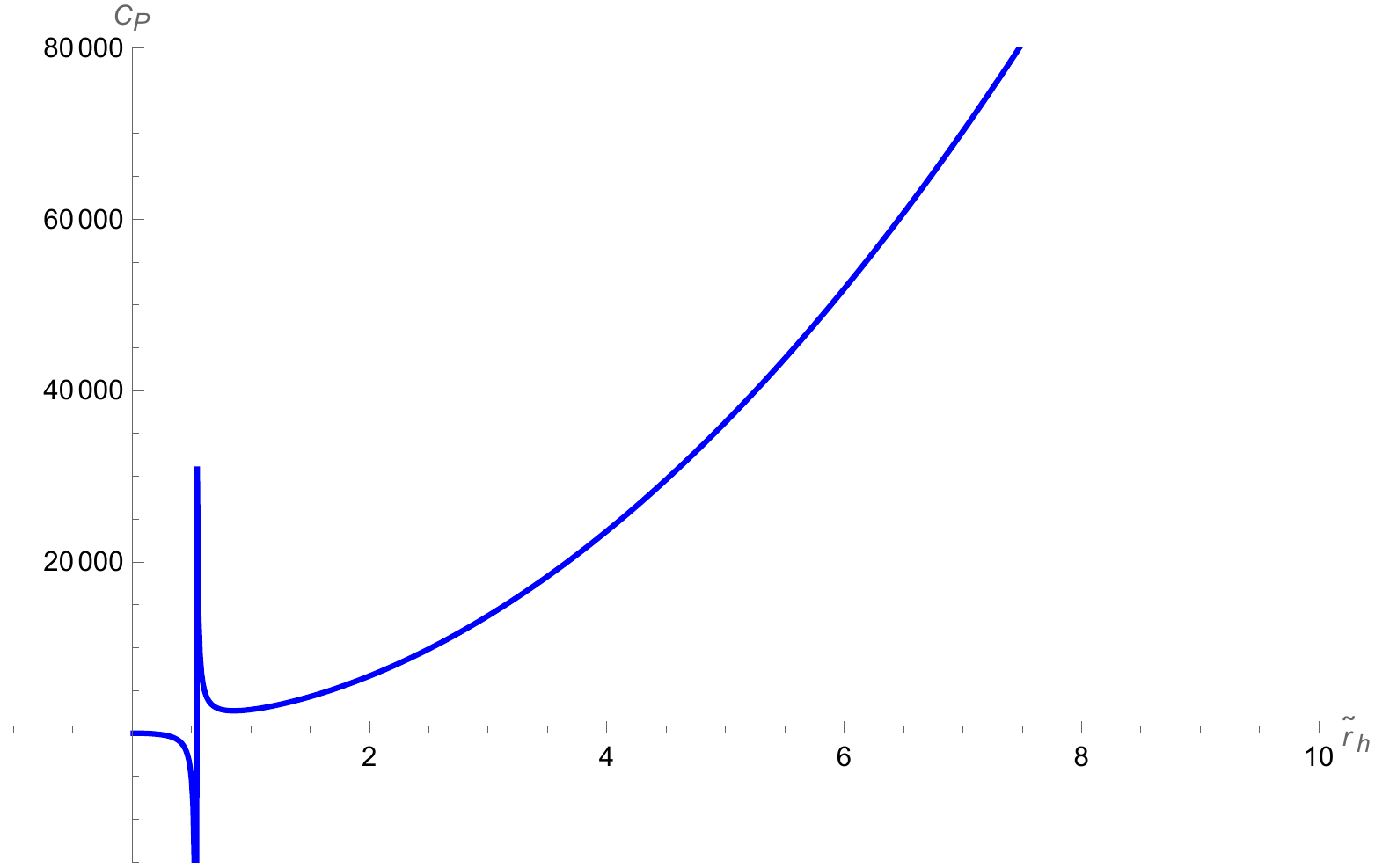}
    \end{subfigure}
    \caption{In this figure, we present (left) the $\tilde{F}$–$\tilde{T}$ plot and (right) the corresponding $C_{P}$–$\tilde{r}_{h}$ plot, both for the parameter set: $\eta = 1$, $l = 15$, and $\xi = 5$. This corresponds to the {\color{blue} blue} curve of Fig \ref{T-rh-new} (Left).}
    \label{beta-1-lit}
\end{figure}

\begin{figure}
    \begin{subfigure}
        \centering
        \includegraphics[width=0.5\linewidth]{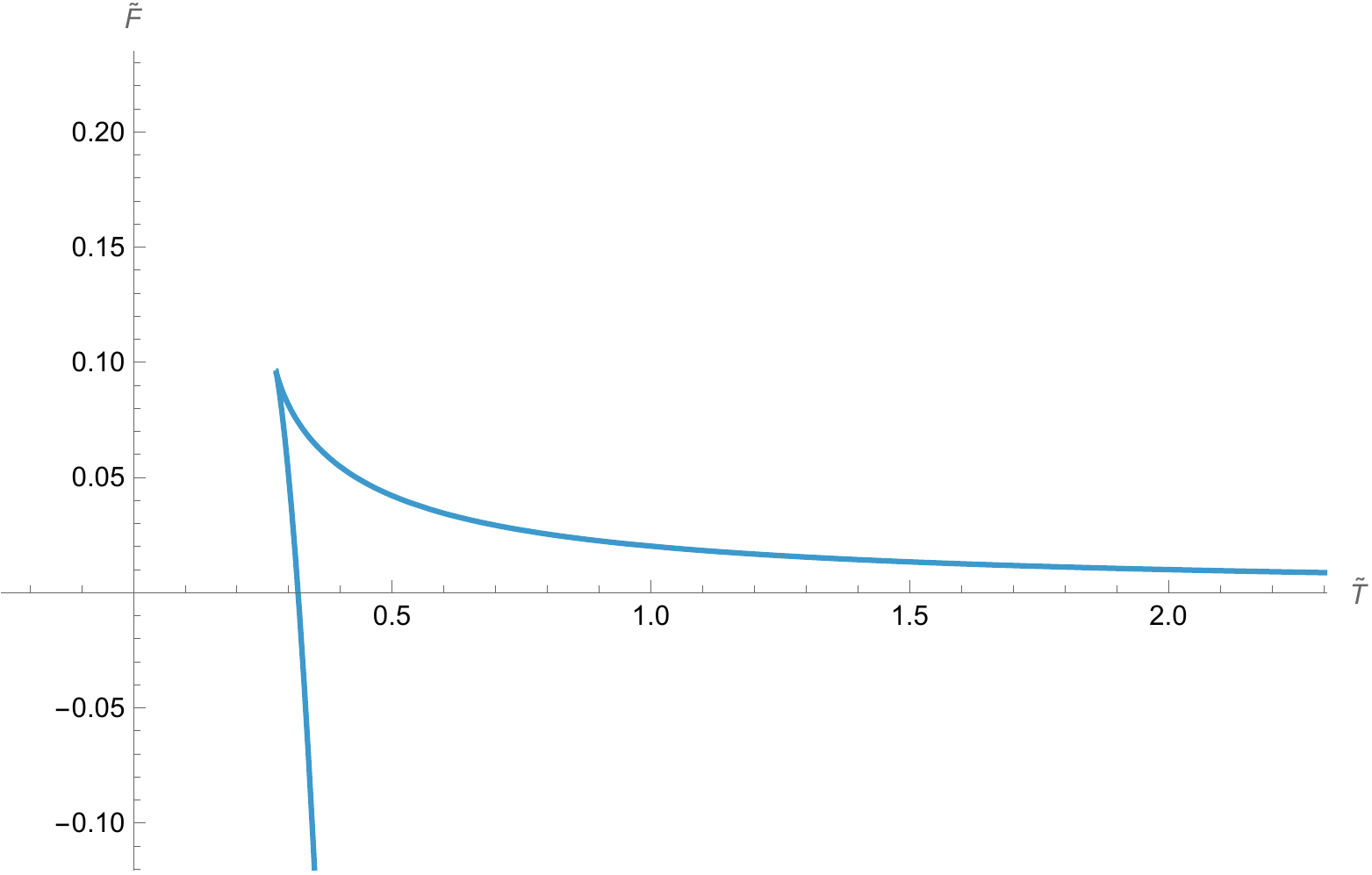}
    \end{subfigure}
    \begin{subfigure}
        \centering
        \includegraphics[width=0.5\linewidth]{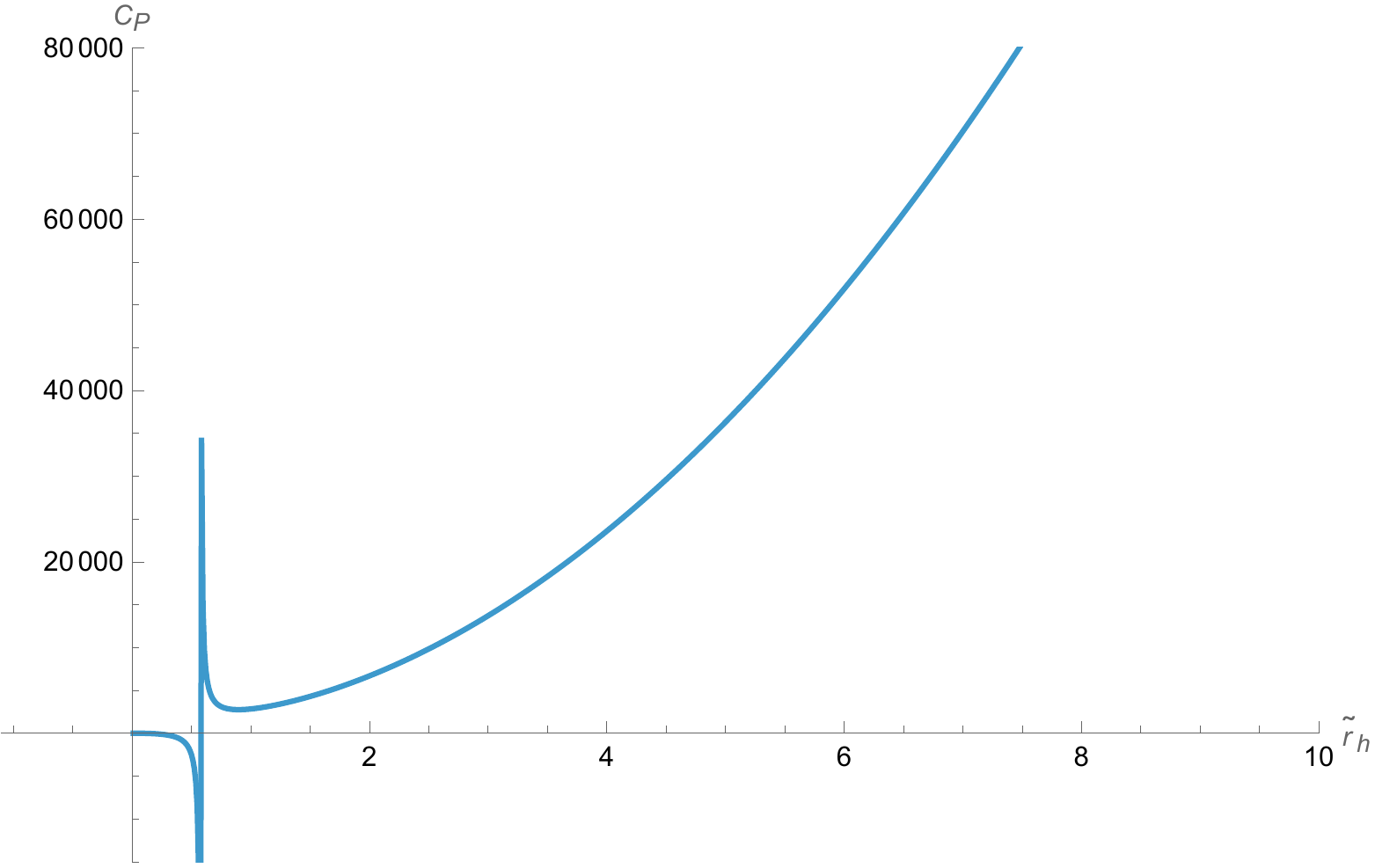}
    \end{subfigure}
    \caption{In this figure, we present (left) the $\tilde{F}$–$\tilde{T}$ plot and (right) the corresponding $C_{P}$–$\tilde{r}_{h}$ plot, both for the parameter set: $\xi = 0$, $l = 15, \eta=1$. There is a Hawking-Page phase structure. }
    \label{alpha-0-bulk}
\end{figure}

\begin{figure}
    \begin{subfigure}
        \centering
        \includegraphics[width=0.5\linewidth]{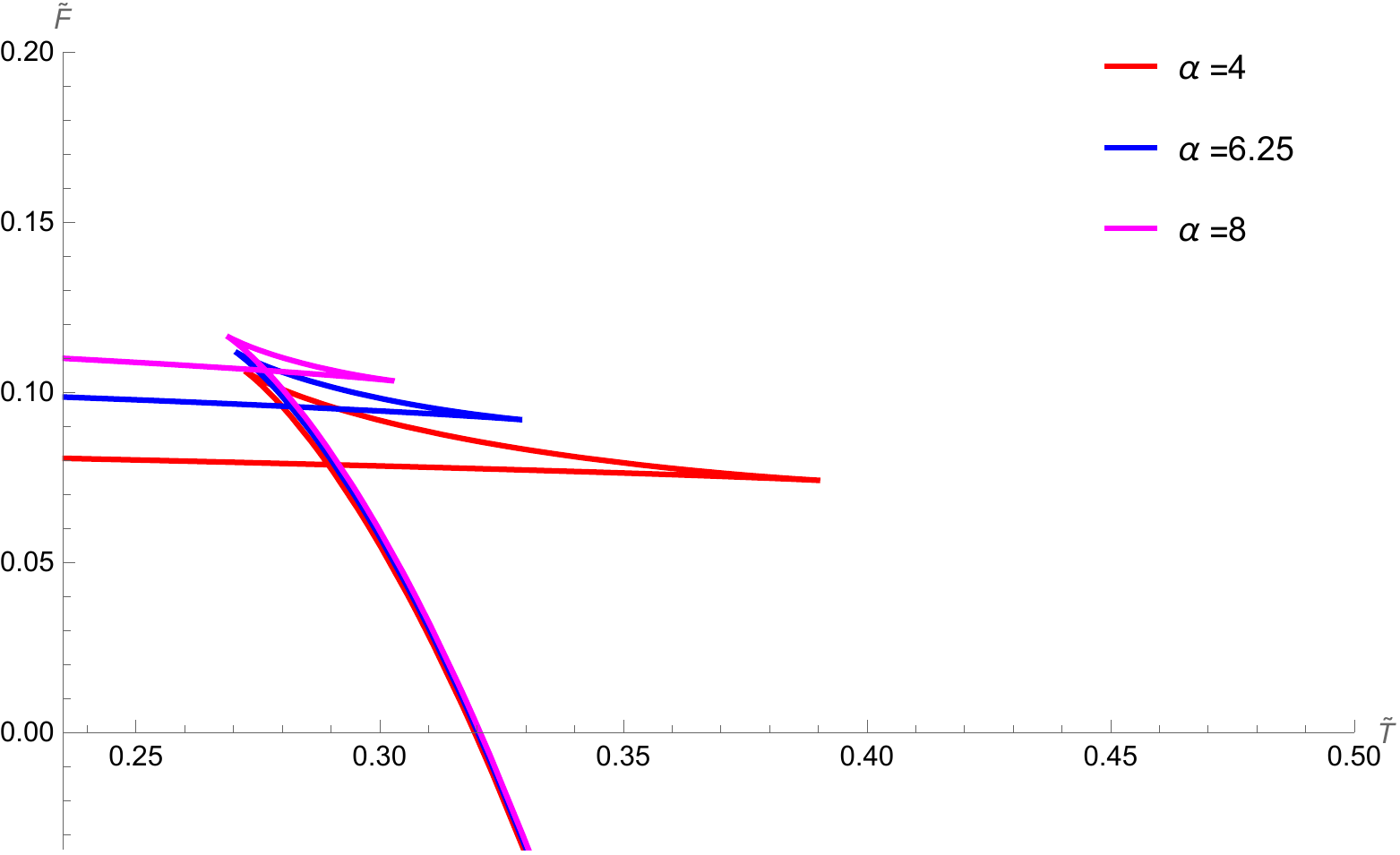}
    \end{subfigure}
    \begin{subfigure}
        \centering
        \includegraphics[width=0.5\linewidth]{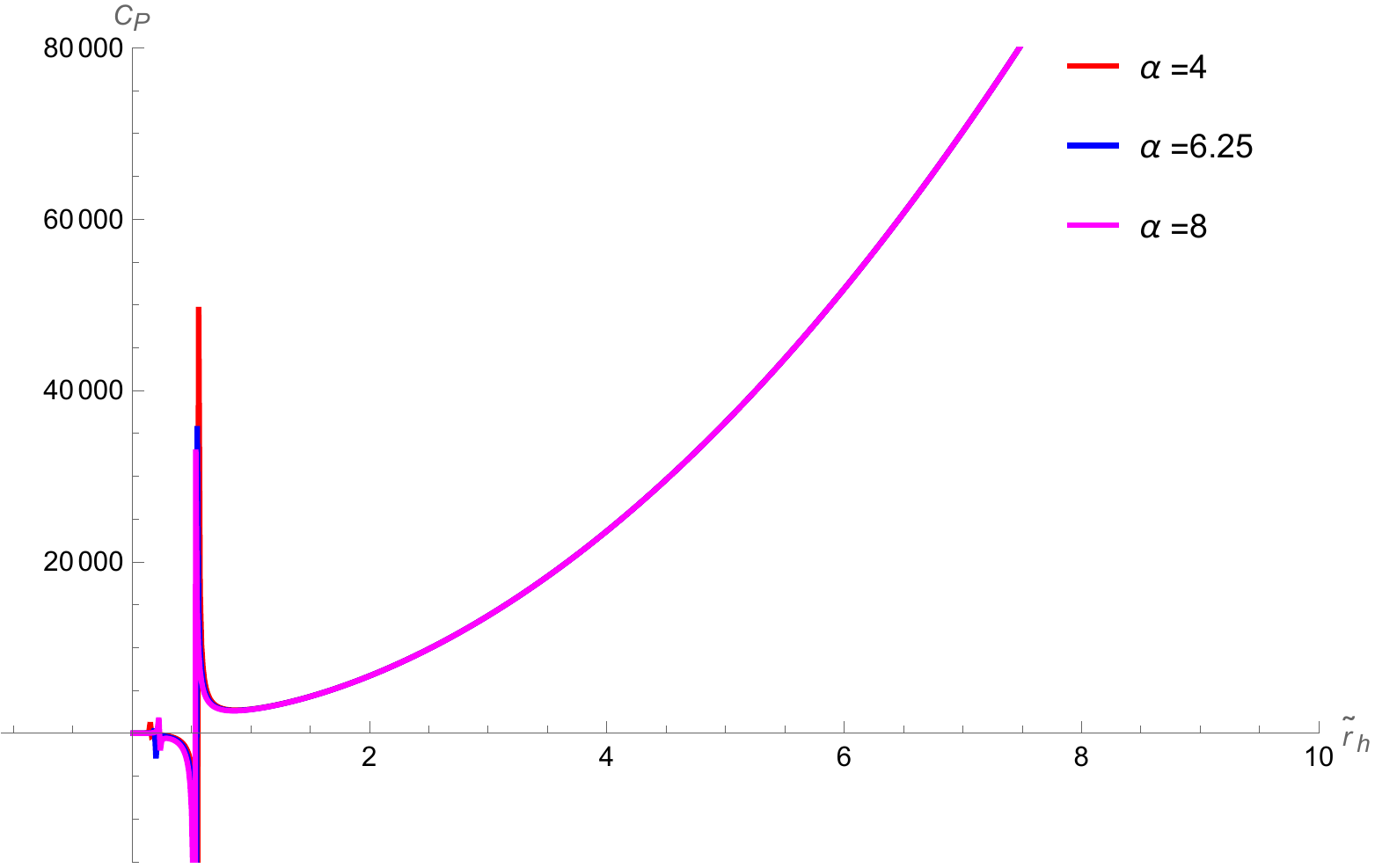}
    \end{subfigure}
    \caption{In this figure, we present (left) the $\tilde{F}$–$\tilde{T}$ plot and (right) the corresponding $C_{P}$–$\tilde{r}_{h}$ plot, both for the parameter set: $\eta = 0$, $l = 15$. All three distinct curves correspond to three curves in Fig \ref{T-rh-new} (Right). }
    \label{beta-0-bulk}
\end{figure}

When $\eta = 0$, the deformed AdS-Schwarzschild black hole reduces to the standard charged AdS-like black hole. In this limit, the critical values are found to be $\xi_{c}=6.25$, $T_{c}=0.0173$, $v_{c}=12.2474$, and $P_{c}=0.000530516$.
In contrast, turning off the deformation parameter $\xi$ eliminates the van der Waals-like phase transition, as shown in Fig.~\ref{P-v-not-vdW} (right).\\
However, for $\eta \ne 0$, numerical evaluation of the universal ratio at the critical point yields the following result (without loss of generality) for $\eta = 1$ and $l = 15$:
\begin{align}
    \xi_{c}=14.8394,\,\,\,v_{c} = 10.3904,\,\,\,T_{c}= 0.016665,\,\,\, P_{c}= 0.000530516
\end{align}
which gives
\begin{align}
    \frac{P_{c} v_{c}}{T_{c}}=0.33077
\end{align}
which is close to as that for the Van der Waals fluid.
Next, we derive the expression for the free energy of the deformed AdS-Schwarzschild black hole,
\begin{align}
    F = M - T_{H} S = \frac{r_h}{4} - \frac{r_h^3}{4 l^2} + \xi \, \frac{2\eta^3 + 8\eta^2 r_h + 12\eta r_h^2 + 9r_h^3}{12 (\eta + r_h)^4}.
\end{align}
To express the thermodynamic quantities in a dimensionless form, we implement the following rescalings:
\begin{align}
    \tilde{\xi} = \frac{\xi}{l^2}, \qquad \tilde{\eta} = \frac{\eta}{l}, \qquad \tilde{r}_h = \frac{r_h}{l}, \qquad \tilde{T} = T_{H} l, \qquad \tilde{M} = \frac{M}{l}.
\end{align}
The resulting $\tilde{F}$–$\tilde{T}$ plot is shown in Fig.~\ref{tilde-F-t}. We observe a characteristic first-order phase transition for $\tilde{\xi_{{c}}}_{1}< \tilde{\xi} < \tilde{\xi_{{c}}}_{2}$. However, this phase transition ceases to exist once the deformation parameter exceeds the critical value $\tilde{\xi}>\tilde{\xi_{c}}_{2}$.
We have shown the heat capacity versus horizon $C_{P}-\tilde{r}_{h}$ plot in Fig \ref{C-rh-bulk}, which shows that when $\xi$ lies between the $\tilde{\xi_{c}}_{1}$ and $\tilde{\xi_{c}}_{2}$ there are four regions associated to the four branches of the $\tilde{F}-\tilde{T}$ curve, out of which there are two stable ($C_{P} > 0$) and two unstable ($C_{P} < 0$). 
An RPT requires the thermodynamically preferred state (global free energy minimum with $C_P>0$) to switch twice (LBH→SBH→LBH) as $\tilde{T}$ varies, producing a finite jump in $F_{\min}$ \cite{Altamirano:2013ane,Dehyadegari:2017hvd}. Tracking $\tilde{F}_{\min}$ across all branches shows it remains continuous, ruling out a zeroth-order transition Fig. \ref{fig:Fmin-T-smooth}. However, a discontinuity in its derivative confirms a standard first-order transition, with only a single branch switch.
\begin{figure}[!h]
        \centering
        \includegraphics[width=0.45\linewidth]{Fmin-T-smooth.png}
        \includegraphics[width=0.45\linewidth]{slope-dFmindT.png}
        \caption{The global minimum of the free energy  $\tilde{F}_{min}$ \textbf{(Left)} and its temperature derivative $d\tilde{F}_{min}/d\tilde{T} $ \textbf{(Right)} as a function of the temperature $\tilde{T}$ at the same parameters as in Fig \ref{tilde-F-t} (Left).   }
        \label{fig:Fmin-T-smooth}
    \end{figure} 
For the $\tilde{\xi}> \tilde{\xi_{c}}_{2}$ there is only single stable branch.
Next, we consider the case $\eta = 1$ with $\tilde{\xi} < \tilde{\xi_{c}}_{1}$, see Fig \ref{beta-1-lit}. In this regime, the system exhibits thermodynamic behaviour similar to that found in earlier studies \cite{Khosravipoor:2023jsl}, featuring a Hawking–Page-like first-order phase transition. The associated $C_{P}$–$\tilde{r}_{h}$ plot reveals two distinct branches, corresponding to stable ($C_{P} > 0$) and unstable ($C_{P} < 0$) phases.
In the undeformed limit, consistent with the findings of~\cite{Hawking:1982dh,Khosravipoor:2023jsl}, the system exhibits a Hawking--Page phase transition, characterized by a single thermodynamically stable branch and one unstable branch, as illustrated in Fig.~\ref{alpha-0-bulk}.
Finally, we examine the case $\eta = 0$, as shown in Fig.~\ref{beta-0-bulk}. In this limit, we observe the van der Waals phase transition.
From the $C_{P}$–$\tilde{r}_{h}$ plot, it is evident that among the three branches, two correspond to thermodynamically stable ($C_{P}>0$) phases, while the remaining one is unstable ($C_{P}<0$). \\

Thus, the deformed AdS-Schwarzschild black hole exhibits a vdW liquid–gas-type phase transition, closely resembling the behaviour observed in the RN-AdS-like black hole. This analogy becomes even more apparent in the limiting case $\eta \to 0$, where the solution smoothly reduces to the RN-AdS-like black hole, known to undergo a similar first-order phase transition.  
Furthermore, as anticipated from Ref.~\cite{Hawking:1982dh,Khosravipoor:2023jsl}, in the limit where the deformation parameter vanishes, the familiar Hawking--Page phase structure is recovered. In the next subsection, we numerically calculate the critical exponents of the mean-field theory for the van der Waals liquid-gas phase transition, and show that they indeed match those of the van der Waals phase transtion. 

\subsection{Critical Exponents} \label{exponent}
In this section, we compute the critical exponents $\alpha$, $\beta$, $\gamma$, and $\delta$ associated with the van der Waals liquid–gas type phase transition.

To compute the universal exponent $\alpha$, we utilize the entropy relation given in Eq.~(\ref{entropy-bulk}). Since the entropy is independent of temperature near the critical point, it follows that $\alpha = 0$.
The computation of the universal critical exponents for the RN-AdS-like black hole, corresponding to the case with control parameter $\eta = 0$, has already been carried out in ref.~\cite{Kubiznak:2012wp}.
 However, the case with both parameters $\xi$ and $\eta$ nonzero is non-trivial and necessitates a numerical treatment. 
 We know for vdW phase, the equation of state is given by 
\begin{align}
    P = \frac{T}{v - b} - \frac{a}{v^2},
\end{align}
where \( v = V/N \) is the specific volume and \( N \) is the number of constituent particles. The critical point is located by imposing the conditions \( \left( \frac{\partial P}{\partial v} \right)_T = 0 \) and \( \left( \frac{\partial^2 P}{\partial v^2} \right)_T = 0 \). Solving these yields
\begin{equation}
P_c = \frac{a}{27b^2}, \quad v_c = 3b, \quad T_c = \frac{8a}{27b}. \label{eq:critical}
\end{equation}
Introducing reduced variables,
\[
\hat{P} = \frac{P}{P_c}, \quad \hat{v} = \frac{v}{v_c}, \quad \hat{T} = \frac{T}{T_c},
\]
the equation of state becomes
\begin{equation}
\hat{P} = \frac{8\hat{T}}{3\hat{v} - 1} - \frac{3}{\hat{v}^2}. \label{eq:reduced-eos}
\end{equation}
Defining reduced number density \( \hat{n} = 1/\hat{v} \), Eq.~\eqref{eq:reduced-eos} is rewritten as
\begin{equation}
\hat{P} = \frac{8\hat{T} \hat{n}}{3 - \hat{n}} - 3\hat{n}^2. \label{eq:reduced-n}
\end{equation}
To investigate critical behaviour, we introduce small deviations from the critical point:
\begin{equation}
t = \hat{T} - 1, \quad \omega = \hat{v} - 1, \quad \pi = \hat{P} - 1. \label{eq:expansion-vars}
\end{equation}
Near the critical point, the difference in number density between the coexisting gas and liquid phases behaves as
\begin{equation}
\Delta n = n_g - n_l \propto (-t)^\beta, \quad t < 0, \label{eq:order-param}
\end{equation}

\begin{figure}[htbp]
    \centering
    \includegraphics[width=0.6\linewidth]{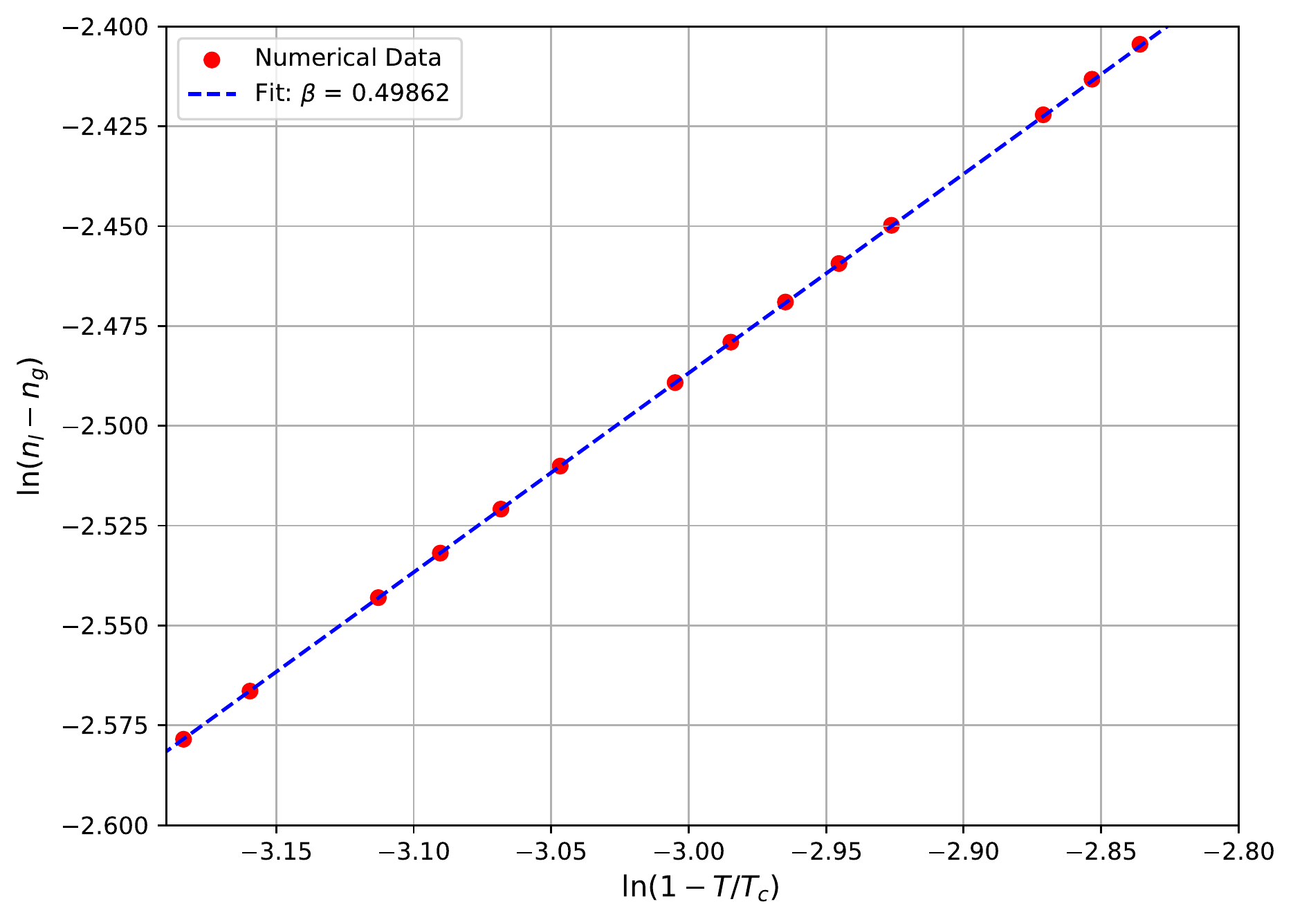} 
    \caption{Plot of $\ln(n_{l}- n_{g})$ versus $\ln(1- \frac{T}{T_{c}})$ across the $\hat{P}$--$\hat{T}$ gas--liquid coexistence curve for a van der Waals fluid. The red dots represent fitted data with slope 
     $\beta \approx 0.499$ and intercept 0.372. Deformation and control parameters: $\xi=13.5$, $\eta=1$.}
    \label{uni-exp-plot}
\end{figure}

In Fig.~\ref{uni-exp-plot}, we plot $\ln(n_{l} - n_{g})$ as a function of $\ln(1 - \frac{T}{T_{c}})$. The numerical data points in the regime $1 - \hat{T} < 10^{-4}$ fit well to a straight line, confirming the scaling $\Delta n \sim (1 - \hat{T})$ with a universal exponent \( \beta = \frac{1}{2} \).

\begin{figure}
    \centering
    \includegraphics[width=0.65\linewidth]{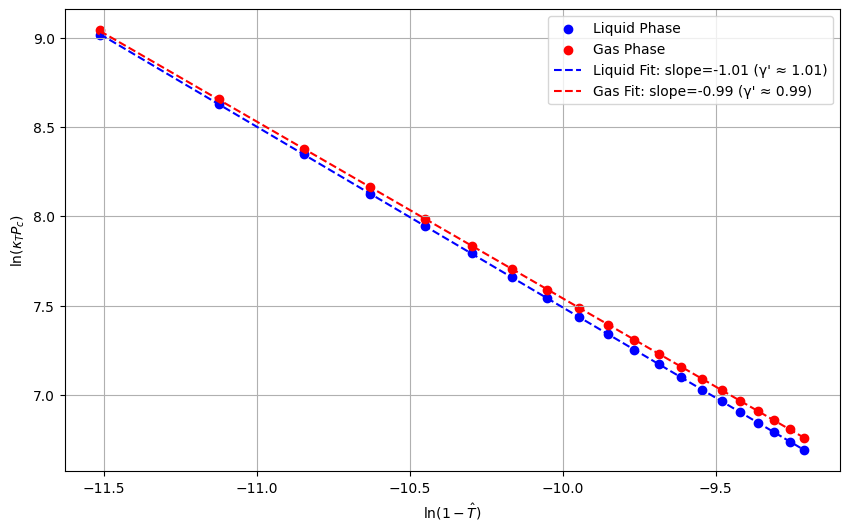}
    \caption{The $\ln$--$\ln$ plot of $\kappa_T P_c$ versus $(1 - \hat{T})$ is shown. The slope of the fitted dashed line for the numerical data is (a) $-1.0111$ along the saturated liquid branch (blue) and (b) $-0.9888$ along the saturated gas branch (red), confirming the critical exponent $\gamma' \approx 1$.}
    \label{kappa-plot}
\end{figure}

The isothermal compressibility in this framework is given by
\begin{equation}
\kappa_T \equiv -\frac{1}{v} \left( \frac{\partial v}{\partial P} \right)_T = \frac{1}{P_c \hat{n}} \left( \frac{\partial \hat{n}}{\partial \hat{P}} \right)_{\hat{T}}. \label{eq:compressibility}
\end{equation}
Substituting Eq.~\eqref{eq:reduced-n} into Eq.~\eqref{eq:compressibility}, one obtains
\begin{equation}
\kappa_T P_c = \frac{(3 - \hat{n})^2}{6\hat{n} \left(4\hat{T} - \hat{n}(3 - \hat{n})^2\right)}. \label{eq:compressibility-explicit}
\end{equation}
The critical exponents $\gamma$ and $\gamma'$ characterize the behavior of the isothermal compressibility $\kappa_{T}$ along the isochoric line $\hat{n} = 1$ and across the $\hat{P}$--$\hat{v}$ coexistence curve, respectively.
\begin{equation}
\kappa_T P_c \propto 
\begin{cases}
t^{-\gamma}, & t > 0, \\
(-t)^{-\gamma'}, & t < 0,
\end{cases}
\label{eq:kappa-critical}
\end{equation}
so on setting $\hat{n}=1$ in Eq. (\ref{eq:compressibility-explicit}), we get \( \gamma = 1 \).
Next, first, we rewrite $\kappa_T P_c$ as follows
\begin{align}
\kappa_T P_c = \frac{1}{\hat{v}} \left( \frac{\partial \hat{T}}{\partial \hat{P}} \right)_{\hat{v}} \left( \frac{\partial \hat{T}}{\partial \hat{v}} \right)_{\hat{P}}^{-1}   
\end{align} 
we plot the $\ln$--$\ln$ curve of $\kappa_T P_c$ versus $(1 - \hat{T})$ along the $\hat{P}$--$\hat{v}$ coexistence line, as shown in Fig.~\ref{kappa-plot}. We get $\gamma' \approx 1$. 
Finally to compute $\delta$, we substitute the reduced variables
\begin{equation}
    \hat{P} = \frac{P}{P_c},\quad 
    \omega= \frac{v-v_c}{v_c},\quad 
    t = \frac{T - T_c}{T_c},
\end{equation}
into the equation of state
\begin{align}
    P= \frac{T}{v} - \frac{1}{2 \pi v^2} + \frac{\xi}{8 \pi \left( \eta + \frac{v}{2} \right)^4},
     \label{P-v}
\end{align}
we get
\begin{equation}
P \;=\; \frac{T_c(1+t)}{v_c(1+\omega)}
\;-\; \frac{1}{2\pi v_c^2 (1+\omega)^2}
\;+\; \frac{\alpha}{8\pi \xi^4 \left(1 + \frac{v_c}{2\xi}\omega\right)^4}
\end{equation}
where we have defined $\zeta = \eta + \frac{v_{c}}{2}$. Now, Taylor expanding each term to third order in $\omega$, 
keeping only leading order in $t$, one obtains
\begin{align}
\hat{P} &= 1 
    + \frac{T_c}{P_c v_c}\,t 
    - \frac{T_c}{P_c v_c}\,t\omega \notag\\
    &\quad + \left.\frac{\partial P}{\partial v}
      \right|_{T_c,v_c}\omega
    + \frac{v_c^2}{2P_c}
      \left.\frac{\partial^2 P}{\partial v^2}
      \right|_{T_c,v_c}\omega^2
    - \frac{v_c^3}{6P_c}
      \left.\frac{\partial^3 P}{\partial v^3}
      \right|_{T_c,v_c}\omega^3
    + \mathcal{O}(\omega^4, t\omega^2).
\end{align}
Since $(v_c, T_c, P_c)$ is the critical point, the 
conditions $(\partial P/\partial v)_{T_c}=0$ and 
$(\partial^2 P/\partial v^2)_{T_c}=0$ hold identically, 
so the $\omega$-linear and $\omega^2$ terms vanish. 
The expansion reduces to
\begin{equation}
    \hat{P} = 1 + At - Bt\omega - D\omega^3 
              + \mathcal{O}(\omega^4,\,t\omega^2),
    \label{eq:reduced_EOS}
\end{equation}
where
\begin{equation}
    A = B = \frac{T_c}{P_c v_c}, \qquad
    D = -\frac{v_c^3}{6P_c}
        \left.\frac{\partial^3 P}{\partial v^3}
        \right|_{T_c,v_c}
      = \frac{v_c^3}{6P_c}
        \left(\frac{3}{\pi v_c^4} 
        - \frac{35\alpha}{128\pi\zeta^7}\right),
\end{equation}
Since $\delta$ is defined by the critical isotherm
\begin{align}
|P - P_c| \propto |v - v_c|^\delta
\quad \text{at } T = T_c
\end{align}
or in reduced variables this becomes
\begin{align}
|\hat{P} - 1| \propto |\omega|^\delta
\quad \text{at } t = 0
\end{align}
Therefore
\begin{align}
|\hat{P} - 1| \propto |\omega|^3
\end{align}
which gives immediately $\delta = 3$.

In the following section, we analyse the boundary thermodynamics. We first construct the dual CFT metric and then examine the associated thermodynamic properties.

\section{CFT thermodynamics with a chemical potential for the central charge} \label{boundary thermo}
In the framework of the AdS/CFT correspondence, a bulk geometry in the AdS spacetime is associated with a conformal boundary that resides in one lower spacetime dimension. The metric of the CFT is identified with the boundary metric of the corresponding asymptotically AdS spacetime, up to a Weyl rescaling factor, as established in \cite{Witten:1998qj,Gubser:1998bc}.  By considering the asymptotic limit $r \xrightarrow{}\infty$, the bulk metric given in Eq (\ref{metric}) simplifies and takes the following form:
\begin{align}
    ds^2 \approx -\frac{r^2}{l^2} dt^2 + \frac{l^2}{r^2} dr^2 + r^2 \left( d\theta^2 + \sin^2\theta d\phi^2 \right).
    \label{CFT-1-metric}
\end{align}
To extract the boundary metric, we use a coordinate transformation \(z = \frac{l^2}{r}\), so \(r = \frac{l^2}{z}\), so that as \(r \to \infty\), \(z \to 0\). Thus, the metric (\ref{CFT-1-metric}) becomes
\begin{align}
    ds^2 = \frac{l^2}{z^2} \left( -dt^2 + dz^2 + l^2 (d\theta^2 + \sin^2\theta d\phi^2) \right).
\end{align}
Following the AdS/CFT prescription, the boundary metric is obtained by rescaling by \(\left( \frac{z}{l} \right)^2\):
\begin{align}
    ds^2_{\text{boundary}} = \lim_{z \to 0} \frac{z^2}{l^2} ds^2 = -dt^2 + l^2( d\theta^2 + \sin^2\theta d\phi^2),
\end{align}
Applying the Weyl scale factor \(\eta\) = \(  \frac{\omega z}{l} \),
the CFT metric is the Einstein static universe ($\mathbf{R} \times S^2$), so up to the constant (Weyl factor), $\omega=\frac{R}{l}$ (where $R$ is the curvature radius of the manifold where the CFT lives), we obtain
\begin{align}
    ds^2 = \omega^2 \left( -dt^2 + l^2 \left( d\theta^2 + \sin^2\theta d\phi^2 \right) \right).
\end{align}
which is the standard AdS-Schwarzschild CFT metric, indicating that the deformation does not alter the boundary geometry. In the subsequent sections, we begin by determining the thermodynamic quantities associated with the dual CFT states corresponding to the bulk geometry. This will be followed by an analysis of the phase transition behaviour exhibited within the dual CFT.  \\

\subsection{Thermodynamic quantities for dual CFT}
The bulk AdS expressions (\ref{mass-bulk})--(\ref{temp}), in conjunction with the holographic dictionary (\ref{bulk-boundary-duality}) and (\ref{C-eq}), yield insights into the (extended) thermodynamical properties of the dual large-$N$, strongly coupled CFT.
We introduce the following dimensionless parameters: 
\[
x = \frac{r_h}{l}, \quad 
\tilde{\xi} = \frac{\xi}{l^2},   
\quad 
b = \frac{\eta}{l}.
\]
With these redefinitions, the thermodynamical quantities in the dual CFT can be expressed as:
\begin{itemize}
    \item \text{Entropy:}
    \begin{equation}
        S = 4\pi C\, x^{2},  
        \label{entropy-bdry}
    \end{equation}

    \item \text{Energy:}
    \begin{equation}
        \mathcal{E} = \frac{2 C x}{R} \left( 1 + x^2 + \frac{\tilde{\xi} (b^2 + 3 x^2 + 3 b x)}{3 x (b + x)^3} \right),  
        \label{energy-bdry}
    \end{equation}

    \item \text{Temperature:}
    \begin{equation}
        T = \frac{1 + x^2 \left( 3 - \frac{\tilde{\xi}}{(b + x)^4} \right)}{4 \pi x R},  
        \label{temp-bdry}
    \end{equation}

    \item \text{Chemical potential:}
    \begin{equation}
        \mu = \frac{x}{R} \left( 1 - x^2 + \frac{\tilde{\xi} x^2}{(b + x)^4} + \frac{2 \tilde{\xi} (b^2 + 3 b x + 3 x^2)}{3 x (b + x)^3} \right).
        \label{bdry-chem-pot}
    \end{equation}
\end{itemize}
The presence of the \(1/R\) dependence in the thermodynamic expressions is a direct consequence of the scale invariance inherent in conformal field theories. Moreover, the proportionality with the central charge \(C\) in Eqs.~(\ref{entropy-bdry})--(\ref{energy-bdry}) reflects the analysis being carried out in the large-\(C\) (or large-\(N\)) limit of the dual conformal field theory. In the subsequent section, we examine the phase structure corresponding to various thermodynamic ensembles in the boundary theory.

\subsection{Thermodynamic ensembles in the dual CFT}
In this section, we investigate the phase structure of various thermodynamic ensembles in the dual CFT, corresponding to thermal states holographically dual to deformed AdS-Schwarzschild black holes. The deformation parameter $\xi$ plays a subtle role and requires a careful interpretation as also discussed in Refs.~\cite{Khosravipoor:2023jsl,Gogoi:2025rcn,Avalos:2023ywb}.
In principle, there exist four distinct thermodynamic ensembles in the boundary CFT, as one may independently choose between two pairs of conjugate thermodynamic variables, namely:
\begin{align*}
\text{Canonical ensemble, fixed } (\mathcal{V}, C): \quad & F \equiv \mathcal{E} - T S, \\
\text{Fixed } (p, C): \quad & G \equiv \mathcal{E} - T S + p \mathcal{V}, \\
\text{Fixed } (p, \mu): \quad & \Phi \equiv \mathcal{E} - T S + p \mathcal{V} - \mu C, \\
\text{Fixed } (\mathcal{V}, \mu): \quad & \Omega \equiv \mathcal{E} - T S - \mu C.
\end{align*}
The final ensemble represents a trivial case, as it follows directly from the Euler relation given in Eq.~(\ref{mu-bdry}). In this work, our focus will be on the remaining three ensembles, each of which exhibits nontrivial and rich phase structure. For each ensemble, we also examine the corresponding heat capacity, which serves as a diagnostic tool for assessing the thermodynamic stability of the system. 
\subsubsection{Phase transition in $(\mathcal{V}, C)$}
\begin{figure}[!h]
    \centering
    \includegraphics[width=0.6\linewidth]{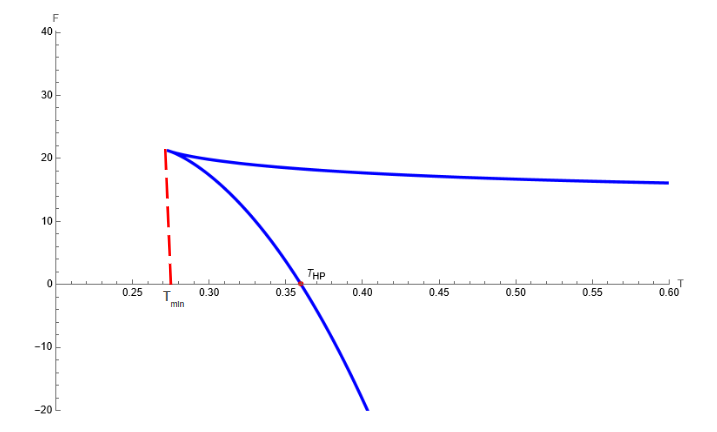}
    \caption{Free-energy-temperature $(F-T)$ curve at the parameter values $R=1, C=20, \tilde{\xi}=5,b=5$. The red dot represents the Hawking temperature $T_{HP}=0.360$ and the minimum temperature is attained at $T_{min}= 0.275.$}
    \label{CFT-F-T-THP}
\end{figure}

\begin{figure}[!h]
   \begin{subfigure}
        \centering
   \includegraphics[width=0.5\linewidth]{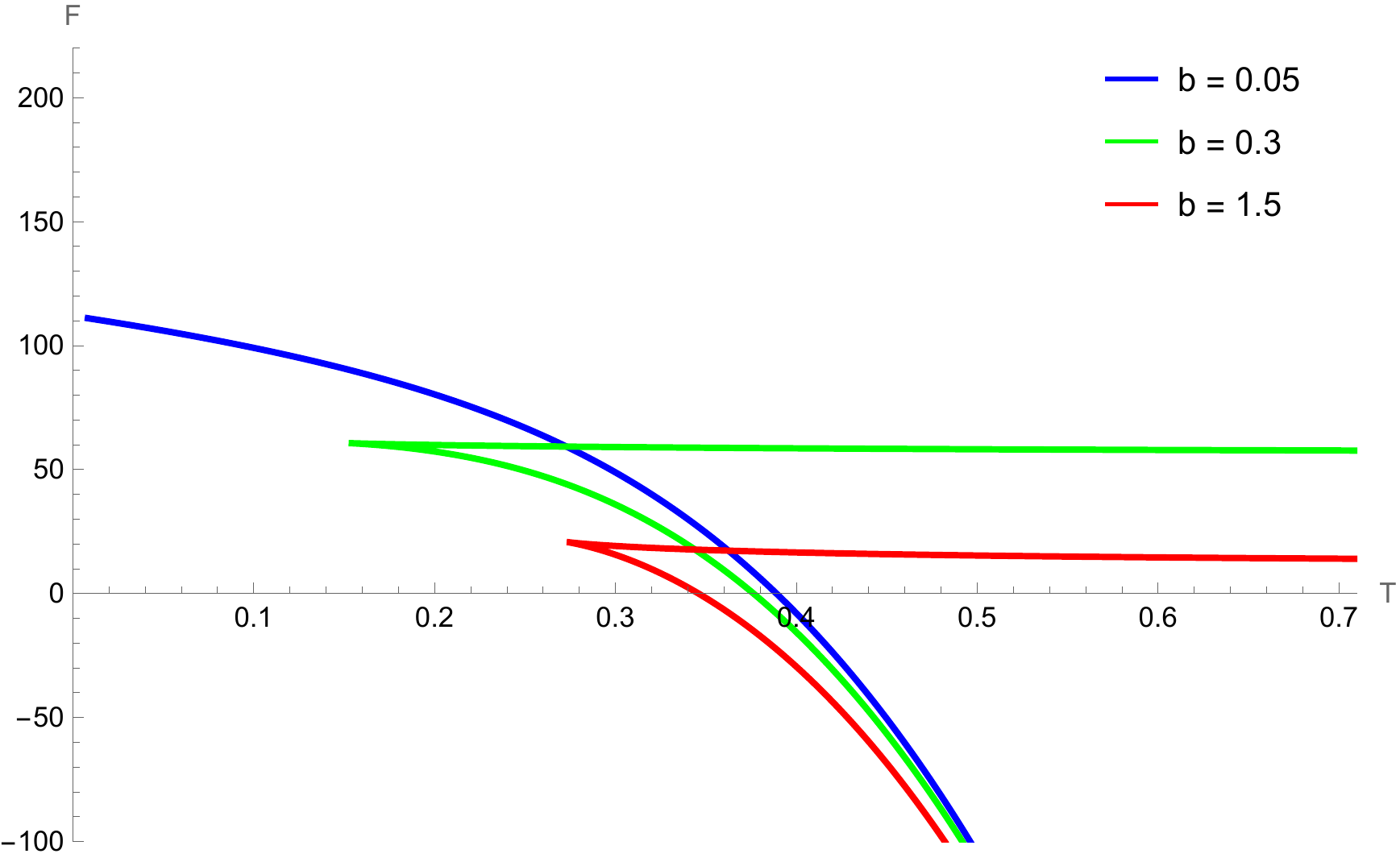}
   \end{subfigure}
   \begin{subfigure}
       \centering
     \includegraphics[width=0.5\linewidth]{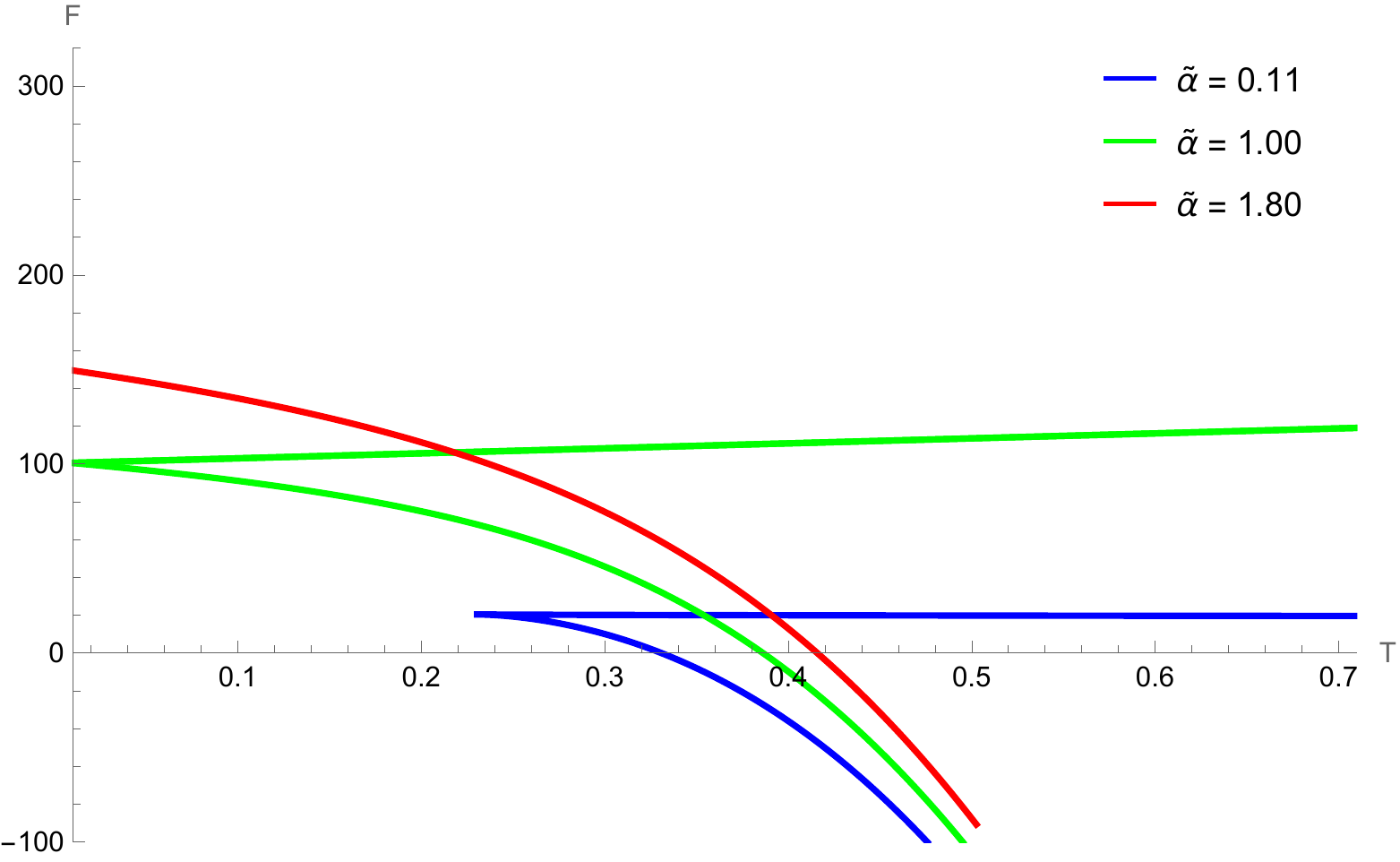}
   \end{subfigure}
    \caption{ $F$–$T$ plots for two distinct parameter choices. \textbf{(Left)}: $b_{c} = 0.1$, $\tilde{\xi} = 1$; \textbf{(Right)}: $\tilde{\xi}_{c} = 1.45$, $b = 0.1$. In both cases, the parameters are fixed as $R = 1$ and $C = 25$. For fixed $\tilde{\xi}$, decreasing $b$ causes the $F$–$T$ curve to transition toward a single-branched structure. In contrast, for fixed $b$, increasing $\tilde{\xi}$ leads to the disappearance of the first-order phase transition in the $F$–$T$ profile. }
    \label{CFT-vary-para-alpha-b}
\end{figure}

\begin{figure}
    \centering
    \includegraphics[width=0.5\linewidth]{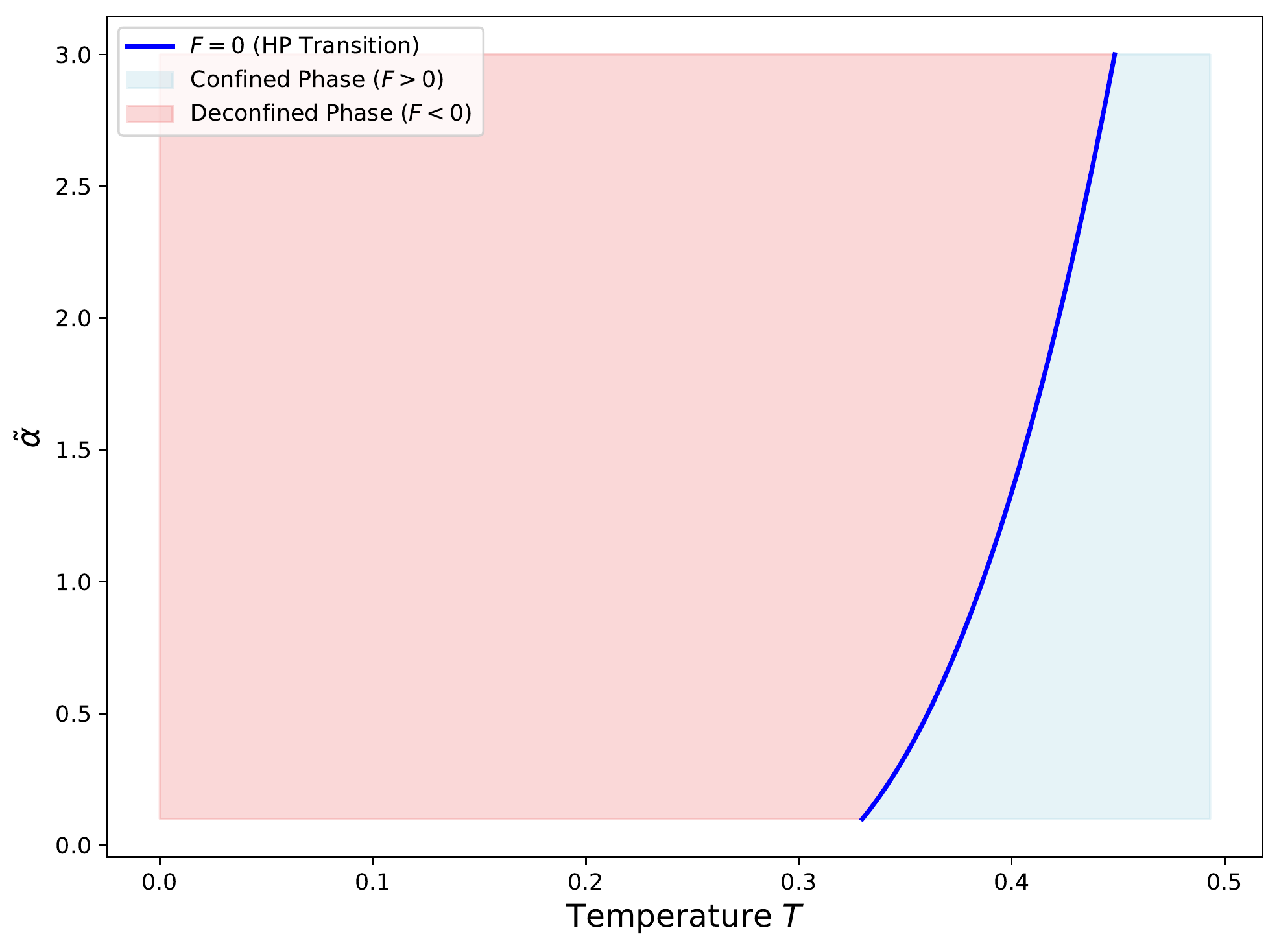} 
    \caption{$\tilde{\xi}$–$T$ diagram for $R = 1$, $C = 25$, and $b = 0.1$, showing the coexistence curve corresponding to the line of (de)confinement phase transitions in the dual CFT. The red-shaded region represents the deconfined phase, whereas the blue-shaded region corresponds to the confined phase.}
    \label{de-confinment}
\end{figure}

\begin{figure}[!h]
    \centering
    \includegraphics[width=0.5\linewidth]{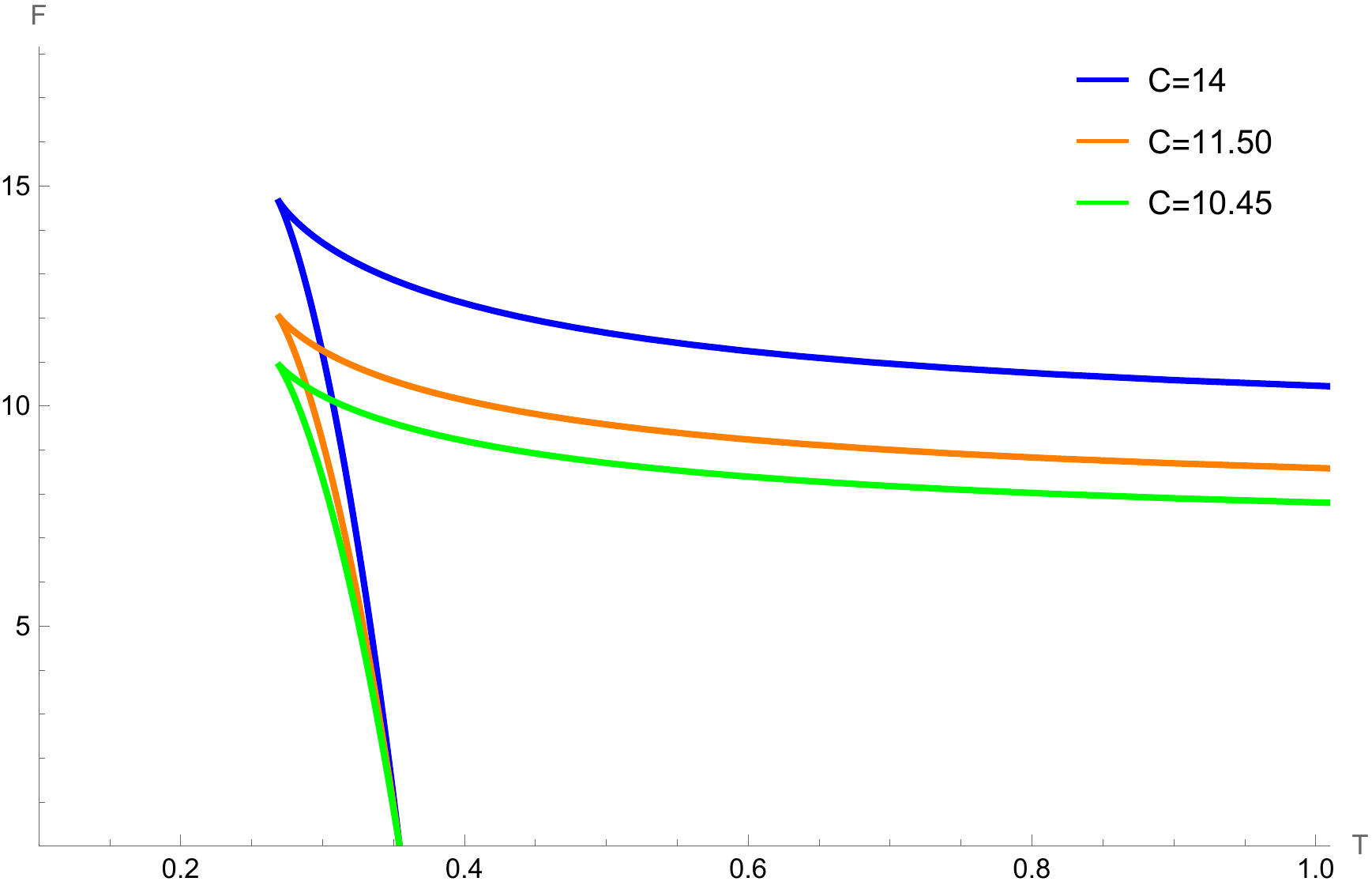}
    \caption{Here, we have set without loss of generality: $\tilde{\xi}=1, b=1, R=1$, whereas $C$ varies as shown in the figure. This shows the Hawking-Page-like phase transition in the dual CFT of the bulk def AdS-Schwarzschild black hole. Each $C$ has the same Hawking-Page temp given by $T_{HP}=0.354$. }
    \label{CFT-FT}
\end{figure}

\begin{figure}[!h]
     \begin{subfigure}
        \centering
    \includegraphics[width=0.5\linewidth]{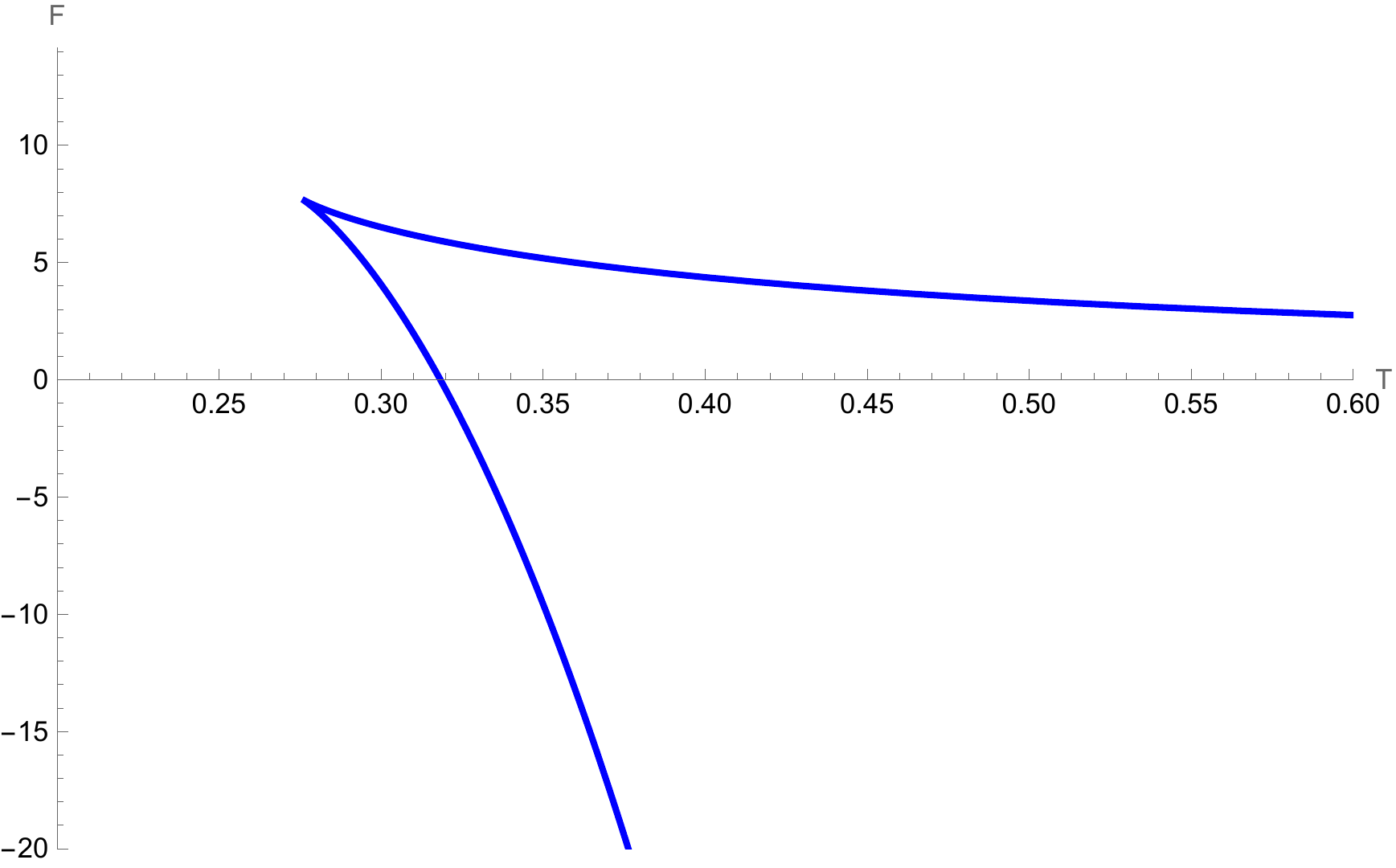}
   \end{subfigure}
   \begin{subfigure}
       \centering
    \includegraphics[width=0.5\linewidth]{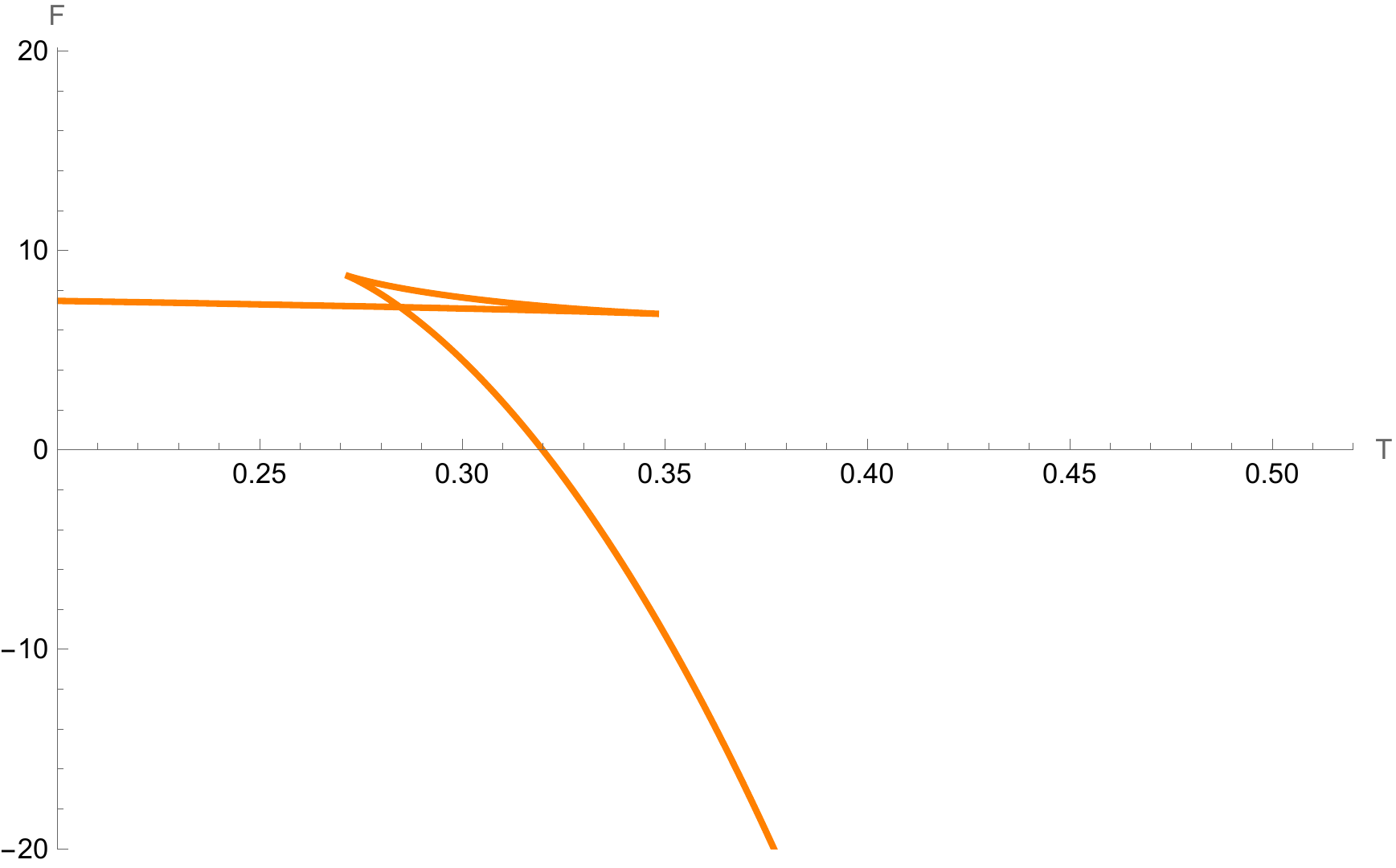}
   \end{subfigure}
    \caption{The $F$--$T$ plots are shown for two representative cases. \textbf{(Left)}: The deformation parameter $\tilde{\xi} = 0$ is set to zero, with $b = 5$, $R = 1$, and $C = 20$. \textbf{(Right)}: The control parameter $b = 0$ is set to zero, with $\tilde{\xi} = 0.01$, $R = 1$, and $C = 20$.}
    \label{CFT-alpha-0-FT}
\end{figure}

In this ensemble, the thermodynamic variables \( \mathcal{V} \) and \( C \) are held fixed, thereby defining the ensemble characterized by the pair \( (\mathcal{V}, C) \). 
We now proceed to investigate the behaviour of the free energy \( F \) as a function of temperature \( T \), for various fixed values of \( (\mathcal{V}, C) \). 
The corresponding free energy is defined as \( F = \mathcal{E} - T S \) 
and its  expression is given by 
\begin{align}
    F = \mathcal{E} - T S = \frac{2 C x}{R} \left( 1 + x^2 + \frac{\tilde{\xi} (b^2 + 3 x^2 + 3 b x)}{3 x (b + x)^3} \right) - \left( \frac{1 + x^2 \left( 3 - \frac{\tilde{\xi}}{(b + x)^4} \right)}{4 \pi x R} \right) (4 \pi C x^{d-2}),
\end{align}
which simplified to
\begin{align}
     F
      &= \frac{C x}{R} \left[ 1 - x^2 + \frac{2 \tilde{\xi} (b^2 + 3 x^2 + 3 b x)}{3 x (b + x)^3} + \frac{\tilde{\xi} x^2}{(b + x)^4} \right] .
\end{align}
In the specific case where the parameter \(\xi = 0\), which corresponds to the Schwarzschild-AdS limit, the function \(F\) is expressed as
\begin{align}
    F = \frac{C x}{R} (1 - x^2).
\end{align}
This result is in full agreement with \cite{Ahmed:2023dnh} for the non-rotating scenario, achieved by setting \(z = 0\) and $d=4$. \\
Given the complexity of the analytic expression for the free energy \( F \), we proceed with the numerical analysis. The dependence on the radius \( R \) is fixed by scale invariance; consequently, the free energy–temperature plots \( F(T) \) corresponding to different values of \( R \) differ only by an overall rescaling. The free energy–temperature (\( F\text{--}T \)) curve is shown in Fig.~\ref{CFT-F-T-THP}, which indicates the presence of a first-order Hawking–Page phase transition.
\\It is found that there exist critical values of the deformation and control parameters, expressed in terms of the dimensionless parameters, denoted as \( \{b_c, \tilde{\xi}_c\} \). The role of these critical points is quite subtle. There arise two situations; first if the $\tilde{\xi}=fixed$ but allowed to vary $b$, then there exists a critical $b_{c}$ above which there is Hawking-Page phase transition. Second, if $b=fixed$ but allowed to vary $\tilde{\xi}$, then there exists critical $\tilde{\xi}_{c}$ below which we observe the Hawking-Page transition. 
From Fig. \ref{CFT-vary-para-alpha-b}, we can infer the behaviour of the free energy--temperature curve, and consequently the Hawking-Page transition temperature \(T_{HP}\), under variations of parameters $\tilde{\xi}$ and $b$. The right panel displays the effect of varying \(\tilde{\xi}\), while the left panel corresponds to varying \(b\). It is evident that, for fixed values of \( (R, C, b) \), increasing \(\tilde{\xi}\) leads to an increase in free energy \(F\) and to a higher Hawking-Page temperature \(T_{HP}\). On the other hand, for fixed \( (R, C, \tilde{\xi}) \), increasing \(b\) results in a decrease in the values of \(F\), accompanied by an decrease in \(T_{HP}\). Nevertheless, they qualitatively exhibit the same underlying thermodynamic phase structure.
In each of these cases, the free energy curve consists of an upper and a lower branch that meet at a cusp, corresponding to low-entropy (small black holes) and high-entropy (large black holes) states, respectively.  
The temperature attains a minimum as a function of $x$ at the cusp $T_{min}$.
In the rest of the cases, the free energy is single valued as a function of the temperature.
Furthermore, the high-entropy deconfined phase becomes dominant in the ensemble when the free energy satisfies \( F < 0 \), whereas the confined phase is thermodynamically favoured when \( F > 0 \). This confinement/deconfinement transition corresponds, in the gravitational dual, to a generalised Hawking–Page phase transition between large AdS black holes and thermal AdS spacetime. Consequently, there exists a line in the $\tilde{\xi}$–$T$ (equivalently in $b-T$) plane along which first-order phase transitions between the confined and deconfined phases occur. This coexistence line can be determined analytically by setting $F = 0$ and eliminating $x$ in favour of the temperature, which gives\footnote{We present the $\tilde{\xi}$–$T$ plane due to its analytical tractability; however, a similar analysis can be carried out for the $b$–$T$ plane as well, although it requires a fully numerical treatment.}
\begin{align}
    \tilde{\xi}(x) = \frac{x^2 -1}{\frac{2 (b^2 + 3 x^2 + 3 b x)}{3 x (b + x)^3} + \frac{x^2}{(b + x)^4}}\,.
\end{align}
In Fig.~\ref{de-confinment}, we plot this line in the $\tilde{\xi}$–$T$ plane. 
These critical values mark boundaries in the parameter space where the thermodynamic behavior of the system undergoes a qualitative change—typically from a regime exhibiting a stable black hole phase (with a Hawking-Page transition) to one where such a transition is absent. Furthermore, since \( \tilde{\xi}_c \) is determined at fixed \(b\), and 
vice-versa, the two critical parameters are interrelated. This interdependence is physically reasonable: while \(\tilde{\xi}\) encodes the geometric deformation, \( b \) governs the scale at which these deformations influence the thermodynamics. Together, they control whether the free energy \( F \) crosses zero at a real, positive temperature, which signifies the presence of a Hawking-Page transition.
\\In general, the Hawking-Page phase transition temperature $T_{HP}$ can be obtained by first computing the value of $x$ from $F(x)=0$ and then substituting this into the expression of temperature Eq (\ref{temp-bdry}). For the vanishing deformation parameter, one gets the value of $T_{HP}$ as
\begin{align}
    T_{HP}= \frac{1}{\pi R},
\end{align}
which is a well-known value for the AdS-Schwarzschild black hole \cite{Ahmed:2023dnh}.
Our observation also suggests that the Hawking-Page transition temperature $ T_{\text{HP}} $ is independent of $ C $ (See Fig \ref{CFT-FT}). 
However, $ C $ affects the magnitude of $ F $, not its critical points, which aligns with $ F $ scaling with $ C $ while $ T_{\text{HP}} $ remains a geometric property. This confirms that deformations alter the gravitational background but not the scaling properties of the dual CFT. 
\\Finally, in the special case where the deformation parameter vanishes, i.e., \(\tilde{\xi} = 0\), we recover a Hawking–Page-like phase transition. This behaviour is depicted in Fig~\ref{CFT-alpha-0-FT}. 
Although this transition was originally identified by Hawking~\cite{Hawking:1982dh}, it has since been recognized that the generalized Hawking–Page transition also occurs in various other black hole spacetimes~\cite{Cong:2021jgb,Tarrio:2011de,Pedraza:2018eey}. 
Furthermore, in the limit of a vanishing control parameter—corresponding to the CFT dual of the charge-like AdS black hole—we observe a van der Waals–like phase structure, consistent with the results reported in~\cite{Cong:2021jgb}.
\subsubsection{Phase transition in ensemble $(p, C)$} 
\begin{figure}[!h]
    \begin{subfigure}
        \centering
   \includegraphics[width=0.5\linewidth]{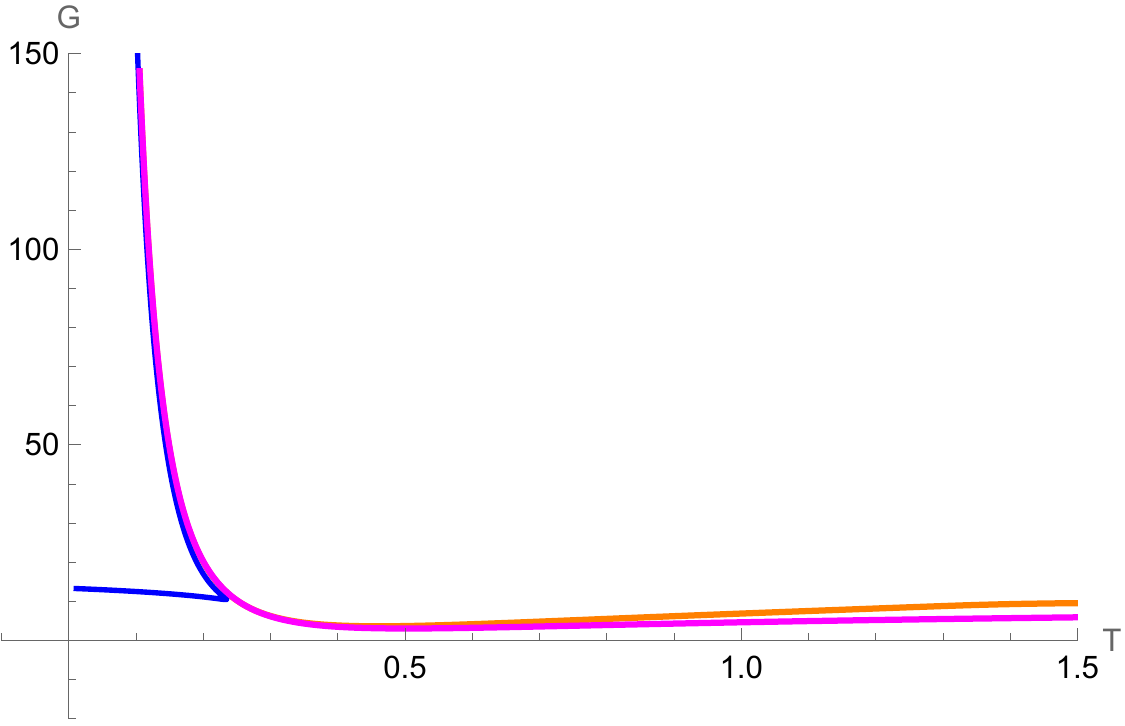}
   \end{subfigure}
     \begin{subfigure}
        \centering
   \includegraphics[width=0.5\linewidth]{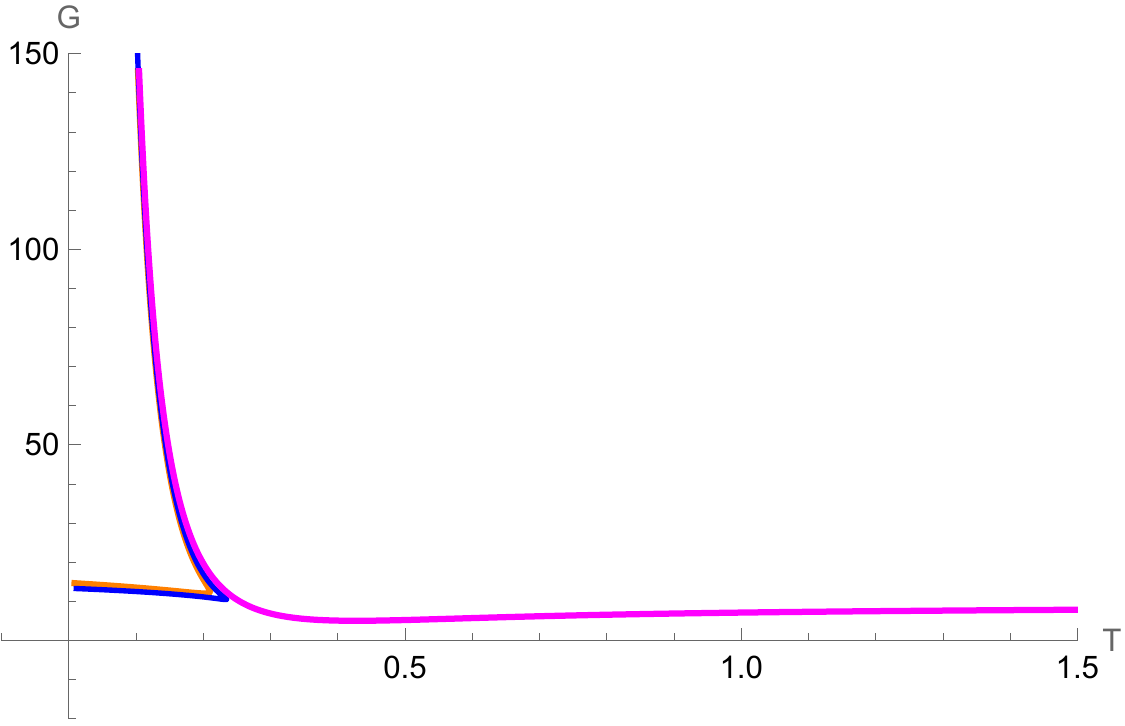}
   \end{subfigure}
    \caption{The $G$--$T$ diagram is displayed for two distinct scenarios. \textbf{(Left)}: The parameter $b=0.1$ is held fixed, while $\tilde{\xi}$ is varied across the following values: (a) $\tilde{\xi} = 0.05$(magenta), (b) $\tilde{\xi} = 1$(blue), and (c) $\tilde{\xi} = 0.105$(orange). \textbf{(Right)}: The deformation parameter $\tilde{\xi}=1$ is kept fixed, and the parameter $b$ is varied for three distinct values: (a) $b =0.01$(orange), (b) $b =0.1$(blue), and (c) $b =1$(magenta). Both qualitatively look similar. $C=2, p=0.4$.}
    \label{new-G-T-plot}
\end{figure}

In this case, we consider the ensemble with fixed pressure and central charge \((p, C)\), for which the appropriate thermodynamic potential is the free energy, defined as $G = E - T S + p \mathcal{V}$. This yields the following expression
\begin{align}
     G = \mathcal{E} - T S + p \mathcal{V} = (T S + \mu C ) - T S + p \mathcal{V} = \mu C + p \mathcal{V}\,.
\end{align}
The chemical potential is given by
\begin{align}
    \mu = \frac{x}{R} \left( 1 - x^2 + \frac{\tilde{\xi} x^2}{(b + x)^4} + \frac{2 \tilde{\xi} (b^2 + 3 b x + 3 x^2)}{3 x (b + x)^3} \right),
\end{align}
which depends explicitly on the dimensionless parameter \( x \) and the scale \( R \).
Since the pressure \( p \) satisfies the constraint  $p = \frac{E}{2\, \mathcal{V}}$, where the spatial volume of the boundary CFT is given by $\mathcal{V} = \Omega_{2} R^{2}$.
Using the expression for the energy \( E \), the pressure becomes
\begin{align}
    p = \frac{\frac{2 C x}{R} \left( 1 + x^2 + \frac{\tilde{\xi} (b^2 + 3 x^2 + 3 b x)}{3 x (b + x)^3} \right)}{2\, \Omega_{2} R^{2}}.
\end{align}
For fixed \( p \) and \( C \), solve for \( R \) as a function of \( x \):

\[ R = \frac{ C x \left( 1 + x^2 + \frac{\tilde{\xi} (b^2 + 3 x^2 + 3 b x)}{3 x (b + x)^3} \right)}{\Omega_{2}~p} \,.\] 
Thus,
\begin{align}
    G(x; p, C) = C \cdot \frac{x}{R(x; p, C)} \left( 1 - x^2 + \frac{\tilde{\xi} x^2}{(b + x)^4} + \frac{2 \tilde{\xi} (b^2 + 3 b x + 3 x^2)}{3 x (b + x)^3} \right) + p (4 \pi R(x;p,C)^2) \,.
\end{align} 
Although a variety of Gibbs free energy versus temperature (\(G\)--\(T\)) curves can be obtained for different parameter choices, not all of them display qualitatively distinct behaviour.
In particular, the two scenarios depicted in the left and right panels of Fig.~\ref{new-G-T-plot}, corresponding to fixed values of \( b \) and \( \tilde{\xi} \), respectively, do not exhibit any qualitative difference in their phase structure.
The central charge \( C \) influences the $G$--$T$ plots by stretching them along the temperature axis; this effect becomes more pronounced as \( C \) decreases. In contrast, increasing the pressure \( p \) stretches the plots along the free energy axis.

\begin{figure}[!h]
   \begin{subfigure}
    \centering
   \includegraphics[width=0.5\linewidth]{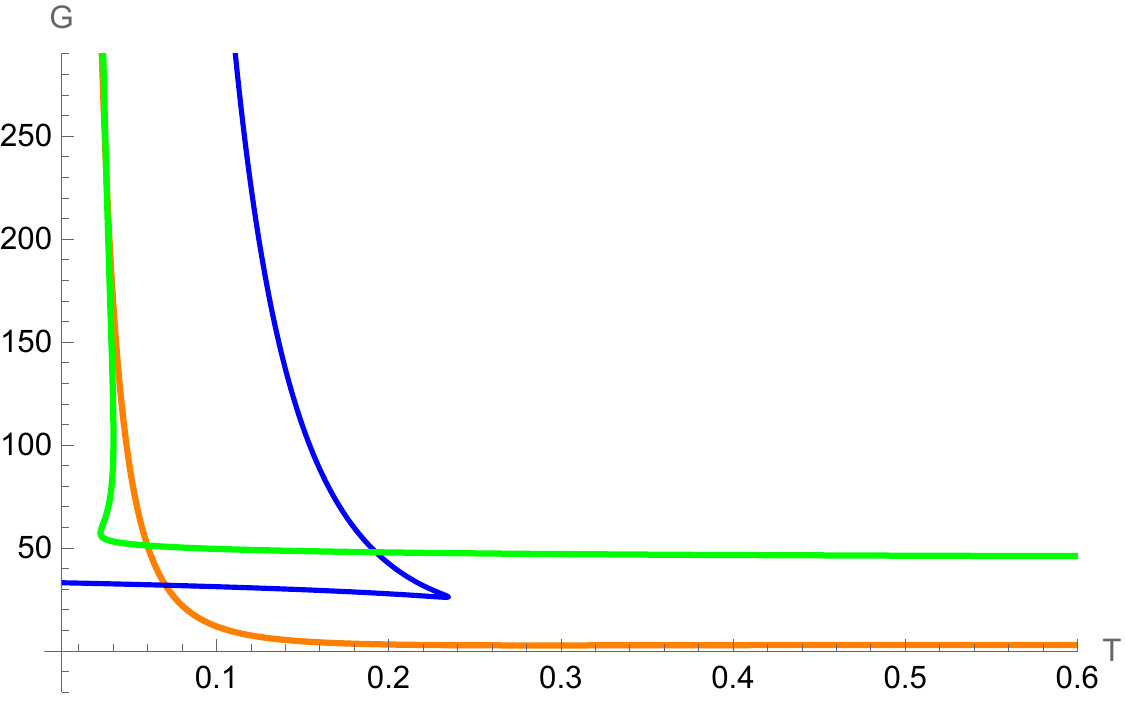} 
   \end{subfigure}
   \begin{subfigure}
    \centering
    \includegraphics[width=0.5\linewidth]{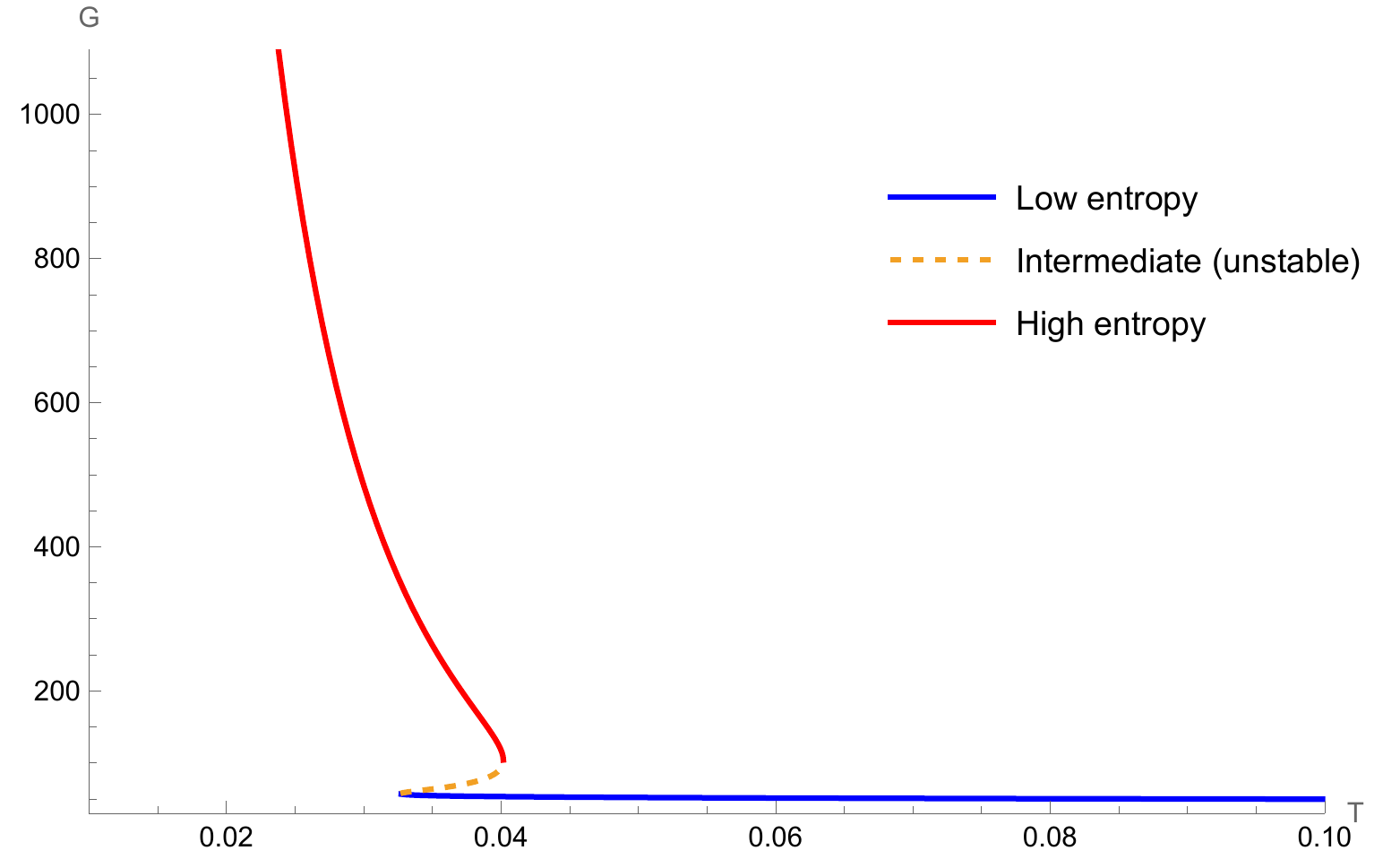}
   \end{subfigure}
    \caption{Left shows plot of Gibbs free energy \( G \) versus temperature \( T \) for three different sets of parameter values.  
Blue curve: \( C = 5,~ p = 1,~ \tilde{\xi} = 1,~ b = 0.1 \);  
Green curve: \( C = 5,~ p = 0.2,~ \tilde{\xi} = 3,~ b = 0.5 \);  
Orange curve: \( C = 4,~ p = 0.1,~ \tilde{\xi} = 0.1,~ b = 0.1 \).  
On the right, the same green curve is shown separately to highlight its three distinct branches corresponding to low entropy (LE), intermediate entropy (IE), and high entropy (HE) phases.}
    \label{G-T-bdry-all-diff}
\end{figure}

The Gibbs free energy–temperature plot exhibits subtle and potentially novel features, with its structure varying significantly depending on the choice of parameters. In Fig \ref{G-T-bdry-all-diff} (Left), we have drawn three distinct $G-T$ plots in which the green curve qualitatively differs from the curves in Fig \ref{new-G-T-plot}. For a suitable set of parameters, the plot displays three distinct branches—low-entropy (LE), intermediate-entropy (IE), and high-entropy (HE). However, in the next section, the specific heat capacities plots reveal that they do not match the traditional plot of the swallowtail-like curve.
\subsubsection{Phase transition in ensemble $(p, \mu)$}
\begin{figure}[!h]
     \begin{subfigure}
        \centering
    \includegraphics[width=0.5\linewidth]{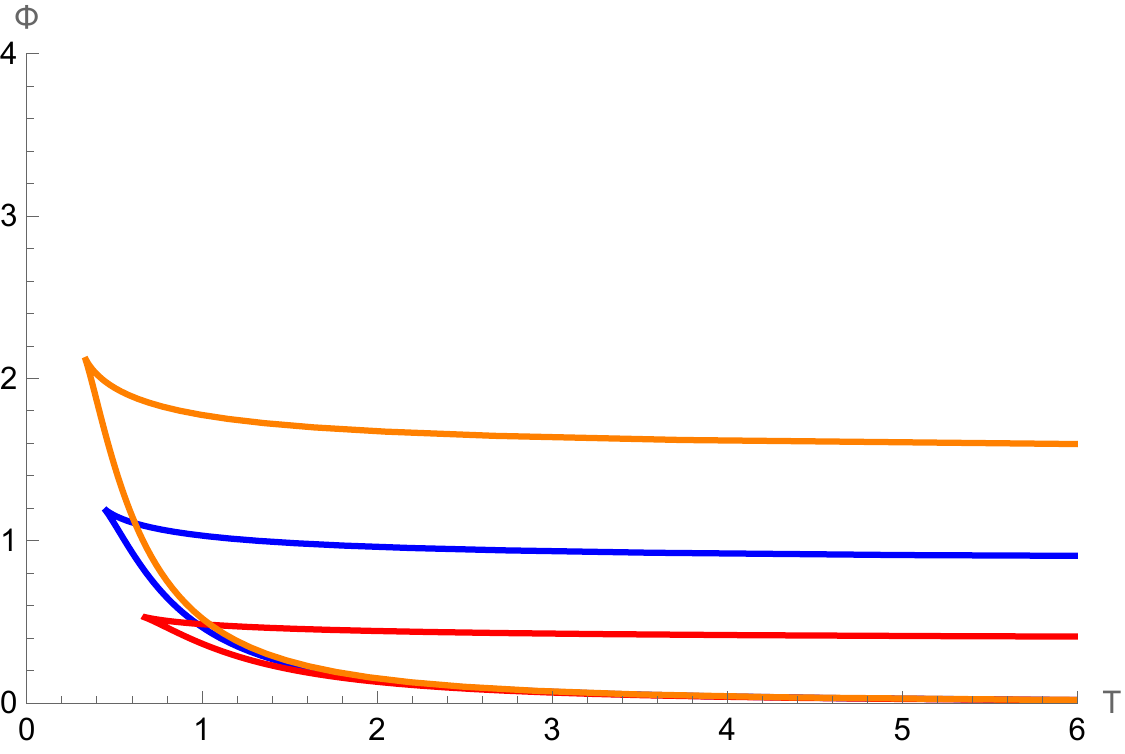}
   \end{subfigure}
   \begin{subfigure}
       \centering
    \includegraphics[width=0.5\linewidth]{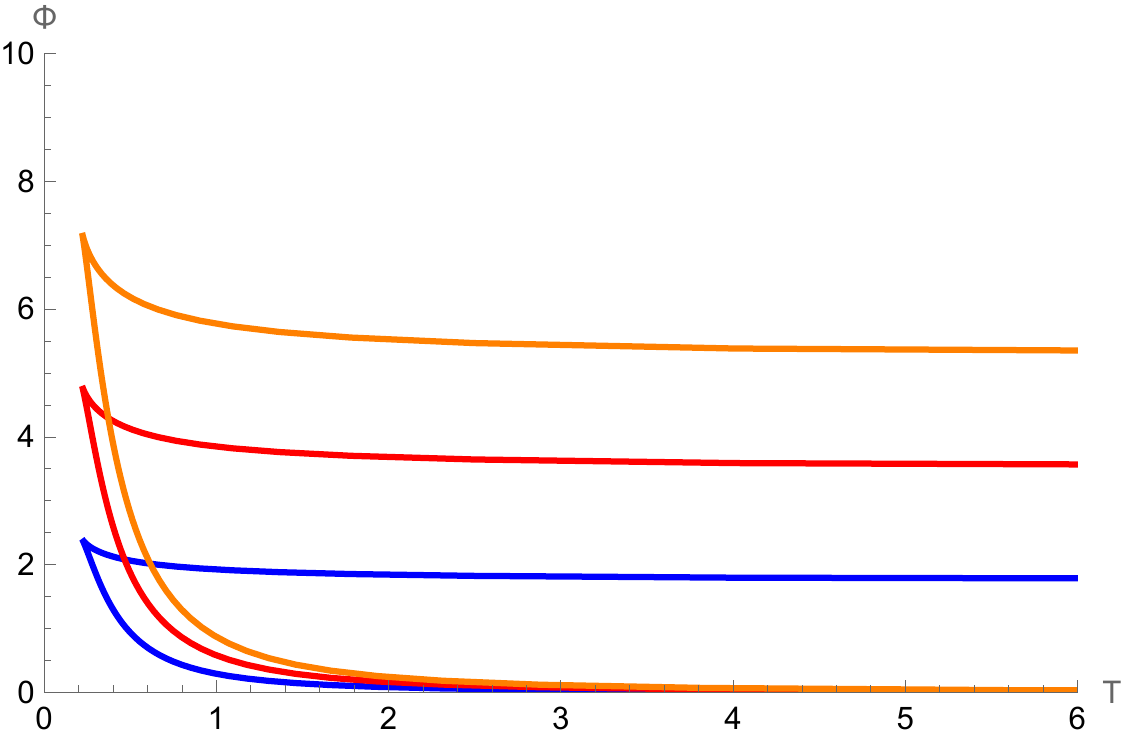}
   \end{subfigure}
    \caption{Plots of the thermodynamic potential $\Phi$ as a function of temperature $T$ for various parameter choices. 
\textbf{(Left)}: Curves are plotted for fixed values $C=2$, $\tilde{\xi}=1$, $b=0.4$, and $p=0.4$, with varying chemical potentials: $\mu=6$ (red), $\mu=4$ (blue), and $\mu=3$ (orange). 
\textbf{(Right)}: Curves correspond to fixed $C=2$, $\tilde{\xi}=1$, $b=0.4$, and $\mu=2$, with varying pressures: $p=0.4$ (red), $p=0.2$ (blue), and $p=0.6$ (orange). These plots do not show the phase transition.}
    \label{CFT-phi-T}
\end{figure}

For the ensemble with fixed \( (p, \mu) \), the corresponding thermodynamic potential is given by $  \Phi = \mathcal{E} - T S + p \mathcal{V} - \mu C$. Invoking the Euler relation,
we find that the free energy simplifies to $\Phi = p \mathcal{V}$.
Hence, in this ensemble, the thermodynamic potential is determined solely by the pressure and volume.
\begin{align}
    \Phi(x; R) = p (4 \pi R(x; \mu)^2)\,,
\end{align} 
where, $R(x;\mu)=  \frac{x}{\mu} \left( 1 - x^2 + \frac{\tilde{\xi} x^2}{(b + x)^4} + \frac{2 \tilde{\xi} (b^2 + 3 b x + 3 x^2)}{3 x (b + x)^3} \right) $. In this case, the thermodynamic ensemble does not exhibit any critical phenomena or phase transitions, as illustrated in Fig.~\ref{CFT-phi-T}. The $\Phi$--$T$ curve displays a double-branched structure with a “tip” that corresponds to a maximum of the thermodynamic potential $\Phi$ and a minimum of the temperature $T$.
\\

In the next section, we compute the critical points and examine the thermodynamic stability. 
\subsection{Critical points and thermodynamic stability}
\begin{figure}
    \centering
    \includegraphics[width=0.5\linewidth]{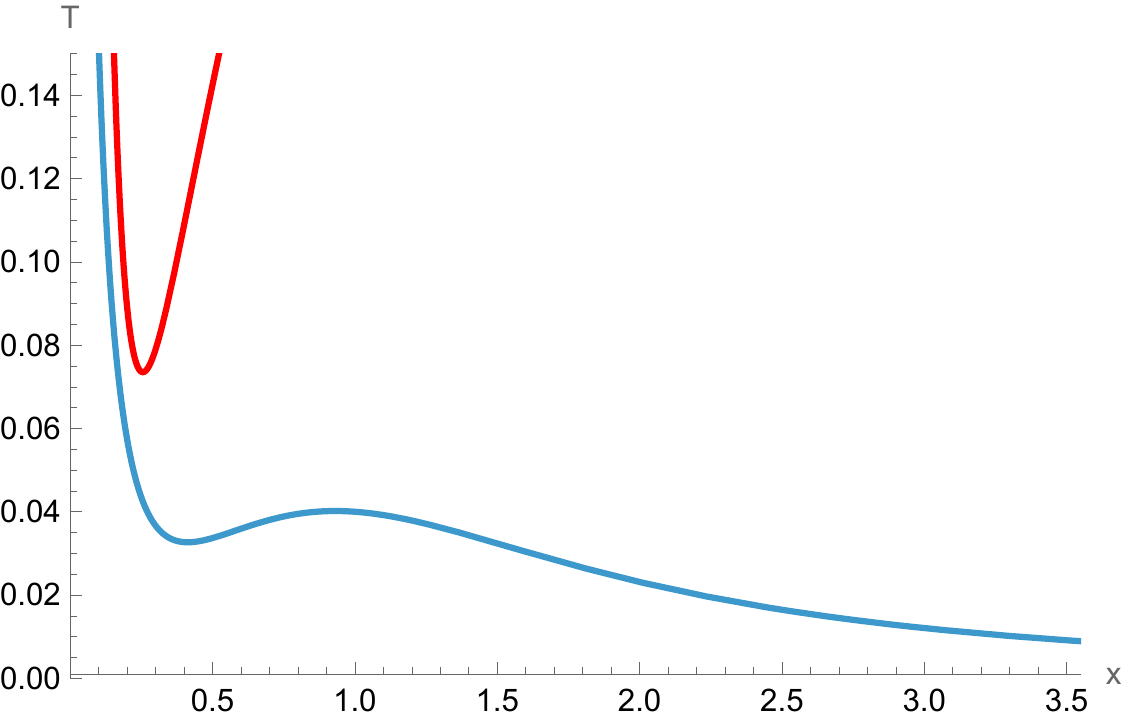}
    \caption{
    $T$--$x$ plot for two different scenarios: the red curve corresponds to the free energy $F$ with parameters $\tilde{\xi} = 1.4$, $b = 0.3$, $C = 25$, and $R = 1$; it shows a minimum, whereas the blue curve corresponds to the free energy $G$ with parameters $p = 0.2$, $\tilde{\xi} = 3$, $b = 0.5$, and $C = 5$, and it shows two extrema.
    }
    \label{T-x-CFT}
\end{figure}

\begin{figure}
     \begin{subfigure}
    \centering
    \includegraphics[width=0.5\linewidth]{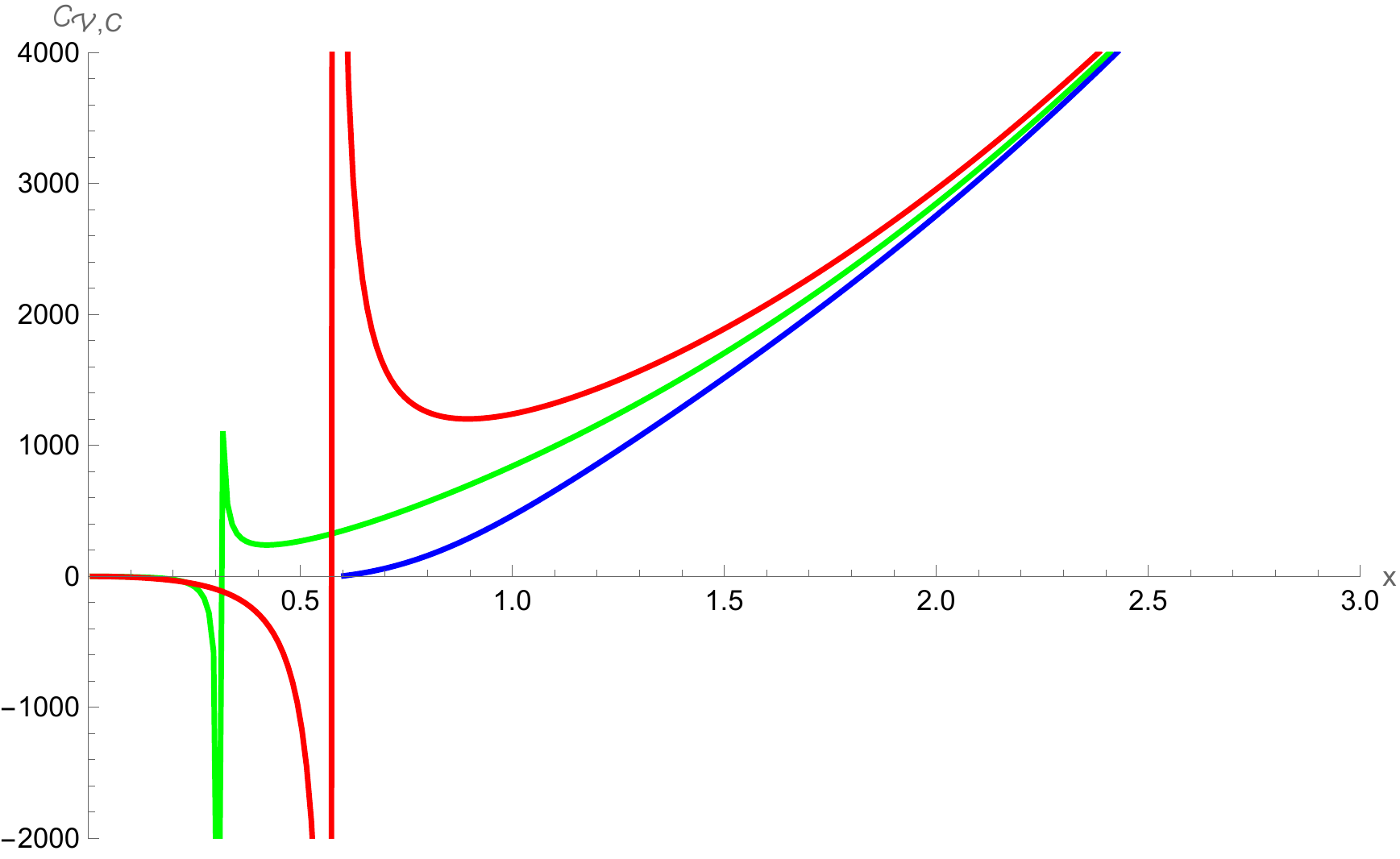}
    \end{subfigure}
    \begin{subfigure}
    \centering
    \includegraphics[width=0.5\linewidth]{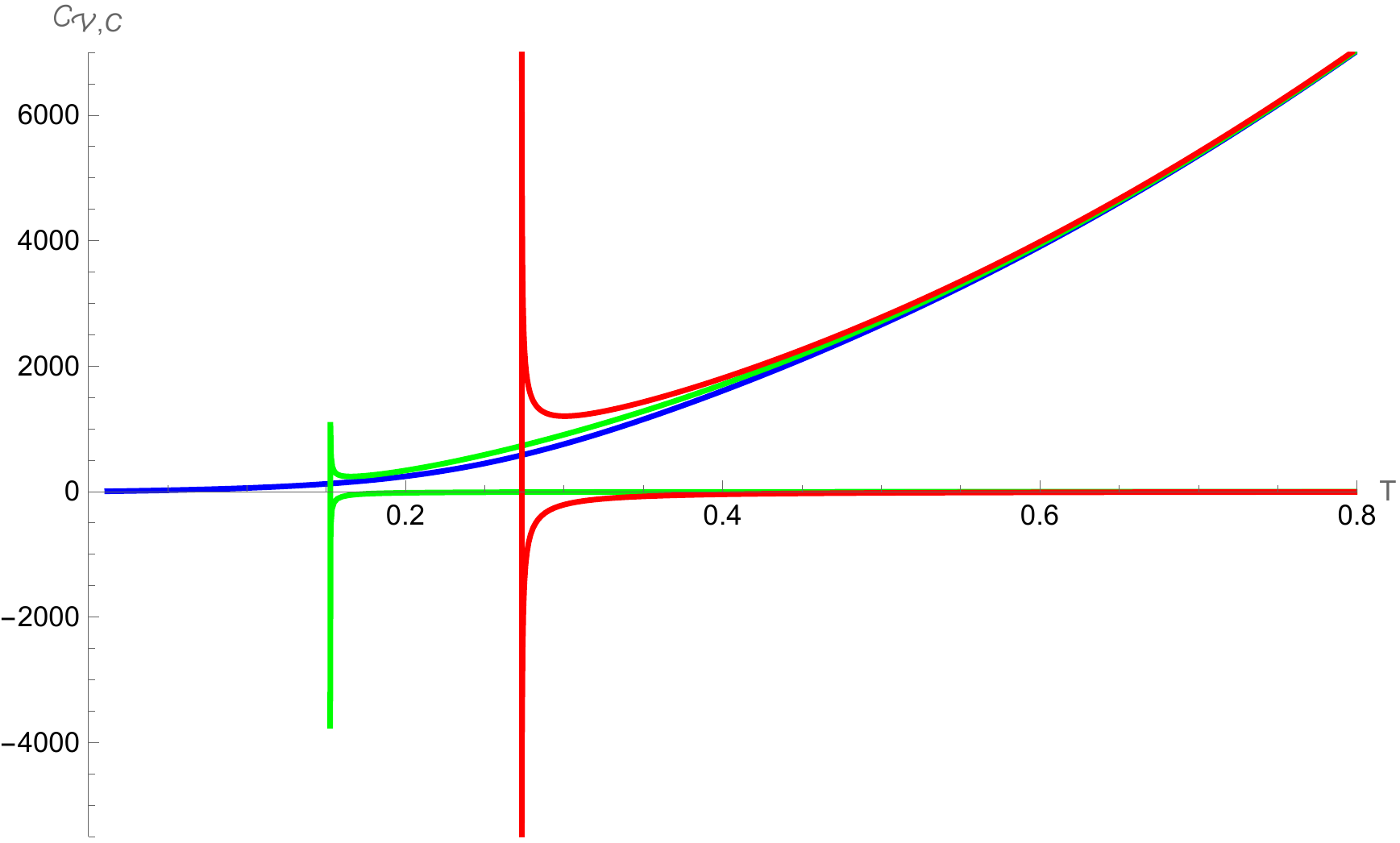}
    \end{subfigure}
\caption{Plots of specific heat \( \mathcal{C}_{C,\mathcal{V}} \) versus \( x \) and \( \mathcal{C}_{C,\mathcal{V}} \) versus temperature \( T \) for the same parameter values as in Fig \ref{CFT-vary-para-alpha-b} (Left).}
\label{heat-cap-fixed-para}
\end{figure}

 In this section, we obtain the critical points of the (extended) CFT thermodynamics corresponding to deformed AdS-Schwarzschild black holes.
 Further, we derive the heat capacities for the ensembles at fixed $(\mathcal{V}, C), ( p, C)$ and $(p, \mu)$, and analyse the thermodynamic stability of the different phases in these ensembles.
In general the critical point is obtained as
\begin{align}
\frac{\partial T}{\partial x} = 0, \quad \frac{\partial^2 T}{\partial x^2} = 0,
\end{align}
which provide two conditions
\begin{align}
    3 x^2 - 1 + \frac{\tilde{\xi} x^2 (3 x - b)}{(b + x)^5} = 0\\
    2 \tilde{\xi} x^3 (2 b - 3 x) = - (b + x)^6 
\end{align}
Fixing $\tilde{\xi}$ (or $b$), one can numerically solve the two governing equations to determine the corresponding values of $x$ and $b$ ( or $\tilde{\xi}$). 
As illustrated in the Fig \ref{T-x-CFT}, the $T$--$x$ curves exhibit critical behaviour, with the presence of extrema indicating the occurrence of phase transitions.
The heat capacity for the thermodynamic stability of the dual CFT system associated with the different ensembles is defined as
\begin{equation}
    \mathcal{C}_{\chi} \equiv T \left( \frac{\partial S}{\partial T} \right)_{\chi}, \chi \equiv (C,\mathcal{V}), (p,C), (p, \mu)\,.
\end{equation}
Since the left and right panels of Fig~\ref{CFT-vary-para-alpha-b} exhibit qualitatively similar behaviour, we present, without loss of generality, only the heat capacity curves corresponding to the left panel. The corresponding plots for the right panel follow the same qualitative trends.
The heat capacity as a function of \(x\) is shown in Fig~\ref{heat-cap-fixed-para}, which includes branches with both positive and negative values of \(\mathcal{C}_{\mathcal{V}C}\). The positive heat capacity branches correspond to the low-entropy solutions shown in Fig~\ref{CFT-vary-para-alpha-b}, while the negative branches are associated with the high-entropy counterparts.
For parameter values satisfying \( b > b_c \) (or equivalently \( \tilde{\xi} < \tilde{\xi}_c \)), which give rise to the double-branch structure in Fig~\ref{CFT-vary-para-alpha-b}, the heat capacity exhibits both positive and negative segments. This indicates the presence of thermodynamic instability, with a first-order phase transition occurring between the two branches.
In contrast, for \( b < b_c \) (or \( \tilde{\xi} > \tilde{\xi}_c \)), the heat capacity displays a single, entirely positive branch, signifying a thermodynamically stable phase throughout.

\begin{figure}[!h]
     \begin{subfigure}
        \centering
    \includegraphics[width=0.5\linewidth]{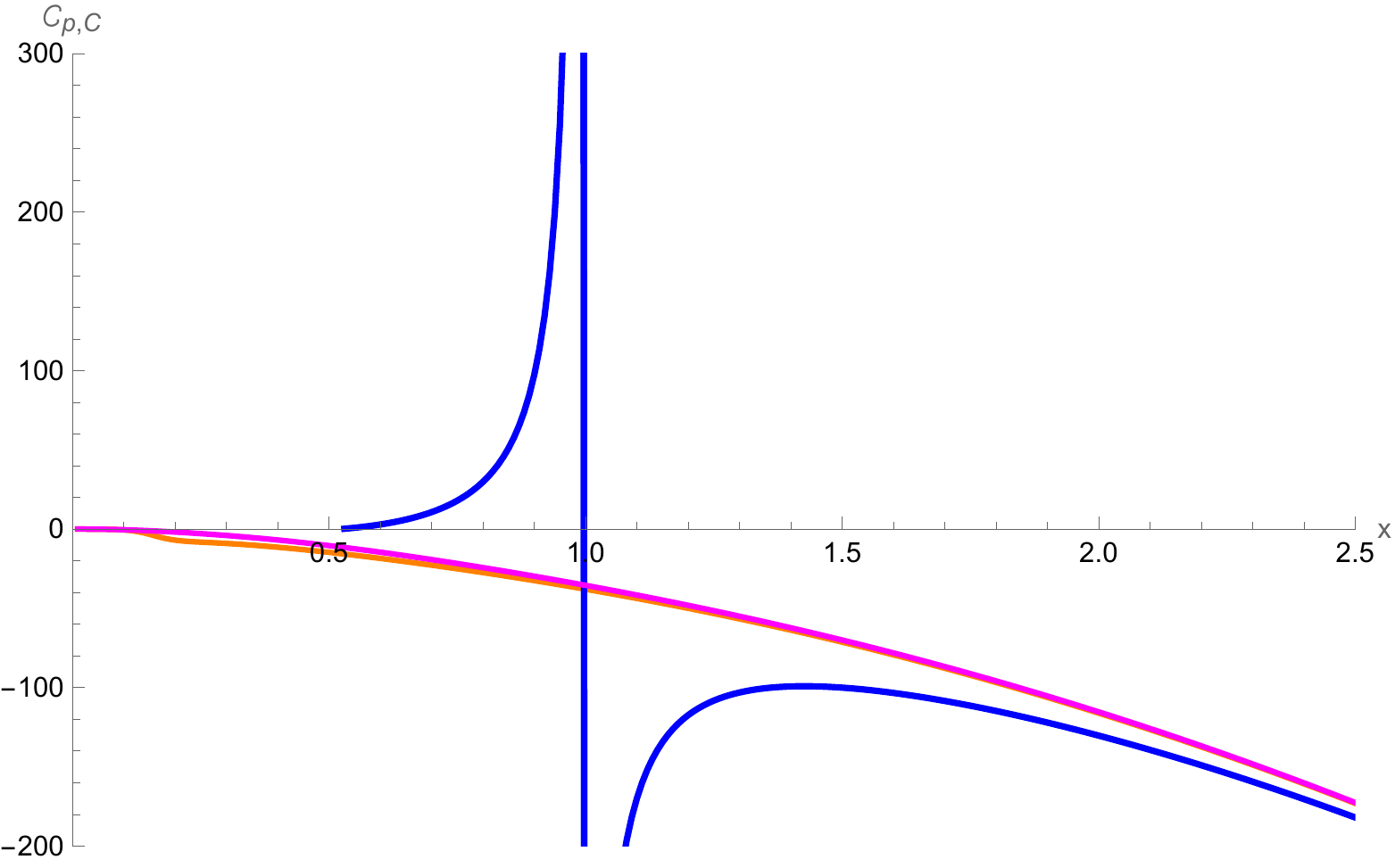}
   \end{subfigure}
   \begin{subfigure}
       \centering
    \includegraphics[width=0.5\linewidth]{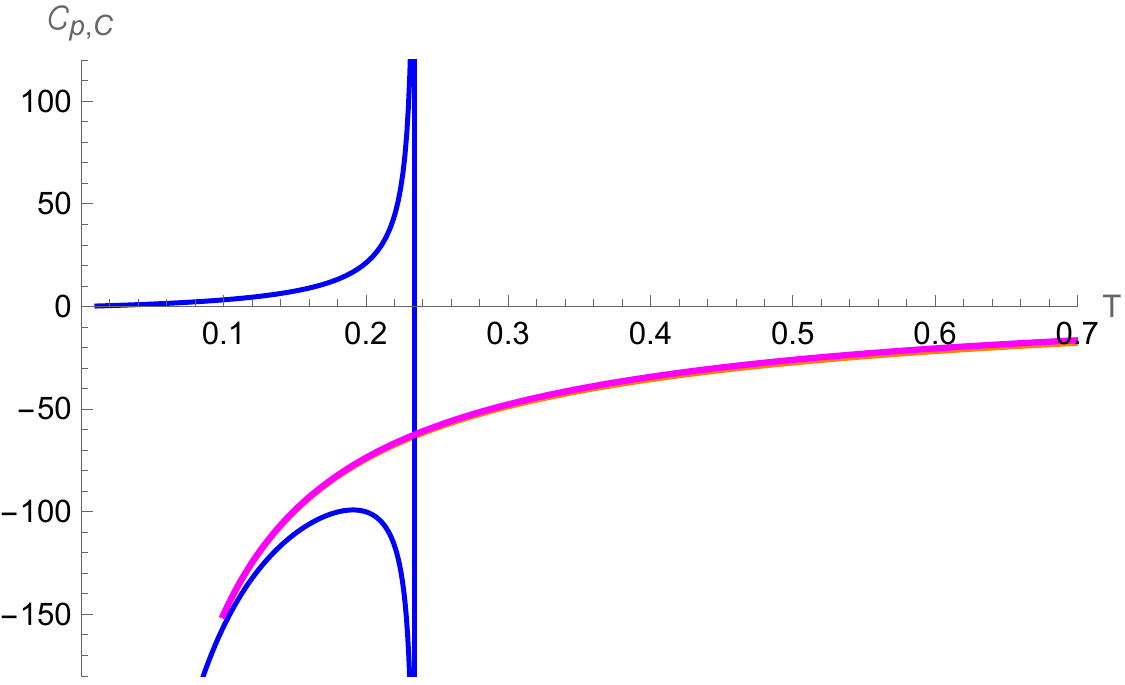}
   \end{subfigure}
    \caption{Plots of specific heat \( \mathcal{C}_{C,\mathcal{V}} \) versus \( x \) and \( \mathcal{C}_{C,\mathcal{V}} \) versus temperature \( T \) for the same parameter values as  in Fig \ref{new-G-T-plot} (Left).}
    \label{spec-heats-for-G-T}
\end{figure}

\begin{figure}[!h]
     \begin{subfigure}
        \centering
    \includegraphics[width=0.5\linewidth]{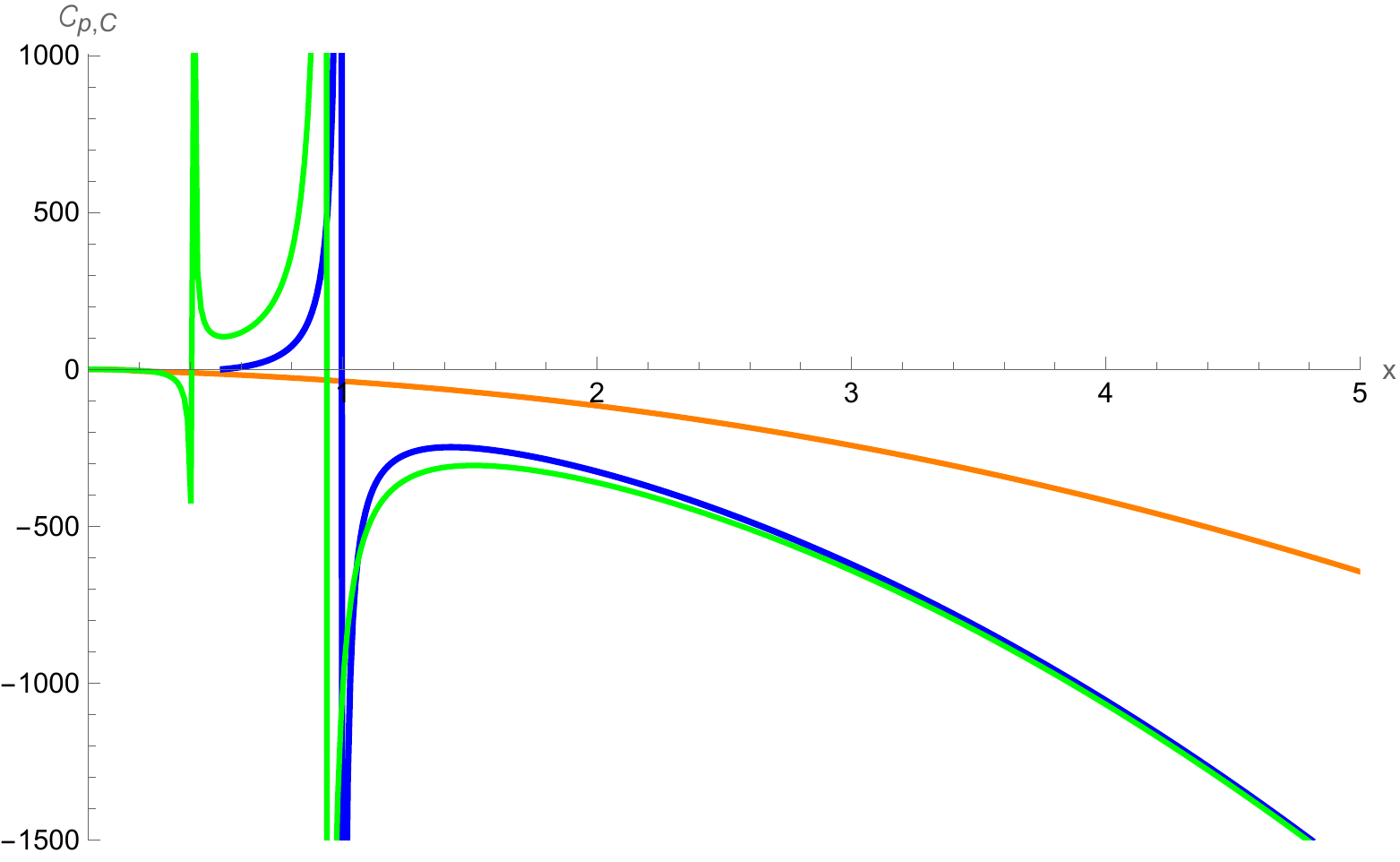}
   \end{subfigure}
   \begin{subfigure}
       \centering
    \includegraphics[width=0.5\linewidth]{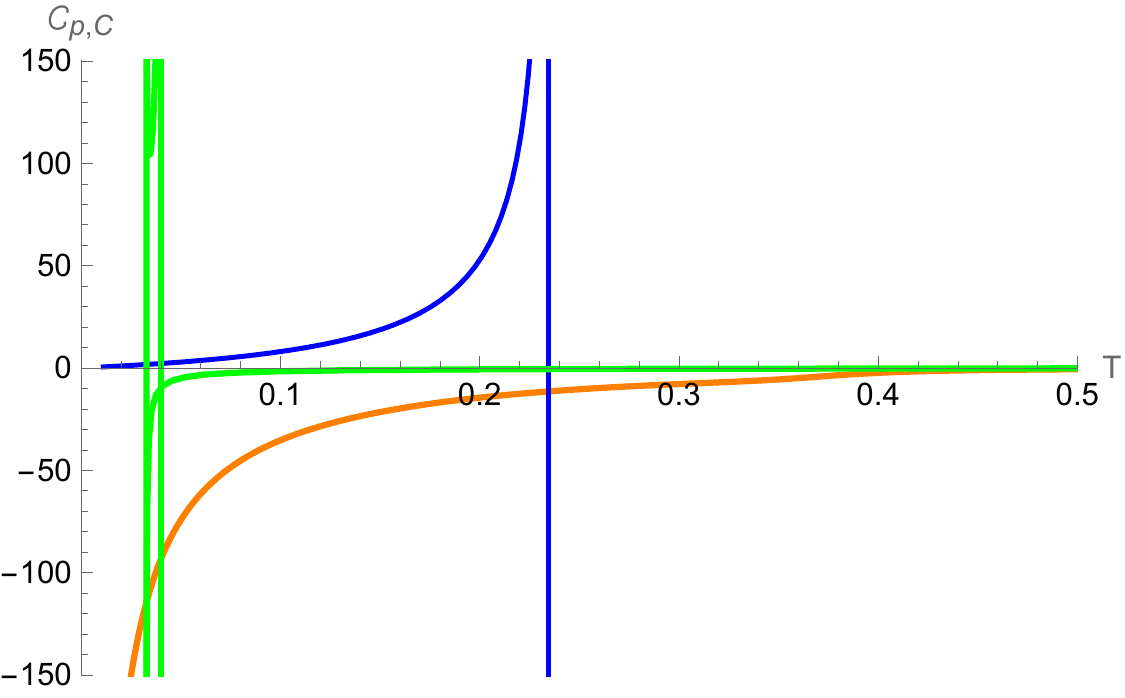}
   \end{subfigure}
\caption{Heat capacities for the same set of parameters as in Fig \ref{G-T-bdry-all-diff}.}
    \label{C-plots-G-T-last}
\end{figure}

\begin{figure}
    \centering
    \includegraphics[width=0.5\linewidth]{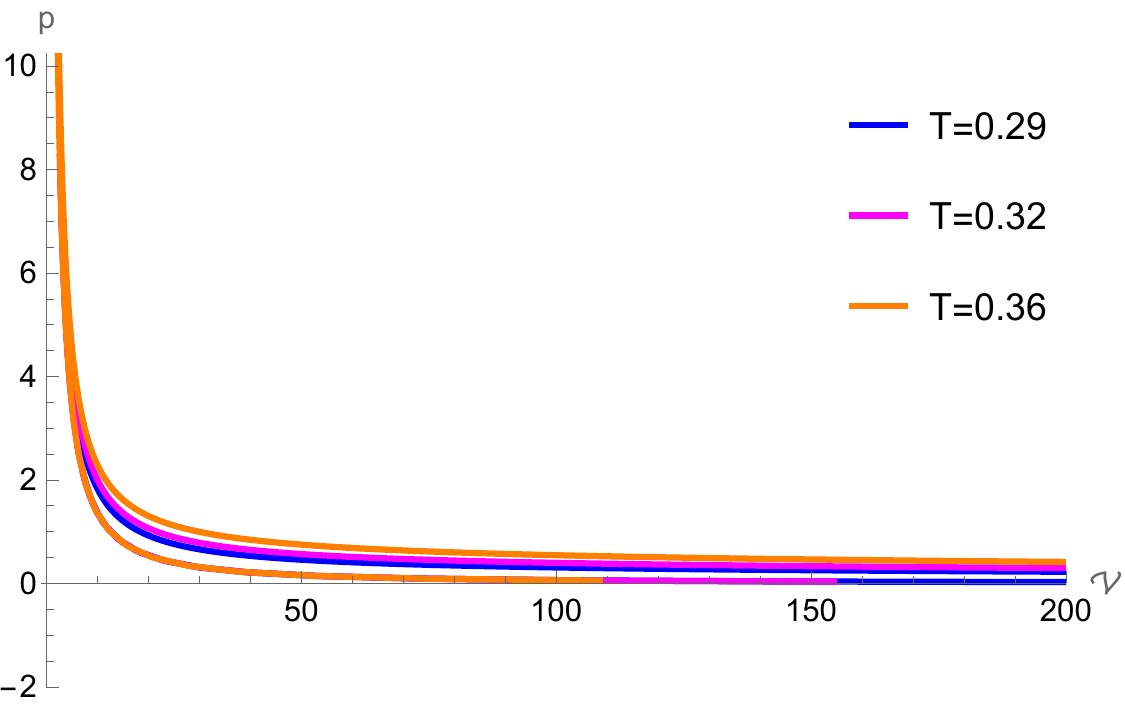}
    \caption{   These \(p\)--\(\mathcal{V}\) curves corresponding to the green curve in Fig.~\ref{G-T-bdry-all-diff} are dual-branched and, unlike the typical van der Waals fluid, do not exhibit a critical temperature characterized by an inflection point in the isotherm.}
    \label{p-V-for-G-T}
\end{figure}

Corresponding to Figs.~\ref{new-G-T-plot}--\ref{G-T-bdry-all-diff}, we present the various possible behaviours of the specific heat capacity in Figs.~\ref{spec-heats-for-G-T}--\ref{C-plots-G-T-last}. In these plots, regions where the specific heat \(\mathcal{C}_{p,C}\) is negative indicate thermodynamically unstable phases, whereas positive values signify stable regions. Notably, for the green curve in Fig.~\ref{G-T-bdry-all-diff}, the corresponding specific heat versus \(x\) plot shows only a single segment with positive specific heat. This implies that only the IE branch corresponds to a thermodynamically stable phase, while the LE and HE branches are unstable. In contrast, in our setup, the deformation parameters lead to a situation where more than one branch becomes unstable, resulting in a unique thermodynamic structure wherein the system favours a single stable phase. This behaviour marks a significant departure from the standard vdW fluid scenario, where a first-order phase transition typically occurs between two stable branches. Also, in order to further compare the swallowtail with the vdW fluid, we have drawn the $p-\mathcal{V}$ for the ensemble $(p,C)$ associated with the Fig \ref{new-G-T-plot}. These plots are shown in the Fig \ref{p-V-for-G-T}. Clearly they do not resemble with the usual vdW fluid behaviour.

\begin{figure}[!h]
     \begin{subfigure}
        \centering
        \includegraphics[width=0.5\linewidth]{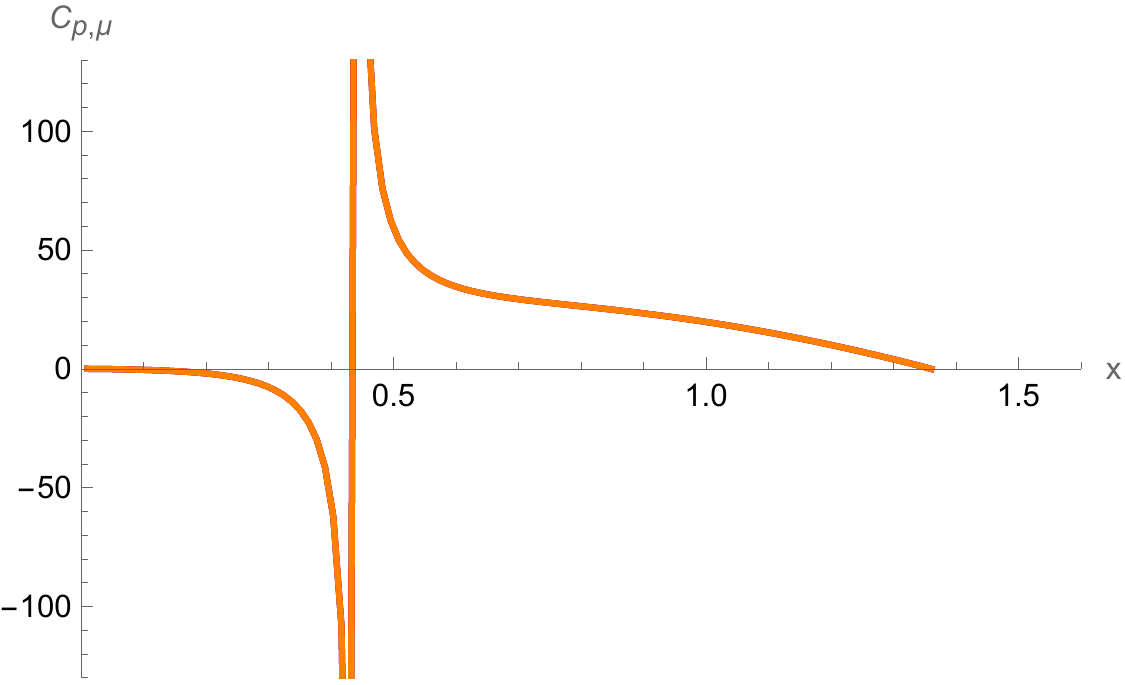}
   \end{subfigure}
   \begin{subfigure}
       \centering
    \includegraphics[width=0.5\linewidth]{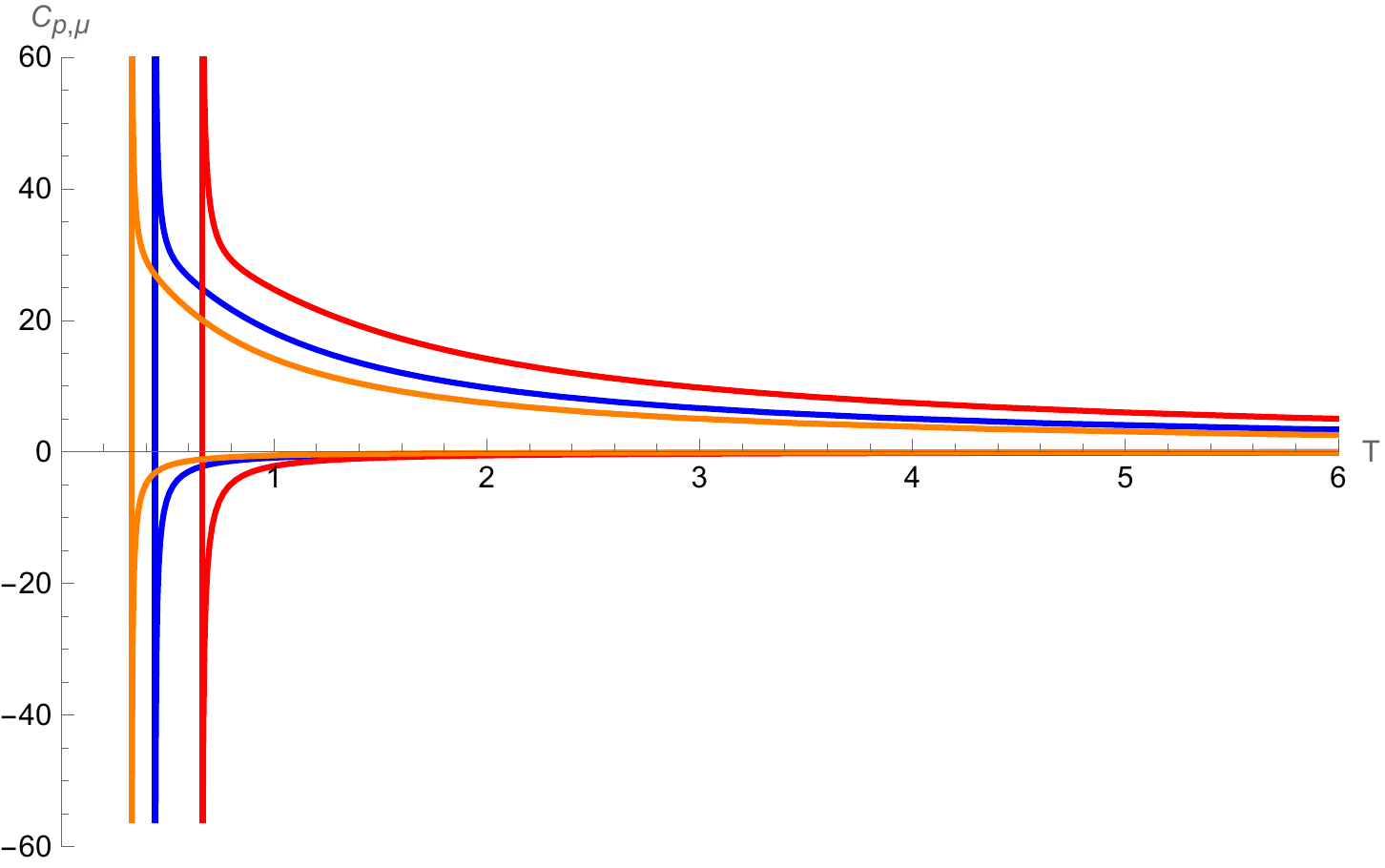}
   \end{subfigure}
    \caption{Heat capacities for the same set of parameters as in Fig. \ref{CFT-phi-T}(Left). In the left panel, the plot shows three overlapping curves, indicating that all three parameter choices yield nearly identical thermodynamic behaviour in this ensemble.}
    \label{C-plots-Phi}
\end{figure}

Finally, for the sake of completeness, we have also included in Fig.~\ref{C-plots-Phi} the specific heat curves for the \((p, \mu)\) ensemble, corresponding to the free energy behaviour shown in Fig.~\ref{CFT-phi-T}. \\
\\
\\
\\
\section{Conclusions} \label{conclusion}
In this work, we have done an extensive investigation of the thermodynamic phase transition of the deformed AdS-Schwarzschild black hole which is obtained from the gravitational decoupling (GD) method \cite{Khosravipoor:2023jsl}, and followed by the study of phase transition of the thermal CFT states dual to the deformed AdS-Schwarzschild black hole in an extended phase space, treating the cosmological constant and its conjugate quantity as thermodynamic variables associated with the pressure and volume, respectively. 

In the analysis of bulk thermodynamics, we have encountered a less trivial two-critical-point structure in the temperature profile, corresponding to two distinct deformation-controlled critical values $\xi_{c_{1}}$ and $\xi_{c_{2}}$. Furthermore, the range $\xi_{c_{1}}<\xi<\xi_{c_{2}}$ supports van der Waals–like transitions absent in both the Schwarzschild-AdS and RN–AdS limits. This thermodynamic behaviour is further supported by computing the universal ratio at the critical point; which is found to be close to the van der Waals value, and by confirming that the mean-field critical exponents are satisfied. On the dual CFT side, we find that the thermodynamic behaviour exhibits a Hawking–Page–like transition in the fixed $(\mathcal{V}, \mathcal{C})$  ensemble, reflecting the confinement–deconfinement phase transition characteristic of the dual CFT.  
Additional ensembles were also considered; notably, the fixed $(p,\mathcal{C})$ ensemble displays unconventional thermodynamic phase structure, warranting further theoretical investigation.

From the bulk perspective, the deformation parameter $\xi$ directly modulates gravitational geometry, creating a rich phase structure.
This bulk behaviour holographically maps to confinement/deconfinement transitions in the boundary $(\mathcal{V}, C)$ ensemble, where $\tilde{\xi} < \tilde{\xi}_c$ triggers deconfinement.
Here the GD deformation introduces new scales ($\xi,\eta$) that independently control confinement regimes.
Such GD couplings represent controlled departures from Einstein gravity that preserve (weak) energy conditions while enriching phase diagrams. Although our analysis focuses on the minimal geometric deformed background, a particular realization of the GD method, similar thermodynamic behavior may also arise in other GD frameworks. An interesting direction for future research would be to explore alternative formulations of gravitational decoupling \cite{Maurya:2022wwa,Brown:2017wpl} and their potential impact on extended holographic thermodynamics.

In future, we would like to extend the analysis for the generic case where the deformation parameters are promoted to the thermodynamic variables. This direction will also shed light on the holographic dictionary, which must be correspondingly generalized, leading to a more intricate structure in the dual field theory thermodynamics.
Future work should also explore higher-dimensional generalizations, particularly since critical exponents exhibit $d$-dependent scaling in analogous deformations \cite{Gheorghiu:2013jha,Estrada:2018vrl}, suggesting dimensionally tunable phase structures. 
\section*{Acknowledgment}
We would like to sincerely thank the anonymous reviewers for their fruitful comments and suggestions. BS acknowledges support from the IIT Kharagpur Institute Fellowship. BS would also like to acknowledge Mr. Sukhdeep Singh Gill for his valuable discussions.

\appendix

\section{Appendix: Smarr relation and deformation conjugates}\label{app:smarr}
In this appendix, we derive the Smarr relation for our deformed 
AdS--Schwarzschild black hole directly from the generalised Euler 
theorem for homogeneous functions, following the Ref \cite{Rodrigues:2022qdp}. 
We also compute the 
deformation conjugates $\Psi_\xi$ and $\Psi_\eta$. \\
\newline
The event horizon $r_h$ is defined by $F(r_h)=0$. Solving for the 
mass $M$ gives
\begin{equation}\label{eq:mass}
M(r_h,l,\xi,\eta)
=\frac{1}{6}\!\left(
3r_h+\frac{3r_h^3}{l^2}
+\xi\,\frac{\eta^2+3\eta r_h+3r_h^2}{(\eta+r_h)^3}
\right)
\end{equation}
or, equivalently, using 
the thermodynamic pressure 
$P=3/(8\pi l^2)\Rightarrow 1/l^2=8\pi P/3$, we may write
\begin{equation}
M\!\left(r_{h},P,\xi,\eta\right)
= \frac{r_{h}}{2}
+ \frac{4\pi P\,r_{h}^{3}}{3}
+ \frac{\xi}{6}\cdot
\frac{\eta^{2}+3\eta r_{h}+3r_{h}^{2}}{\left(\eta+r_{h}\right)^{3}}\,.
\label{M_rh_app}
\end{equation}
viewing the mass as a function of the four independent natural 
thermodynamic variables $(S,P,\xi,\eta)$. 
Explicitly, substituting $r_{+}=\sqrt{S/\pi}$ 
into \eqref{M_rh_app}:
\begin{equation}
M(S,\xi,\eta,P)
= \frac{1}{2}\sqrt{\frac{S}{\pi}}
+ \frac{4\pi P}{3}\left(\frac{S}{\pi}\right)^{\!3/2}
+ \frac{\xi}{6}\cdot
\frac{\eta^{2}+3\eta\sqrt{S/\pi}+3S/\pi}
     {\!\left(\eta+\sqrt{S/\pi}\right)^{\!3}}\,.
\label{eq:Mfull}
\end{equation}
All subsequent 
thermodynamic quantities and the Smarr formula are derived from 
this single relation.\\
\newline
Next, we determine the scaling dimensions of all thermodynamic 
variables by requiring $M$ to be a generalised homogeneous 
function. 
Consider the one-parameter
family of rescalings
\begin{equation}
r_{h} \;\longrightarrow\; l^{\,a}\,r_{h},\qquad
\eta   \;\longrightarrow\; l^{\,a}\,\eta,\qquad
\xi    \;\longrightarrow\; l^{\,b}\,\xi,\qquad
P      \;\longrightarrow\; l^{\,h}\,P,
\label{eq:rescaling}
\end{equation}
and require $M\to l^{n}M$ for some degree $n$.
Since $\eta$ and $r_h$ appear only 
in the combination $(\eta+r_h)^n$, they must carry identical 
length dimensions, and since $l$ is the AdS radius it also 
carries dimension of length $[\eta]=[r_h]=a$. 
Under \eqref{eq:rescaling}, each term in \eqref{M_rh_app} transforms as
follows. 
\begin{align}
\frac{r_h}{2}
&\;\longrightarrow\;
\frac{l^a r_{h}}{2}\,,
\label{eq:term1}
\\[6pt]
\frac{4\pi P\,r_h^3}{3}
&\;\longrightarrow\;
\frac{4\pi\, l^h P\cdot l^{3a}r_{h}^3}{3}
= l^{h+3a}\cdot\frac{4\pi P r_{h}^3}{3}\,,
\label{eq:term2}
\\[6pt]
\frac{\xi}{6}\cdot
\frac{\eta^2+3\eta r_++3r_+^2}{(\eta+r_+)^3}
&\;\longrightarrow\;
\frac{l^b\xi}{6}\cdot
\frac{l^{2a}(\eta^2+3\eta r_++3r_+^2)}{l^{3a}(\eta+r_+)^3}
= l^{b-a}\cdot
\frac{\xi}{6}\cdot
\frac{\eta^2+3\eta r_++3r_+^2}{(\eta+r_+)^3}\,.
\label{eq:term3}
\end{align}
For all three terms to transform with the same overall power of $l$
(equal to $n$), we require
\begin{equation}
n = a = h + 3a = b - a\,.
\end{equation}
This gives the three conditions:
\begin{equation}
n = a,\qquad h = -2a,\qquad b = 2a\,.
\label{eq:conditions}
\end{equation}
Setting the free parameter $a = 1$, we obtain the scaling dimensions:
\begin{equation}
[r_{+}] = 1,\quad
[\eta]   = 1,\quad
[\xi]    = 2,\quad
[P]      = -2,\quad
[M]      = 1\,.
\label{dimensions}
\end{equation}
 Equivalently, in terms of $S=\pi r_+^2$:
\begin{equation}
[S] = 2,\quad
[\eta] = 1,\quad
[\xi] = 2,\quad
[P] = -2,\quad
[M] = 1\,.
\label{dimensions}
\end{equation}
The mass is therefore a generalised homogeneous function of
degree $n=1$:
\begin{equation}
M\!\left(l^{2}S,\;l^{2}\xi,\;l\,\eta,\;l^{-2}P\right)
= l^{1}\cdot M(S,\xi,\eta,P)\,.
\label{eq:homogeneity}
\end{equation}
\newline
According to Euler's identity for a homogeneous function with degree $n$, we have
\begin{equation}\label{euler}
n\cdot f \;=\; \sum_{i=1}^{m} a_i\,x_i\,
\frac{\partial f}{\partial x_i}\,.
\end{equation}

\medskip

Applying \eqref{euler} to $M(S,P,\xi,\eta)$ with $n=1$ and 
scaling exponents $(a_S,a_P,a_\xi,a_\eta)=(2,-2,2,1)$ from 
\eqref{dimensions}:
\begin{equation}\label{euler_applied}
M
= 2\,S\,\frac{\partial M}{\partial S}
\;+\;(-2)\,P\,\frac{\partial M}{\partial P}
\;+\; 2\,\xi\,\frac{\partial M}{\partial\xi}
\;+\; 1\cdot\eta\,\frac{\partial M}{\partial\eta}\,.
\end{equation}
This is a pure mathematical identity valid for all 
$(S,P,\xi,\eta)$ and identifying each partial derivative with a 
thermodynamic conjugate would produce the Smarr formula.
\newline
The first law of black hole thermodynamics for this system is
\begin{equation}\label{firstlaw}
dM = T_H\,dS + V\,dP + \Psi_\xi\,d\xi + \Psi_\eta\,d\eta\,,
\end{equation}
where the conjugates are defined as partial derivatives of 
$M$ at constant remaining variables:
\begin{equation}\label{conjugates_def}
T_H = \left(\frac{\partial M}{\partial S}\right)_{P,\xi,\eta},\qquad
V   = \left(\frac{\partial M}{\partial P}\right)_{S,\xi,\eta},\qquad
\Psi_\xi = \left(\frac{\partial M}{\partial\xi}\right)_{S,P,\eta},\qquad
\Psi_\eta= \left(\frac{\partial M}{\partial\eta}\right)_{S,P,\xi}.
\end{equation}

\medskip
\noindent\textbf{Temperature.}
From the surface gravity $T_H = F'(r_h)/4\pi$, using 
\eqref{M_rh_app} we get
\begin{align}
T_{H} = \left.\frac{\partial M}{\partial S}\right|_{\xi,\eta,P}
  &= \frac{F'(r_h)}{4\pi}
  = \frac{1}{2\pi r_h}\frac{\partial M}{\partial r_h}\bigg|_{\xi,\eta,P}\, \\
  &=\frac{1}{4\pi r_h}
    \left[
    1 + 8\pi P\,r_h^2
    - \frac{\xi\,r_h^2}{(\eta+r_h)^4}
    \right].
\label{TH_app}
\end{align}

\medskip
\noindent\textbf{Thermodynamic volume.}
Differentiating $M$ with respect to $P$ at fixed 
$(S,\xi,\eta)$, i.e., at fixed $r_h$:
\begin{equation}\label{V_app}
V = \left.\frac{\partial M}{\partial P}\right|_{S,\xi,\eta}
  = \frac{4\pi r_h^3}{3}\,,
\end{equation}
which is the standard geometric volume. Together with 
$S=\pi r_h^2$ this gives
\begin{equation}\label{S_V_app}
S = \pi r_h^2,\qquad
V = \frac{4\pi}{3}r_h^3,\qquad
P = \frac{3}{8\pi l^2}\,.
\end{equation}

\medskip
\noindent\textbf{Deformation conjugates.}
Differentiating \eqref{M_rh_app} with respect to $\xi$ and 
$\eta$ at fixed $r_h$ (equivalently, at fixed $S$):
\begin{align}
\Psi_\xi
&= \left.\frac{\partial M}{\partial \xi}\right|_{S,\eta,P}
 = \frac{1}{6}\,
   \frac{\eta^2+3\eta r_h+3r_h^2}{(\eta+r_h)^3}\,,
\label{Psi_xi_app}
\\[8pt]
\Psi_\eta
&=  \left.\frac{\partial M}{\partial \eta}\right|_{S,\xi,P}
= -\frac{\xi\!\left(\eta^2+4\eta r_h+6r_h^2\right)}
        {6\left(\eta+r_h\right)^4}\,.
\label{Psi_eta_app}
\end{align}
\medskip
\newline
\textbf{ The Smarr Formula}\\
\noindent Substituting the conjugates 
\eqref{TH_app}, \eqref{V_app}, \eqref{Psi_xi_app}, 
\eqref{Psi_eta_app} into the Euler identity 
\eqref{euler_applied}:
\begin{equation}\label{Smarr_general}
M \;=\; 2\,T_H\,S \;-\; 2\,P\,V 
    \;+\; 2\,\xi\,\Psi_\xi \;+\; \eta\,\Psi_\eta\,.
\end{equation}
This is the Smarr formula for the deformed 
AdS--Schwarzschild black hole. 
 (We verified the equality symbolically on Mathematica; the algebraic cancellations are straightforward but somewhat long, and we therefore omit the intermediate expanded polynomial.)
 
\section*{Declaration}
During the preparation of this work, the authors used ChatGPT (OpenAI) to assist in drafting and debugging preliminary analysis code. All generated code was carefully reviewed, tested, and validated by the authors. The authors take full responsibility for the correctness of the code and the results presented in this manuscript.

 \bibliographystyle{JHEP}
 \bibliography{defads.bib}

@article{Khosravipoor:2023jsl,
    author = "Khosravipoor, Mohammad Reza and Farhoudi, Mehrdad",
    title = "{Thermodynamics of deformed AdS-Schwarzschild black hole}",
    eprint = "2311.02456",
    archivePrefix = "arXiv",
    primaryClass = "gr-qc",
    doi = "10.1140/epjc/s10052-023-12222-2",
    journal = "Eur. Phys. J. C",
    volume = "83",
    number = "11",
    pages = "1045",
    year = "2023"
}

@article{Witten:1998qj,
    author = "Witten, Edward",
    title = "{Anti de Sitter space and holography}",
    eprint = "hep-th/9802150",
    archivePrefix = "arXiv",
    reportNumber = "IASSNS-HEP-98-15",
    doi = "10.4310/ATMP.1998.v2.n2.a2",
    journal = "Adv. Theor. Math. Phys.",
    volume = "2",
    pages = "253--291",
    year = "1998"
}

@article{Gubser:1998bc,
    author = "Gubser, S. S. and Klebanov, Igor R. and Polyakov, Alexander M.",
    title = "{Gauge theory correlators from noncritical string theory}",
    eprint = "hep-th/9802109",
    archivePrefix = "arXiv",
    reportNumber = "PUPT-1767",
    doi = "10.1016/S0370-2693(98)00377-3",
    journal = "Phys. Lett. B",
    volume = "428",
    pages = "105--114",
    year = "1998"
}

@article{Altamirano:2013ane,
    author = "Altamirano, Natacha and Kubiznak, David and Mann, Robert B.",
    title = "{Reentrant phase transitions in rotating anti\textendash{}de Sitter black holes}",
    eprint = "1306.5756",
    archivePrefix = "arXiv",
    primaryClass = "hep-th",
    reportNumber = "PI-STRONGGRV-332",
    doi = "10.1103/PhysRevD.88.101502",
    journal = "Phys. Rev. D",
    volume = "88",
    number = "10",
    pages = "101502",
    year = "2013"
}

@article{Ahmed:2023dnh,
    author = "Ahmed, Moaathe Belhaj and Cong, Wan and Kubiznak, David and Mann, Robert B. and Visser, Manus R.",
    title = "{Holographic CFT phase transitions and criticality for rotating AdS black holes}",
    eprint = "2305.03161",
    archivePrefix = "arXiv",
    primaryClass = "hep-th",
    doi = "10.1007/JHEP08(2023)142",
    journal = "JHEP",
    volume = "08",
    pages = "142",
    year = "2023"
}

@article{Cong:2021jgb,
    author = "Cong, Wan and Kubiznak, David and Mann, Robert B. and Visser, Manus R.",
    title = "{Holographic CFT phase transitions and criticality for charged AdS black holes}",
    eprint = "2112.14848",
    archivePrefix = "arXiv",
    primaryClass = "hep-th",
    doi = "10.1007/JHEP08(2022)174",
    journal = "JHEP",
    volume = "08",
    pages = "174",
    year = "2022"
}

@article{Hawking:1982dh,
    author = "Hawking, S. W. and Page, Don N.",
    title = "{Thermodynamics of Black Holes in anti-De Sitter Space}",
    reportNumber = "PRINT-83-0019 (CAMBRIDGE)",
    doi = "10.1007/BF01208266",
    journal = "Commun. Math. Phys.",
    volume = "87",
    pages = "577",
    year = "1983"
}

@article{Witten:1998zw,
    author = "Witten, Edward",
    editor = "Bergstrom, L. and Lindstrom, U.",
    title = "{Anti-de Sitter space, thermal phase transition, and confinement in gauge theories}",
    eprint = "hep-th/9803131",
    archivePrefix = "arXiv",
    reportNumber = "IASSNS-HEP-98-21",
    doi = "10.4310/ATMP.1998.v2.n3.a3",
    journal = "Adv. Theor. Math. Phys.",
    volume = "2",
    pages = "505--532",
    year = "1998"
}

@article{Kubiznak:2016qmn,
    author = "Kubiznak, David and Mann, Robert B. and Teo, Mae",
    title = "{Black hole chemistry: thermodynamics with Lambda}",
    eprint = "1608.06147",
    archivePrefix = "arXiv",
    primaryClass = "hep-th",
    doi = "10.1088/1361-6382/aa5c69",
    journal = "Class. Quant. Grav.",
    volume = "34",
    number = "6",
    pages = "063001",
    year = "2017"
}

@article{Kastor:2009wy,
    author = "Kastor, David and Ray, Sourya and Traschen, Jennie",
    title = "{Enthalpy and the Mechanics of AdS Black Holes}",
    eprint = "0904.2765",
    archivePrefix = "arXiv",
    primaryClass = "hep-th",
    doi = "10.1088/0264-9381/26/19/195011",
    journal = "Class. Quant. Grav.",
    volume = "26",
    pages = "195011",
    year = "2009"
}

@article{Dolan:2011xt,
    author = "Dolan, Brian P.",
    title = "{Pressure and volume in the first law of black hole thermodynamics}",
    eprint = "1106.6260",
    archivePrefix = "arXiv",
    primaryClass = "gr-qc",
    doi = "10.1088/0264-9381/28/23/235017",
    journal = "Class. Quant. Grav.",
    volume = "28",
    pages = "235017",
    year = "2011"
}

@article{Dolan:2010ha,
    author = "Dolan, Brian P.",
    title = "{The cosmological constant and the black hole equation of state}",
    eprint = "1008.5023",
    archivePrefix = "arXiv",
    primaryClass = "gr-qc",
    reportNumber = "DIAS-STP-10-10",
    doi = "10.1088/0264-9381/28/12/125020",
    journal = "Class. Quant. Grav.",
    volume = "28",
    pages = "125020",
    year = "2011"
}

@article{Cvetic:2010jb,
    author = "Cvetic, M. and Gibbons, G. W. and Kubiznak, D. and Pope, C. N.",
    title = "{Black Hole Enthalpy and an Entropy Inequality for the Thermodynamic Volume}",
    eprint = "1012.2888",
    archivePrefix = "arXiv",
    primaryClass = "hep-th",
    reportNumber = "DAMTP-2010-125, MIFPA-10-56",
    doi = "10.1103/PhysRevD.84.024037",
    journal = "Phys. Rev. D",
    volume = "84",
    pages = "024037",
    year = "2011"
}

@article{Kubiznak:2014zwa,
    author = "Kubiznak, David and Mann, Robert B.",
    editor = "Dasgupta, Arundhati",
    title = "{Black hole chemistry}",
    eprint = "1404.2126",
    archivePrefix = "arXiv",
    primaryClass = "gr-qc",
    doi = "10.1139/cjp-2014-0465",
    journal = "Can. J. Phys.",
    volume = "93",
    number = "9",
    pages = "999--1002",
    year = "2015"
}

@article{Kastor:2010gq,
    author = "Kastor, David and Ray, Sourya and Traschen, Jennie",
    title = "{Smarr Formula and an Extended First Law for Lovelock Gravity}",
    eprint = "1005.5053",
    archivePrefix = "arXiv",
    primaryClass = "hep-th",
    doi = "10.1088/0264-9381/27/23/235014",
    journal = "Class. Quant. Grav.",
    volume = "27",
    pages = "235014",
    year = "2010"
}

@article{Zou:2013owa,
    author = "Zou, De-Cheng and Zhang, Shao-Jun and Wang, Bin",
    title = "{Critical behavior of Born-Infeld AdS black holes in the extended phase space thermodynamics}",
    eprint = "1311.7299",
    archivePrefix = "arXiv",
    primaryClass = "hep-th",
    doi = "10.1103/PhysRevD.89.044002",
    journal = "Phys. Rev. D",
    volume = "89",
    number = "4",
    pages = "044002",
    year = "2014"
}

@article{Altamirano:2013uqa,
    author = "Altamirano, Natacha and Kubiz{\v{n}}{\'a}k, David and Mann, Robert B. and Sherkatghanad, Zeinab",
    title = "{Kerr-AdS analogue of triple point and solid/liquid/gas phase transition}",
    eprint = "1308.2672",
    archivePrefix = "arXiv",
    primaryClass = "hep-th",
    doi = "10.1088/0264-9381/31/4/042001",
    journal = "Class. Quant. Grav.",
    volume = "31",
    pages = "042001",
    year = "2014"
}

@article{Cai:2013qga,
    author = "Cai, Rong-Gen and Cao, Li-Ming and Li, Li and Yang, Run-Qiu",
    title = "{P-V criticality in the extended phase space of Gauss-Bonnet black holes in AdS space}",
    eprint = "1306.6233",
    archivePrefix = "arXiv",
    primaryClass = "gr-qc",
    doi = "10.1007/JHEP09(2013)005",
    journal = "JHEP",
    volume = "09",
    pages = "005",
    year = "2013"
}

@article{Xu:2014tja,
    author = "Xu, Hao and Xu, Wei and Zhao, Liu",
    title = "{Extended phase space thermodynamics for third order Lovelock black holes in diverse dimensions}",
    eprint = "1405.4143",
    archivePrefix = "arXiv",
    primaryClass = "gr-qc",
    doi = "10.1140/epjc/s10052-014-3074-1",
    journal = "Eur. Phys. J. C",
    volume = "74",
    number = "9",
    pages = "3074",
    year = "2014"
}

@article{Dolan:2014vba,
    author = "Dolan, Brian P. and Kostouki, Anna and Kubiznak, David and Mann, Robert B.",
    title = "{Isolated critical point from Lovelock gravity}",
    eprint = "1407.4783",
    archivePrefix = "arXiv",
    primaryClass = "hep-th",
    doi = "10.1088/0264-9381/31/24/242001",
    journal = "Class. Quant. Grav.",
    volume = "31",
    number = "24",
    pages = "242001",
    year = "2014"
}

@article{Zou:2016sab,
    author = "Zou, De-Cheng and Yue, Ruihong and Zhang, Ming",
    title = "{Reentrant phase transitions of higher-dimensional AdS black holes in dRGT massive gravity}",
    eprint = "1612.08056",
    archivePrefix = "arXiv",
    primaryClass = "gr-qc",
    doi = "10.1140/epjc/s10052-017-4822-9",
    journal = "Eur. Phys. J. C",
    volume = "77",
    number = "4",
    pages = "256",
    year = "2017"
}

@article{Ovalle:2017fgl,
    author = "Ovalle, Jorge",
    title = "{Decoupling gravitational sources in general relativity: from perfect to anisotropic fluids}",
    eprint = "1704.05899",
    archivePrefix = "arXiv",
    primaryClass = "gr-qc",
    doi = "10.1103/PhysRevD.95.104019",
    journal = "Phys. Rev. D",
    volume = "95",
    number = "10",
    pages = "104019",
    year = "2017"
}

@article{Ovalle:2018gic,
    author = "Ovalle, J.",
    title = "{Decoupling gravitational sources in general relativity: The extended case}",
    eprint = "1812.03000",
    archivePrefix = "arXiv",
    primaryClass = "gr-qc",
    doi = "10.1016/j.physletb.2018.11.029",
    journal = "Phys. Lett. B",
    volume = "788",
    pages = "213--218",
    year = "2019"
}

@article{Ovalle:2023ref,
    author = "Ovalle, Jorge and Casadio, Roberto and Giusti, Andrea",
    title = "{Regular hairy black holes through Minkowski deformation}",
    eprint = "2304.03263",
    archivePrefix = "arXiv",
    primaryClass = "gr-qc",
    doi = "10.1016/j.physletb.2023.138085",
    journal = "Phys. Lett. B",
    volume = "844",
    pages = "138085",
    year = "2023"
}

@article{Kubiznak:2012wp,
    author = "Kubiznak, David and Mann, Robert B.",
    title = "{P-V criticality of charged AdS black holes}",
    eprint = "1205.0559",
    archivePrefix = "arXiv",
    primaryClass = "hep-th",
    doi = "10.1007/JHEP07(2012)033",
    journal = "JHEP",
    volume = "07",
    pages = "033",
    year = "2012"
}

@article{Gogoi:2025rcn,
    author = "Gogoi, Dhruba Jyoti and Hazarika, Poppy and Bora, Jyatsnasree and Changmai, Ranjan",
    title = "{Thermodynamics of Deformed AdS-Schwarzschild Black Holes in the Presence of Thermal Fluctuations}",
    eprint = "2501.15629",
    archivePrefix = "arXiv",
    primaryClass = "hep-th",
    doi = "10.1002/prop.70004",
    journal = "Fortsch. Phys.",
    volume = "73",
    number = "6",
    pages = "e70004",
    year = "2025"
}

@article{Avalos:2023ywb,
    author = "Avalos, R. and Bargue{\~n}o, Pedro and Contreras, E.",
    title = "{A Static and Spherically Symmetric Hairy Black Hole in the Framework of the Gravitational Decoupling}",
    eprint = "2303.04119",
    archivePrefix = "arXiv",
    primaryClass = "gr-qc",
    doi = "10.1002/prop.202200171",
    journal = "Fortsch. Phys.",
    volume = "71",
    number = "4-5",
    pages = "2200171",
    year = "2023"
}

@article{Tarrio:2011de,
    author = "Tarrio, Javier and Vandoren, Stefan",
    title = "{Black holes and black branes in Lifshitz spacetimes}",
    eprint = "1105.6335",
    archivePrefix = "arXiv",
    primaryClass = "hep-th",
    reportNumber = "ITF-UU-11-19, SPIN-11-14",
    doi = "10.1007/JHEP09(2011)017",
    journal = "JHEP",
    volume = "09",
    pages = "017",
    year = "2011"
}

@article{Pedraza:2018eey,
    author = "Pedraza, Juan F. and Sybesma, Watse and Visser, Manus R.",
    title = "{Hyperscaling violating black holes with spherical and hyperbolic horizons}",
    eprint = "1807.09770",
    archivePrefix = "arXiv",
    primaryClass = "hep-th",
    doi = "10.1088/1361-6382/ab0094",
    journal = "Class. Quant. Grav.",
    volume = "36",
    number = "5",
    pages = "054002",
    year = "2019"
}

@article{Karch:2015rpa,
    author = "Karch, Andreas and Robinson, Brandon",
    title = "{Holographic Black Hole Chemistry}",
    eprint = "1510.02472",
    archivePrefix = "arXiv",
    primaryClass = "hep-th",
    doi = "10.1007/JHEP12(2015)073",
    journal = "JHEP",
    volume = "12",
    pages = "073",
    year = "2015"
}

@article{Visser:2021eqk,
    author = "Visser, Manus R.",
    title = "{Holographic thermodynamics requires a chemical potential for color}",
    eprint = "2101.04145",
    archivePrefix = "arXiv",
    primaryClass = "hep-th",
    doi = "10.1103/PhysRevD.105.106014",
    journal = "Phys. Rev. D",
    volume = "105",
    number = "10",
    pages = "106014",
    year = "2022"
}

@article{Cong:2021fnf,
    author = "Cong, Wan and Kubiznak, David and Mann, Robert B.",
    title = "{Thermodynamics of AdS Black Holes: Critical Behavior of the Central Charge}",
    eprint = "2105.02223",
    archivePrefix = "arXiv",
    primaryClass = "hep-th",
    doi = "10.1103/PhysRevLett.127.091301",
    journal = "Phys. Rev. Lett.",
    volume = "127",
    number = "9",
    pages = "091301",
    year = "2021"
}

@article{Kastor:2014dra,
    author = "Kastor, David and Ray, Sourya and Traschen, Jennie",
    title = "{Chemical Potential in the First Law for Holographic Entanglement Entropy}",
    eprint = "1409.3521",
    archivePrefix = "arXiv",
    primaryClass = "hep-th",
    reportNumber = "ACFI-T14-17",
    doi = "10.1007/JHEP11(2014)120",
    journal = "JHEP",
    volume = "11",
    pages = "120",
    year = "2014"
}

@article{Maldacena:1997re,
    author = "Maldacena, Juan Martin",
    title = "{The Large $N$ limit of superconformal field theories and supergravity}",
    eprint = "hep-th/9711200",
    archivePrefix = "arXiv",
    reportNumber = "HUTP-97-A097, HUTP-98-A097",
    doi = "10.4310/ATMP.1998.v2.n2.a1",
    journal = "Adv. Theor. Math. Phys.",
    volume = "2",
    pages = "231--252",
    year = "1998"
}

@article{Susskind:1994vu,
    author = "Susskind, Leonard",
    title = "{The World as a hologram}",
    eprint = "hep-th/9409089",
    archivePrefix = "arXiv",
    reportNumber = "SU-ITP-94-33",
    doi = "10.1063/1.531249",
    journal = "J. Math. Phys.",
    volume = "36",
    pages = "6377--6396",
    year = "1995"
}

@article{Bekenstein:1973ur,
    author = "Bekenstein, Jacob D.",
    title = "{Black holes and entropy}",
    doi = "10.1103/PhysRevD.7.2333",
    journal = "Phys. Rev. D",
    volume = "7",
    pages = "2333--2346",
    year = "1973"
}

@article{Bardeen:1973gs,
    author = "Bardeen, James M. and Carter, B. and Hawking, S. W.",
    title = "{The Four laws of black hole mechanics}",
    doi = "10.1007/BF01645742",
    journal = "Commun. Math. Phys.",
    volume = "31",
    pages = "161--170",
    year = "1973"
}

@article{Witten:2024upt,
    author = "Witten, Edward",
    title = "{Introduction to black hole thermodynamics}",
    eprint = "2412.16795",
    archivePrefix = "arXiv",
    primaryClass = "hep-th",
    doi = "10.1140/epjp/s13360-025-06288-y",
    journal = "Eur. Phys. J. Plus",
    volume = "140",
    number = "5",
    pages = "430",
    year = "2025"
}

@article{Wald:1999vt,
    author = "Wald, Robert M.",
    title = "{The thermodynamics of black holes}",
    eprint = "gr-qc/9912119",
    archivePrefix = "arXiv",
    doi = "10.12942/lrr-2001-6",
    journal = "Living Rev. Rel.",
    volume = "4",
    pages = "6",
    year = "2001"
}

@article{Gheorghiu:2013jha,
    author = "Gheorghiu, Tamara and Vacaru, Olivia and Vacaru, Sergiu I.",
    title = "{Off-Diagonal Deformations of Kerr Black Holes in Einstein and Modified Massive Gravity and Higher Dimensions}",
    eprint = "1312.4844",
    archivePrefix = "arXiv",
    primaryClass = "gr-qc",
    doi = "10.1140/epjc/s10052-014-3152-4",
    journal = "Eur. Phys. J. C",
    volume = "74",
    number = "12",
    pages = "3152",
    year = "2014"
}

@article{Estrada:2018vrl,
    author = "Estrada, Milko and Prado, Reginaldo",
    title = "{The Gravitational decoupling method: the higher dimensional case to find new analytic solutions}",
    eprint = "1809.03591",
    archivePrefix = "arXiv",
    primaryClass = "gr-qc",
    doi = "10.1140/epjp/i2019-12555-8",
    journal = "Eur. Phys. J. Plus",
    volume = "134",
    number = "4",
    pages = "168",
    year = "2019"
}

@article{Maurya:2022wwa,
    author = "Maurya, S. K. and Newton Singh, Ksh. and Lohakare, Santosh V. and Mishra, B.",
    title = "{Anisotropic Strange Star Model Beyond Standard Maximum Mass Limit by Gravitational Decoupling in f(Q)$f(Q)$ Gravity}",
    eprint = "2208.04735",
    archivePrefix = "arXiv",
    primaryClass = "gr-qc",
    doi = "10.1002/prop.202200061",
    journal = "Fortsch. Phys.",
    volume = "70",
    number = "11",
    pages = "2200061",
    year = "2022"
}

@article{Brown:2017wpl,
    author = "Brown, Jon and Cole, Alex and Shiu, Gary and Cottrell, William",
    title = "{Gravitational decoupling and the Picard-Lefschetz approach}",
    eprint = "1710.04737",
    archivePrefix = "arXiv",
    primaryClass = "hep-th",
    reportNumber = "MAD-TH-17-06",
    doi = "10.1103/PhysRevD.97.025002",
    journal = "Phys. Rev. D",
    volume = "97",
    number = "2",
    pages = "025002",
    year = "2018"
}

@article{Ahmed:2023snm,
    author = "Ahmed, Moaathe Belhaj and Cong, Wan and Kubiz{\v{n}}{\'a}k, David and Mann, Robert B. and Visser, Manus R.",
    title = "{Holographic Dual of Extended Black Hole Thermodynamics}",
    eprint = "2302.08163",
    archivePrefix = "arXiv",
    primaryClass = "hep-th",
    doi = "10.1103/PhysRevLett.130.181401",
    journal = "Phys. Rev. Lett.",
    volume = "130",
    number = "18",
    pages = "181401",
    year = "2023"
}

@article{Rodrigues:2022qdp,
    author = "Rodrigues, Manuel E. and de S. Silva, Marcos V. and Vieira, Henrique A.",
    title = "{Bardeen-Kiselev black hole with a cosmological constant}",
    eprint = "2203.04965",
    archivePrefix = "arXiv",
    primaryClass = "gr-qc",
    doi = "10.1103/PhysRevD.105.084043",
    journal = "Phys. Rev. D",
    volume = "105",
    number = "8",
    pages = "084043",
    year = "2022"
}

@article{Dehyadegari:2017hvd,
    author = "Dehyadegari, Amin and Sheykhi, Ahmad",
    title = "{Reentrant phase transition of Born-Infeld-AdS black holes}",
    eprint = "1711.01151",
    archivePrefix = "arXiv",
    primaryClass = "gr-qc",
    doi = "10.1103/PhysRevD.98.024011",
    journal = "Phys. Rev. D",
    volume = "98",
    number = "2",
    pages = "024011",
    year = "2018"
}
\end{document}